\documentclass[iop]{emulateapj}

\usepackage{natbib}
\usepackage{graphicx}
\usepackage{subfigure}
\usepackage{multirow}
\usepackage{amsmath}
\mathchardef\mhyphen="2D 

\newcommand{\Msun}{\mbox{$M_{\odot}$}}
\newcommand{\msun}{\mbox{$M_{\odot}$}}

\newcommand{\Teff}{\mbox{$T_{\rm eff}$}}
\newcommand{\teff}{\mbox{$T_{\rm eff}$}}

\newcommand{\logg}{\mbox{$\log g$}}

\newcommand{\feh}{\mbox{[Fe/H]}}
\newcommand{\narratio}{\mbox{$N_{\rm TP-AGB}/N_{\rm RGB}$}}
\newcommand{\OCT}{\mbox{$\dot{M}_{\rm pre\mhyphen dust}^{\rm R75}$}}
\newcommand{\NOV}{\mbox{$\dot{M}_{\rm pre\mhyphen dust}^{\rm mSC05}$}}
\newcommand{\NOVeta}{\mbox{$\dot{M}_{\rm pre\mhyphen dust}^{\eta=0}$}}
\newcommand{\SC}{\mbox{$\dot{M}_{\rm pre\mhyphen dust}^{\rm SC05}$}}

\defcitealias{Girardi2010}{G10}

\slugcomment{Draft \date}

\shorttitle{Constraining TP-AGB models with HST data}
\shortauthors{Rosenfield et al.}

\begin{document}

\title{Evolution of Thermally Pulsing Asymptotic Giant Branch Stars IV. Constraining Mass-Loss \& Lifetimes of Low Mass, Low Metallicity AGB Stars. \footnote{Based on observations made with the NASA/ESA Hubble Space Telescope, obtained from the Data Archive at the Space Telescope Science Institute, which is operated by the Association of Universities for Research in Astronomy, Inc., under NASA contract NAS 5-26555.}}

\author{
Philip Rosenfield\altaffilmark{1, 2},
Paola Marigo\altaffilmark{2},
L\'eo Girardi\altaffilmark{3},
Julianne J.\ Dalcanton\altaffilmark{1},
Alessandro Bressan\altaffilmark{4},
Marco Gullieuszik\altaffilmark{3},
Daniel Weisz \altaffilmark{1, 5, 6},
Benjamin F.\ Williams\altaffilmark{1},
Andrew Dolphin \altaffilmark{7},
Bernhard Aringer  \altaffilmark{8}
}
\altaffiltext{1}{Department of Astronomy, University of Washington, Box 351580, Seattle, WA 98195, USA}
\altaffiltext{2}{Department of Physics and Astronomy G. Galilei, University of Padova, Vicolo dell’Osservatorio 3, I-35122 Padova, Italy}
\altaffiltext{3}{Osservatorio Astronomico di Padova -- INAF, Vicolo dell'Osservatorio 5, I-35122 Padova, Italy}
\altaffiltext{4}{Astrophysics Sector, SISSA, Via Bonomea 265, I-34136 Trieste, Italy}
\altaffiltext{5}{Department of Astronomy, University of California at Santa Cruz, 1156 High Street, Santa Cruz, CA, 95064 USA}
\altaffiltext{6}{Hubble Fellow}
\altaffiltext{7}{Raytheon Company, 1151 East Hermans Road, Tucson, AZ 85756, USA}
\altaffiltext{8}{University of Vienna, Department of Astrophysics, Turkenschanzstra\ss e 17, A-1180 Wien, Austria}

\begin{abstract}
The evolution and lifetimes of thermally pulsating asymptotic giant branch (TP-AGB) stars suffer from significant uncertainties. In this work, we analyze the numbers and luminosity functions of TP-AGB stars in six quiescent, low metallicity  ([Fe/H] $\lesssim -0.86$) galaxies taken from the ANGST sample, using HST photometry in both optical and near-infrared filters. The galaxies contain over 1000 TP-AGB stars (at least 60 per field). We compare the observed TP-AGB luminosity functions and relative numbers of TP-AGB and RGB stars, \narratio, to models generated from different suites of TP-AGB evolutionary tracks after adopting star formation histories (SFH) derived from the HST deep optical observations. We test various mass-loss prescriptions that differ in their treatments of mass-loss before the onset of dust-driven winds (pre-dust). These comparisons confirm that pre-dust mass-loss is important, since models that neglect pre-dust mass-loss fail to explain the observed \narratio\ ratio or the luminosity functions. In contrast, models with more efficient pre-dust mass-loss produce results consistent with observations.  We find that for [Fe/H]$\lesssim-0.86$, lower mass TP-AGB stars ($M\lesssim1\msun$) must have lifetimes of $\sim0.5$ Myr and higher masses ($M\lesssim 3\msun$) must have lifetimes $\lesssim 1.2$ Myr. In addition, assuming our best-fitting mass-loss prescription, we show that the third dredge up has no significant effect on TP-AGB lifetimes in this mass and metallicity range.
\end{abstract}

\keywords{stars: general}

\section{Introduction}

Understanding the asymptotic giant branch (AGB) phase of stellar evolution is critical for various aspects of galaxy evolution. It affects the interpretation of the integrated light of distant galaxies,  as well the origin of cosmic dust and the chemical enrichment of the interstellar medium. Although AGB lifetimes are generally very short (less than a few Myr), their high luminosities may contribute significantly to the integrated spectral energy distribution of galaxies \citep[e.g.,][]{Maraston2005, Conroy2009, Melbourne2012}, particularly at redder wavelengths.

Despite its importance, the AGB phase is the most uncertain evolutionary phase of low and intermediate mass ($\sim1-8 \msun$) stars, primarily due its complexity \citep[for a review, see][]{Herwig2005}. AGB stars have two burning shells around an electron-degenerate core. As they evolve, AGB stars undergo a series of He-shell flashes (called thermal pulses, TP), which cause the base of the convective envelope to deepen within the star. Depending on the stellar mass and metallicity, the base of the convective envelope may reach past the intershell region, bringing nuclearly-processed material from the interior up to the surface (the so-called third dredge-up),  with consequent enrichment in $^{4}$He, $^{12}$C, $^{19}$F, $^{22}$Ne, $^{25}$Mg, $^{26}$Al and other isotopes, plus the heavy elements produced by the slow-neutron capture nucleosynthesis \citep[e.g.,][]{Forestini97, Busso99, Herwig2005}. Additionally, the most massive and luminous AGB stars (with initial masses $\gtrsim 3\, M_{\odot}$ depending on metallicity) experience a process known as hot-bottom burning (HBB) during the quiescent interpulse periods. The innermost layers of the convective envelope become hot enough for the hydrogen-burning reactions to be activated, which results in a further modification of the surface  chemical composition (mainly via the CNO, NeNa, MgAl cycles) and a sizable steepening of the brightening rate along the AGB \citep[e.g.,][]{Bloecker91, Ventura05}. Eventually, the chemically enriched envelope is expelled into the interstellar medium through stellar winds  \citep{Marigo01, Marigo03a, Karakas09}.

At present, though a general picture of the evolution during the AGB phase is reasonably clear, the quantitative details of all key physical processes at work (third dredge-up, strength and nucleosynthesis of HBB, mass-loss, pulsation, and dust formation) are still out of focus. The problem can be tackled from various perspectives and with different strategies.  One approach is more linked to the underlying physics, involving detailed analysis of the physical processes and their control parameters (e.g. the dependence of HBB on the convection theory and nuclear network adopted, the role of overshooting for the efficiency of the third dredge-up, the sensitivity of pulsation models to the treatment of the subphotospheric convection, etc.). The nature of the other approach is more phenomenological, wherein one constrains  uncertain parameters through statistical studies that compare observational data of AGB samples with population synthesis simulations.

These two approaches are complementary. On one side, the results of population synthesis analysis can be extremely useful to the physics-based studies, as they provide the correct analysis for the development of improved theory. The calibration of the efficiency of the third dredge-up based on the observed luminosity functions of carbon stars is a fitting example \citep[][]{Groenewegen93, Marigo1999, Marigo2007}. On the other side, any relevant improvement in the physical description of the AGB  should be incorporated among the ingredients within population synthesis studies to enhance their predicting power. The adoption of variable molecular opacities to correctly describe the Hayashi lines of carbon stars, and hence interpret their position in the observed infrared color-magnitude diagrams, nicely illustrates this point \citep{Marigo2002, Marigo03b}.

As part of the ambitious goal to reach a satisfactory understanding of the AGB phase, in this study we opt for the statistical population synthesis approach. The major aim is to obtain  a calibration of the TP-AGB lifetimes as a function of the initial stellar mass in the low-metallicity regime. This calibration is important not only for the practical purposes of generating improved stellar tools useful to the community, but also for unveiling the behavior of uncertain processes, such as the mass-loss at low metallicity, outside of the regime probed in Milky Way (MW) and Magellanic Clouds (MCs) studies. 

\subsection{Recent efforts to calibrate TP-AGB lifetimes}
The most popular method to constrain TP-AGB lifetimes is through determining the relative number densities of evolutionary phases on a color-magnitude diagram (CMD). However, small numbers of observed TP-AGB stars can hamper the usefulness of the derived model constraints \citep[e.g.,][]{Frogel1990, VanLoon2005, Girardi2007, Boyer2009}, particularly in stellar clusters.

Several studies have carried out detailed analysis of individual clusters in the Large and Small Magellanic Clouds (LMC, SMC) and the MW \citep[e.g.,][]{Lebzelter2007, Kamath2010, vanLoon2006, Boyer2010}. \citet{Girardi2007} combined star counts from LMC and SMC clusters in age and metallicity bins to measure TP-AGB lifetimes with reduced uncertainties due to small number statistics. These studies have proven very useful for TP-AGB model calibrations in the specific ages and metallicities of the MW and MCs cluster populations. In the intermediate-metallicity regime ($Z=0.002 - 0.008$) \citet{Marigo2008} calibrated TP-AGB lifetimes based on the observed C-star and M-star luminosity functions in the MC fields and C- and M-star star counts in MC clusters \citep[using data from][]{Groenewegen2002, Girardi2007}. 

Recently, \citet{Girardi2013} presented a substantial issue regarding the calibration of the lifetimes of AGB stars in MC clusters. It is known that 1) the lifetimes of core He-burning stars as a function of mass have a discontinuity and sharp increase at and above $M_{\rm HeF}\sim 1.75\Msun$ where stars become able to burn Helium in a non-degenerate core \citep[e.g.,][]{Girardi2000} and 2) the lifetimes of AGB stars peak for masses of $\sim2\msun$ \citep[e.g,][]{Marigo2008, Weiss2009, Karakas2002}. These two effects conspire together to ``boost'' the number of TP-AGB stars observed in MC clusters of ages $\sim$1.5~Gyr, providing numbers far above those predicted by the fuel consumption theorem \citep{Renzini1986}.  In other words, the number of TP-AGB stars found on the CMD \emph{are not equally} proportional to the lifetime of stars in that phase \emph{ at all ages}. The issue becomes particularly serious given that {\em most} of the TP-AGB stars in MC clusters are found in those clusters with ages close to 1.5 Gyr.

Perhaps related to this issue, the derived lifetimes from TP-AGB stars in MC clusters were found to be overestimated when extrapolated to metallicities found in dwarf galaxies \citep{Gullieuszik2008, Held2010, Melbourne2010}. In other words, TP-AGB models calibrated only with MC observations do not necessarily describe AGB star populations at low metallicities. Not only must the MC calibration be reassessed, but model constraints are needed at even lower metallicities than the MCs.

In an attempt to extend the TP-AGB calibration to low metallicities, \citet[][hereafter, G10]{Girardi2010} further constrained the \citet{Marigo2008} TP-AGB models using optical observations of low-metallicity nearby, non-Local Group galaxies. These samples were from the ACS Nearby Galaxy Survey Treasury \citep[ANGST;][]{Dalcanton2009} and the Archival Nearby Galaxies: Reduce, Reuse, Recycle \citep[ANGRRR;][]{Gilbert2013} database. ANGST is a volume limited sample of $\sim 70$ non-Local Group  galaxies to $\lesssim 4$ Mpc, in which deep optical photometry allowed the measurement of their star formation histories (SFH). Using a subset of 12 dwarf galaxies from the ANGST sample that showed little to no recent star formation, \citetalias{Girardi2010} compared the luminosity function (LF) of each galaxy with those of simulations using the measured SFHs, and updated the mass-loss prescriptions for low-mass, low-metallicity, TP-AGB stars before the onset of dust driven winds (pre-dust winds). This correction resulted in good data-model agreement. Indeed, the revised TP-AGB models reproduced the initial-to-final mass relationship of the Galactic globular cluster M4 \citep{Kalirai2009}.

Using a completely different approach, \citet{Kalirai14} put constraints on the lifetimes and the core mass growth of intermediate-age TP-AGB stars (with initial masses in the range 1.6 $M_{\odot} - 3.8 M_{\odot}$) with slightly super-solar initial metallicity ($Z=0.02$), combining recent accurate measurements of white dwarf masses in the Hyades and Praesepe star clusters and new TP-AGB models \citep{Marigo2013}. 

\subsection{This work in context}
\label{sec_wic}
The method applied by \citetalias{Girardi2010} represents a promising approach given the increasing availability of deep photometric data from which the detailed SFHs of nearby galaxies can be derived. Its range of applicability, however, was initially limited by the optical filter-set available in ANGST.  The reddest filter in the ANGST survey is $F814W$, which is more affected by circumstellar dust than redder filters and which also requires higher bolometric corrections for cooler TP-AGB stars \citep[][]{Girardi2008}. This limitation allows for a complete AGB sample only for relatively hot ($\teff\gtrsim4000$~K) TP-AGB stars. Therefore, the method was applied only to a few very metal-poor dwarf galaxies without any sign of recent star formation.

To generate a complete sample of TP-AGB stars in galaxies with more recent star formation and higher metallicities (and hence cooler $\teff$),  near-infrared (NIR) filters are necessary. In a series of two papers, we build upon the method of \citetalias{Girardi2010} by calculating the LFs of modeled TP-AGB stars using the near-infrared $WFC3/HST$ ``snapshot'' follow-up survey to ANGST \citep[][hereafter AGB-SNAP]{Dalcanton2012}.  This sample includes 26 NIR fields of 23 ANGST galaxies, with 10,000's of TP-AGB stars. In principle, this filter set allows us to probe TP-AGB masses from 1 to 5 \msun\ with $\teff  > 3000$K. However, we are still limited by the NIR filterset for which it is not possible to robustly separate C-rich from O-rich AGB stars \citep[][]{Dalcanton2012, Boyer2013}.

\begin{figure*}
\begin{center}
\includegraphics[width=0.23\textwidth]{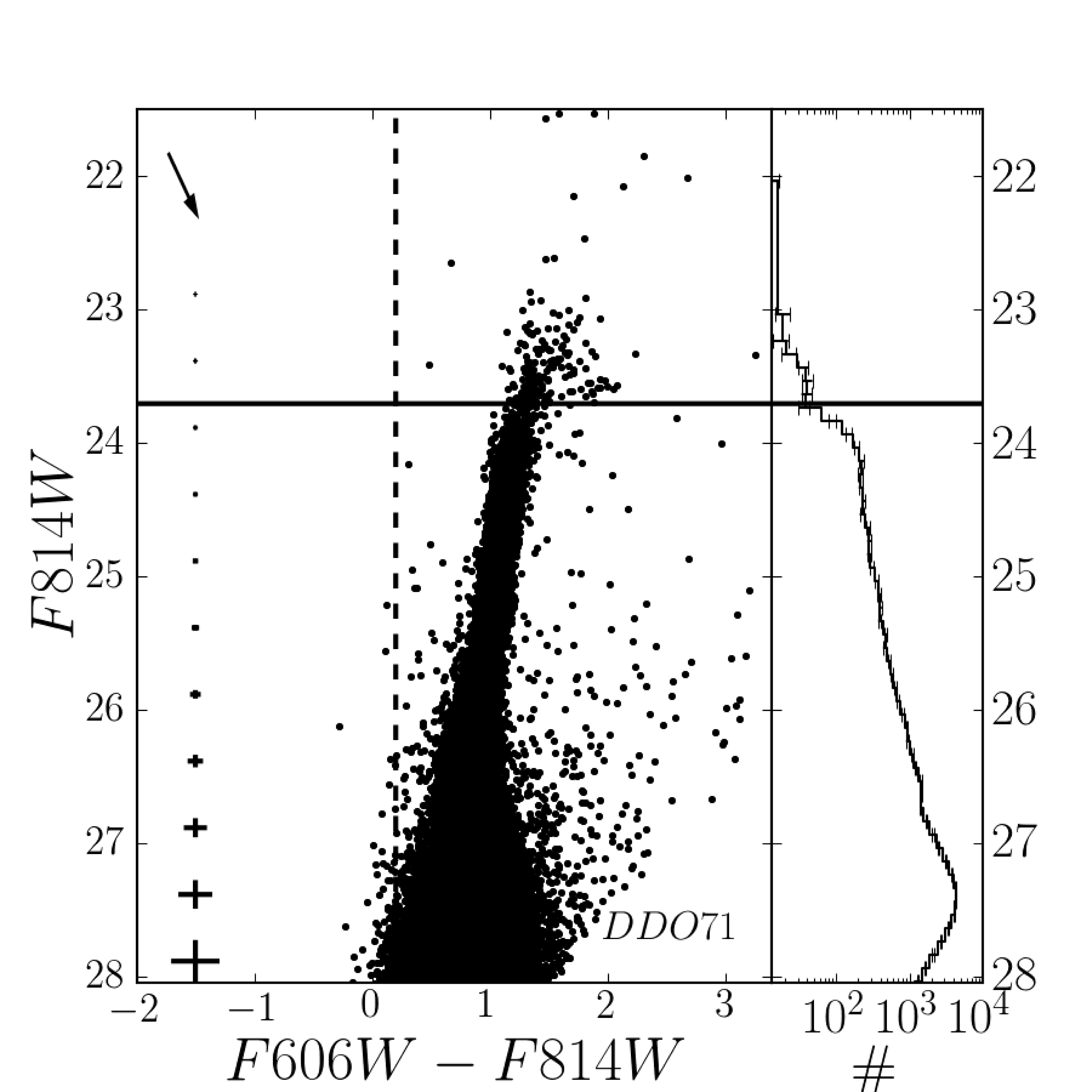}
\includegraphics[width=0.23\textwidth]{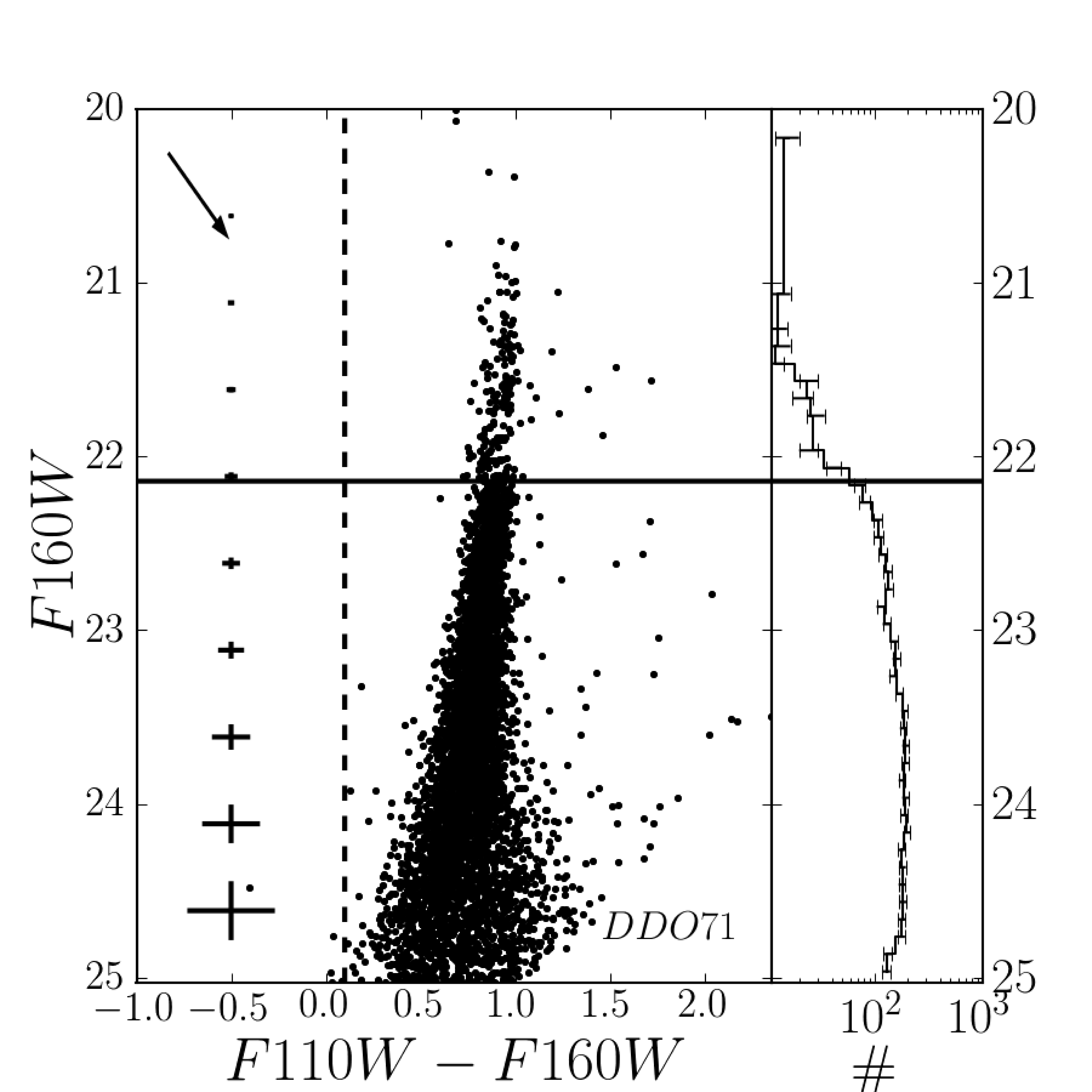}
\hspace{2 pc}
\includegraphics[width=0.23\textwidth]{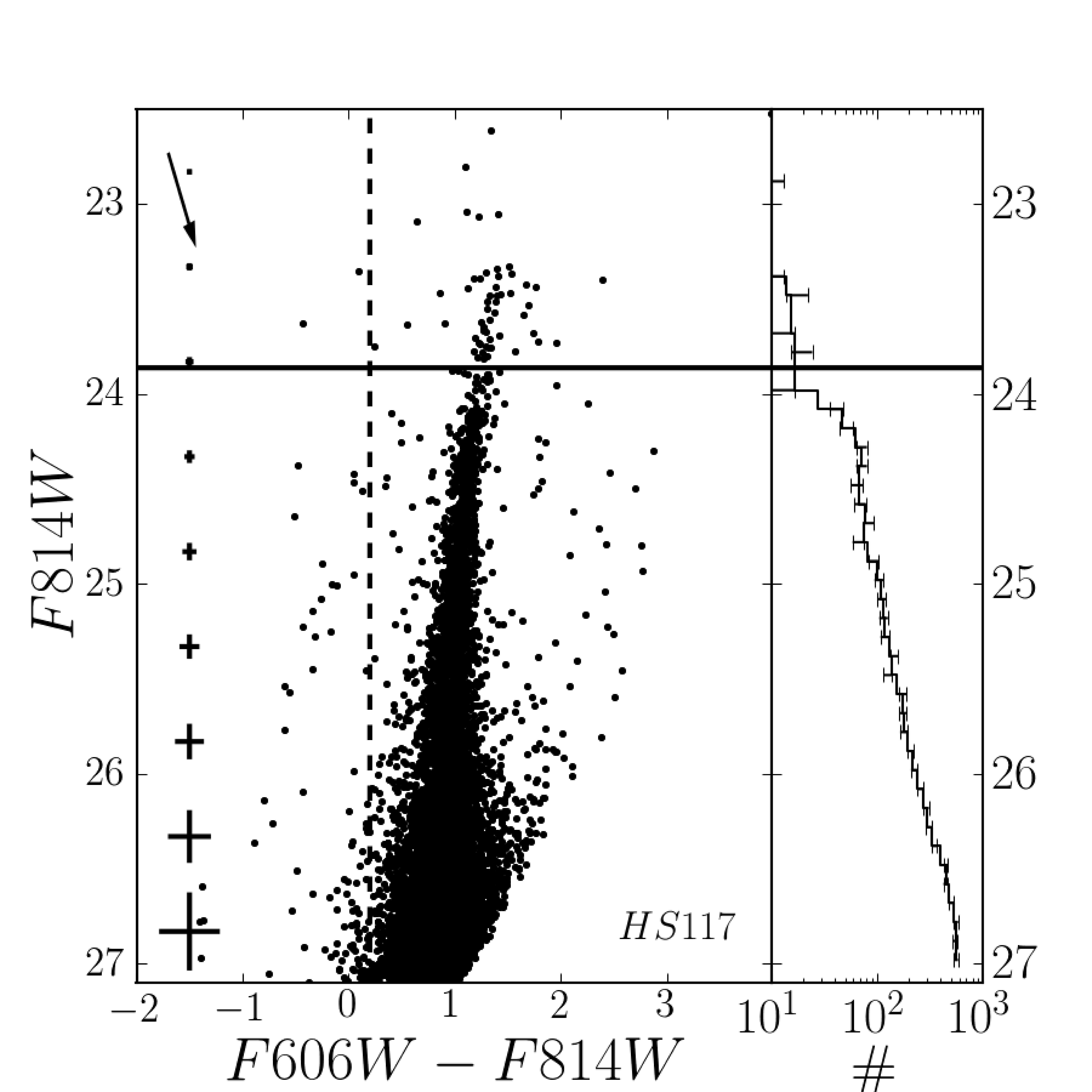}
\includegraphics[width=0.23\textwidth]{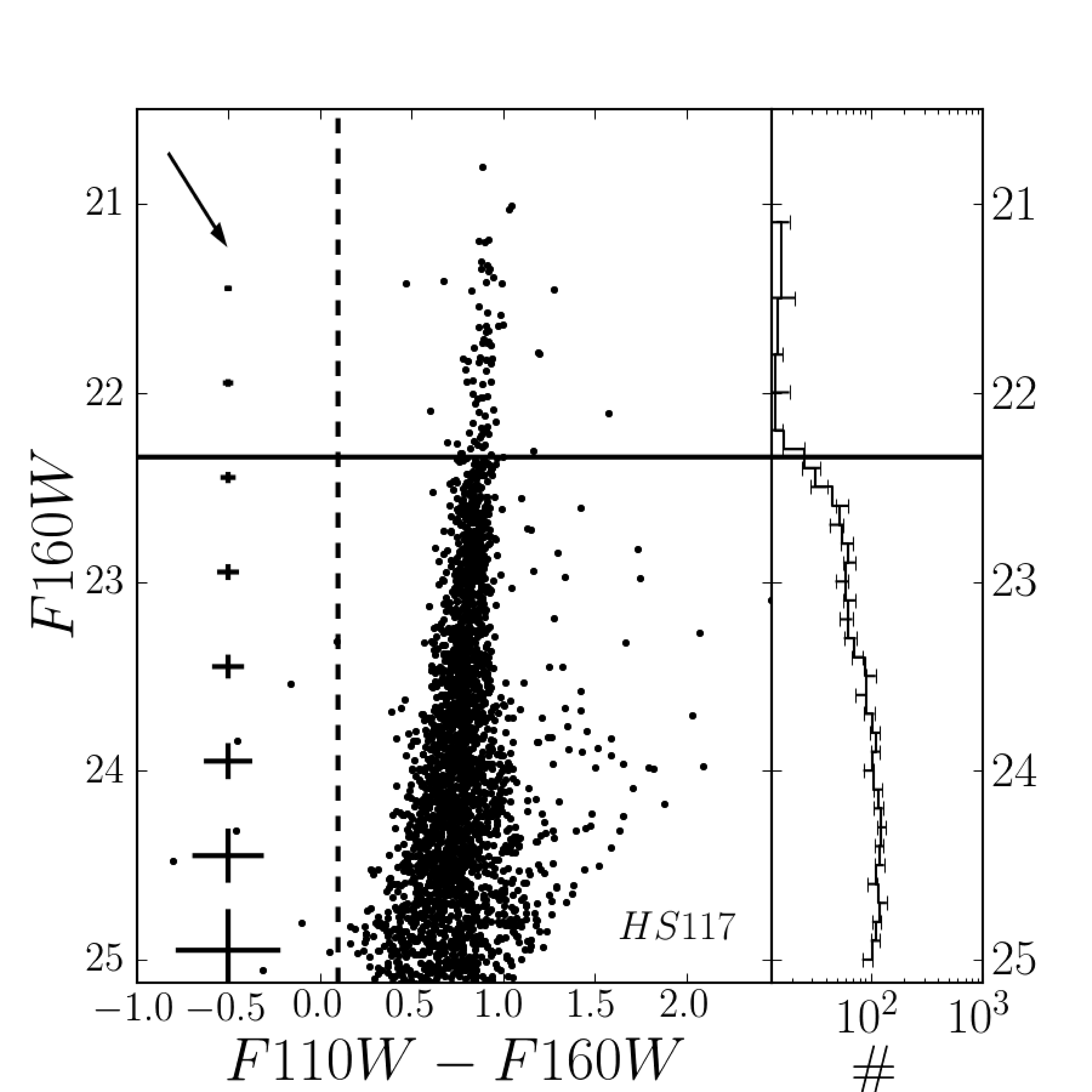}
\\
\includegraphics[width=0.23\textwidth]{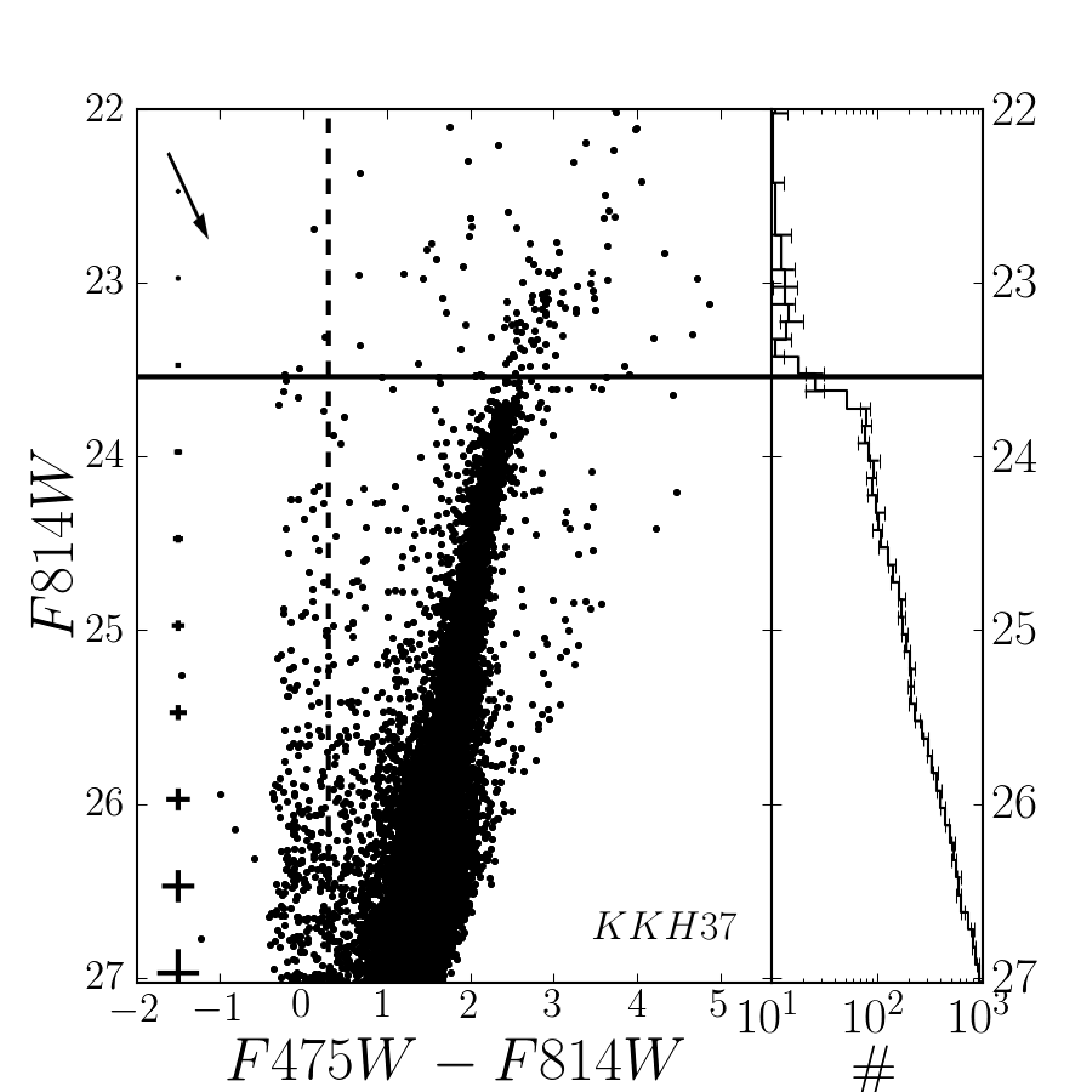} 
\includegraphics[width=0.23\textwidth]{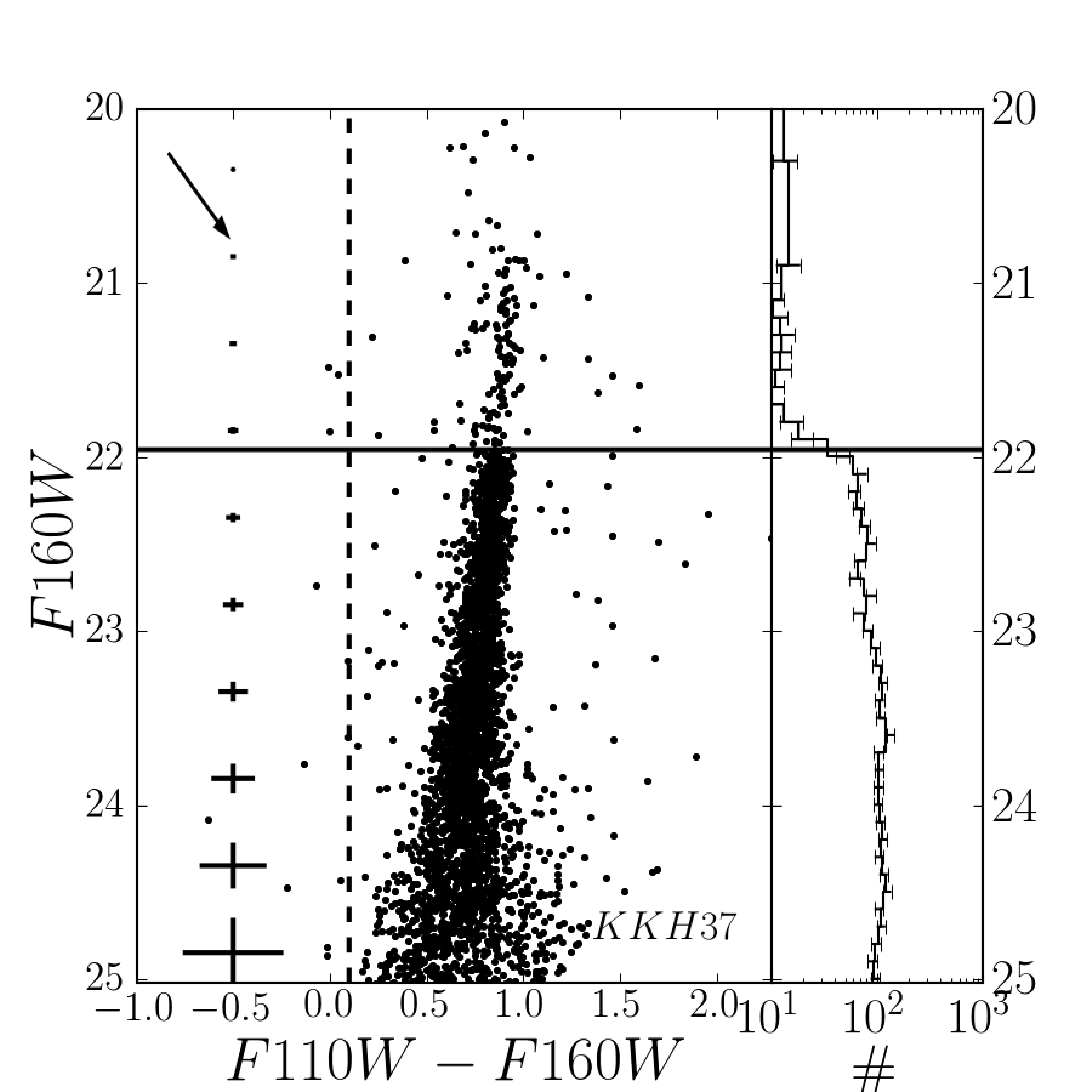}
\hspace{2 pc}
\includegraphics[width=0.23\textwidth]{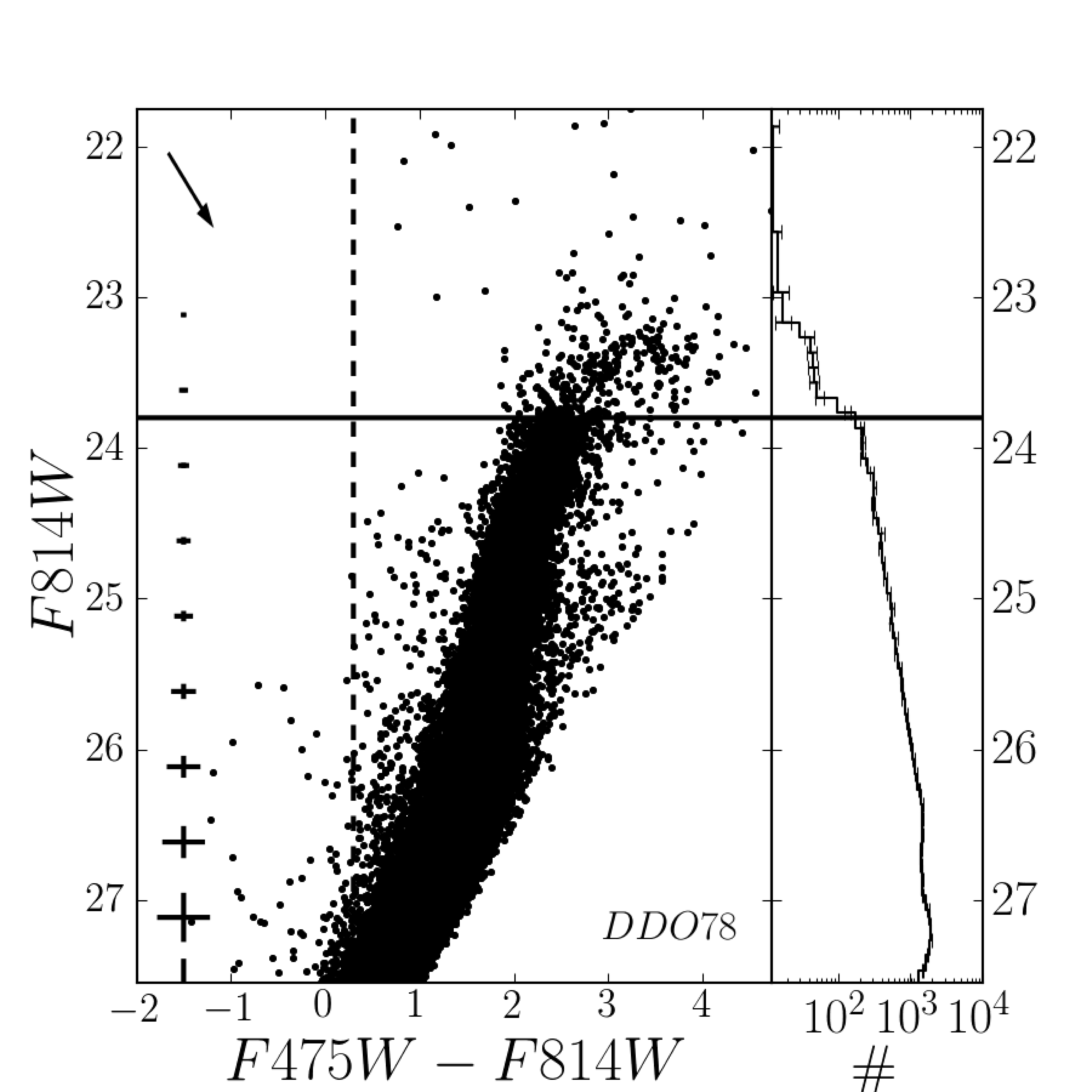}
\includegraphics[width=0.23\textwidth]{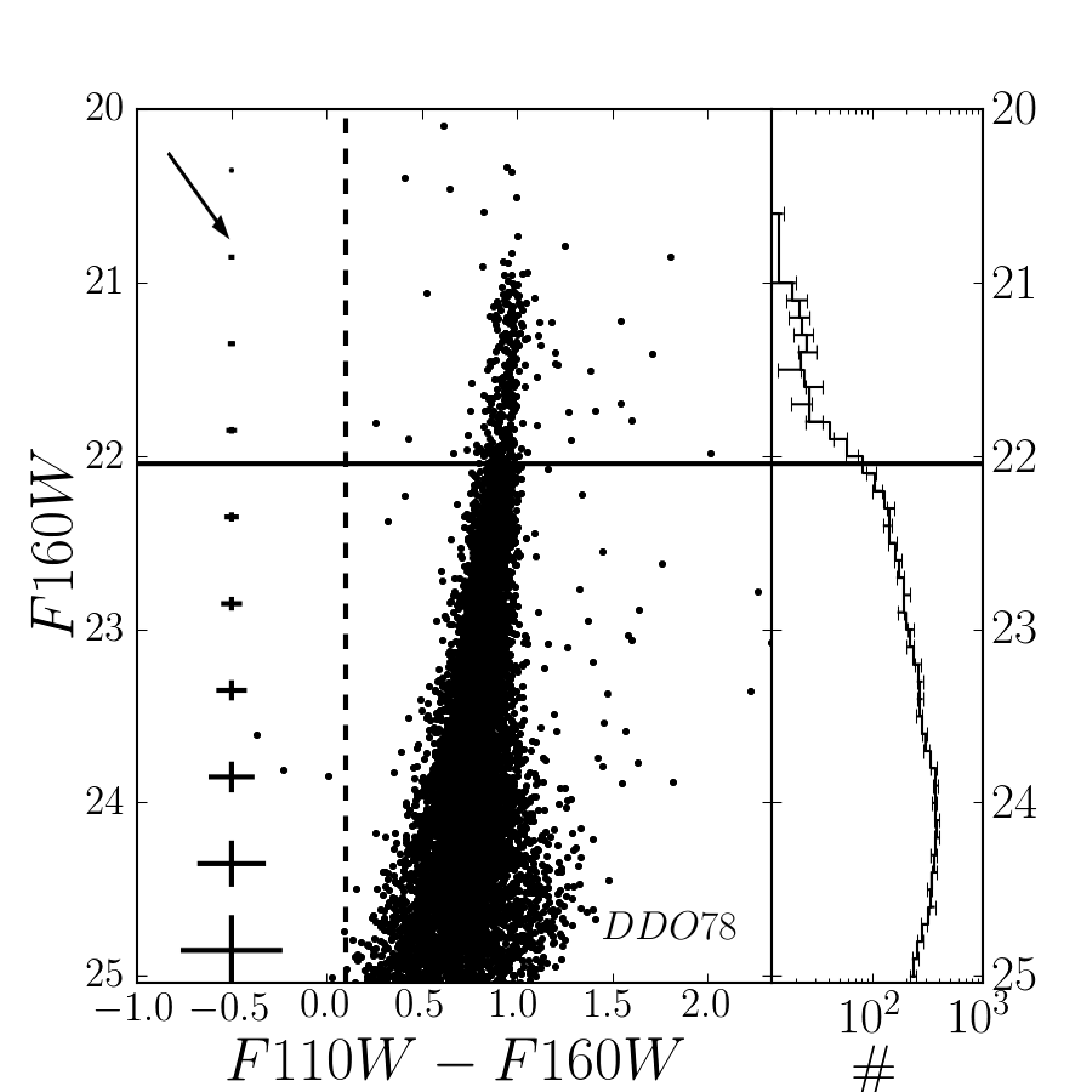}
\\
\includegraphics[width=0.23\textwidth]{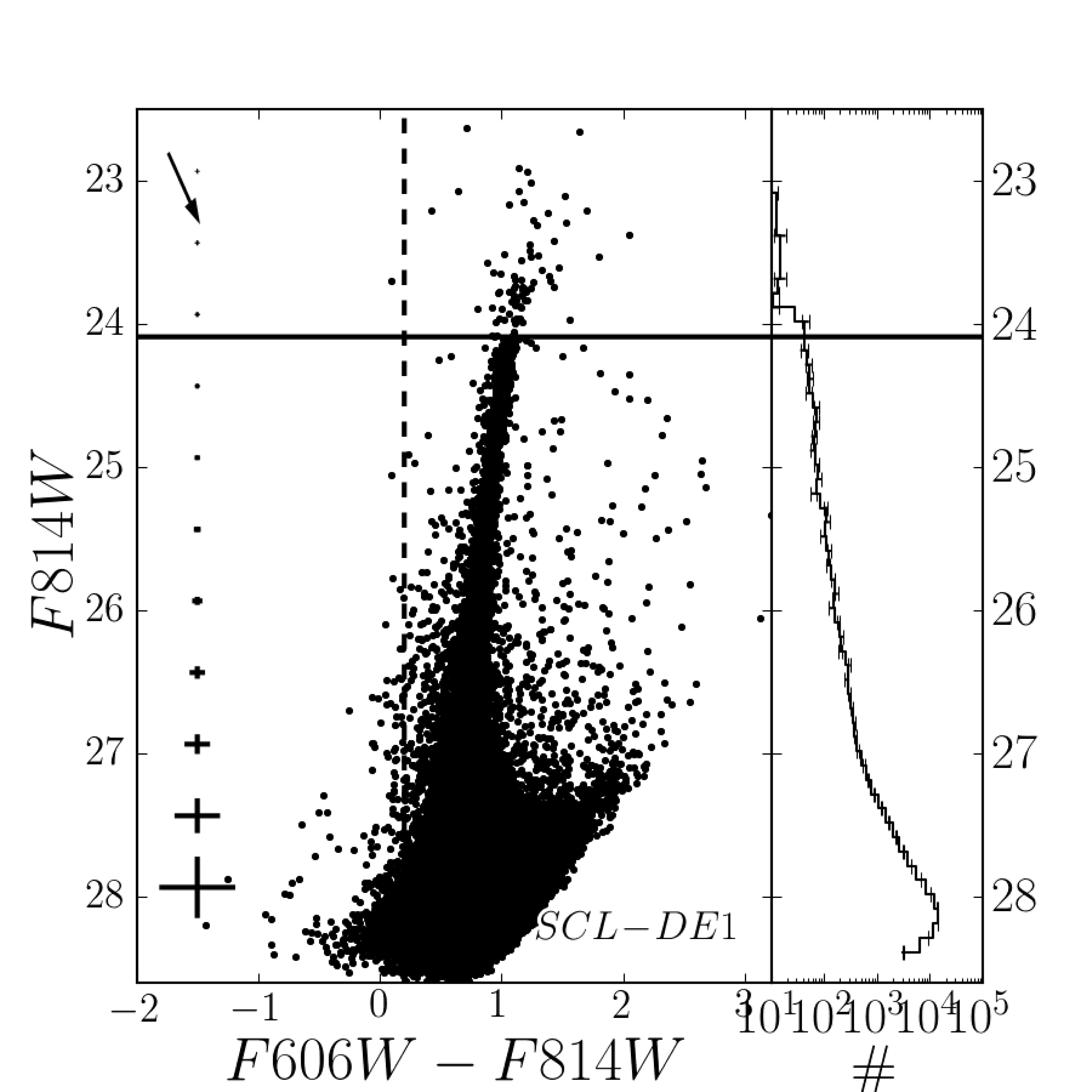}
\includegraphics[width=0.23\textwidth]{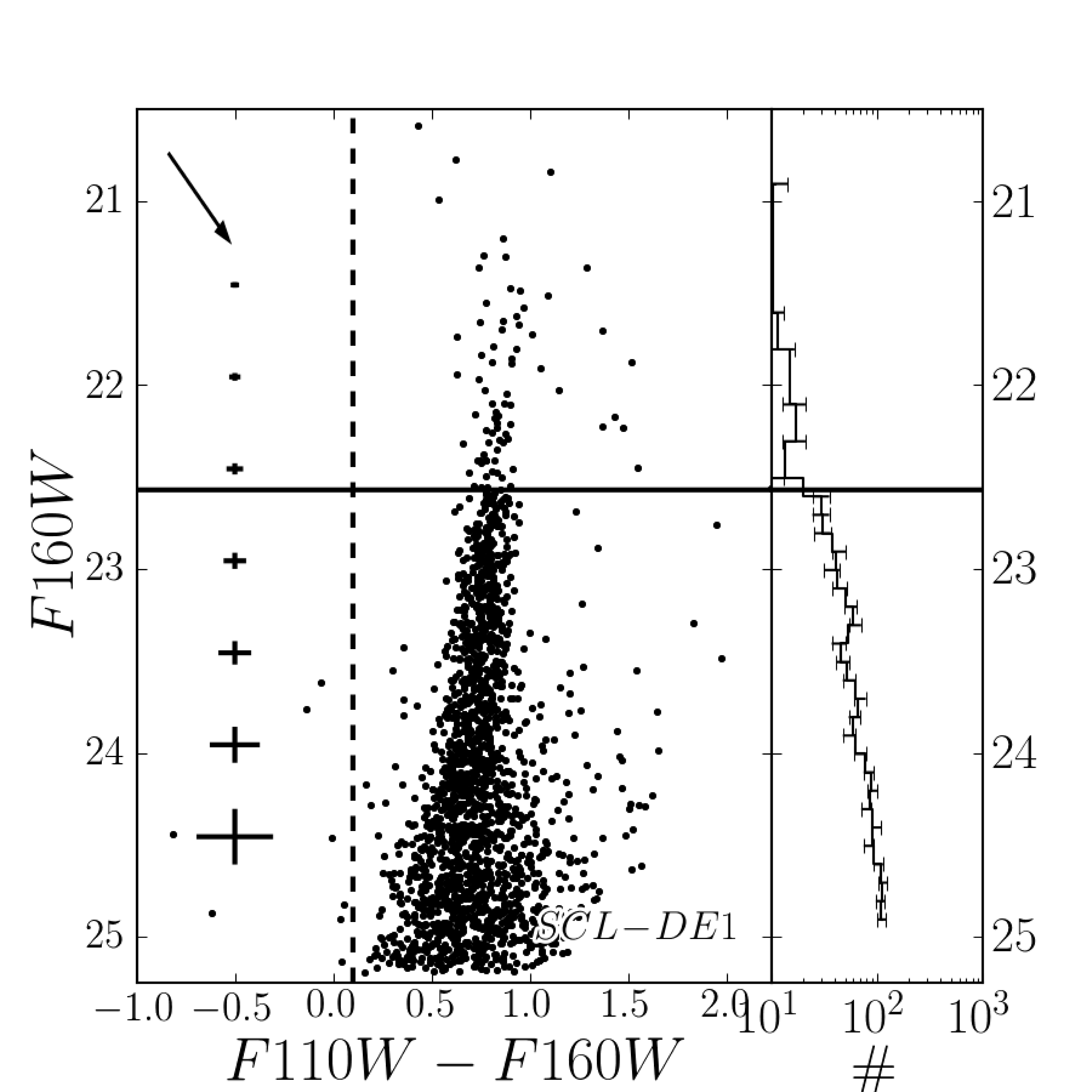}
\hspace{2 pc}
\includegraphics[width=0.23\textwidth]{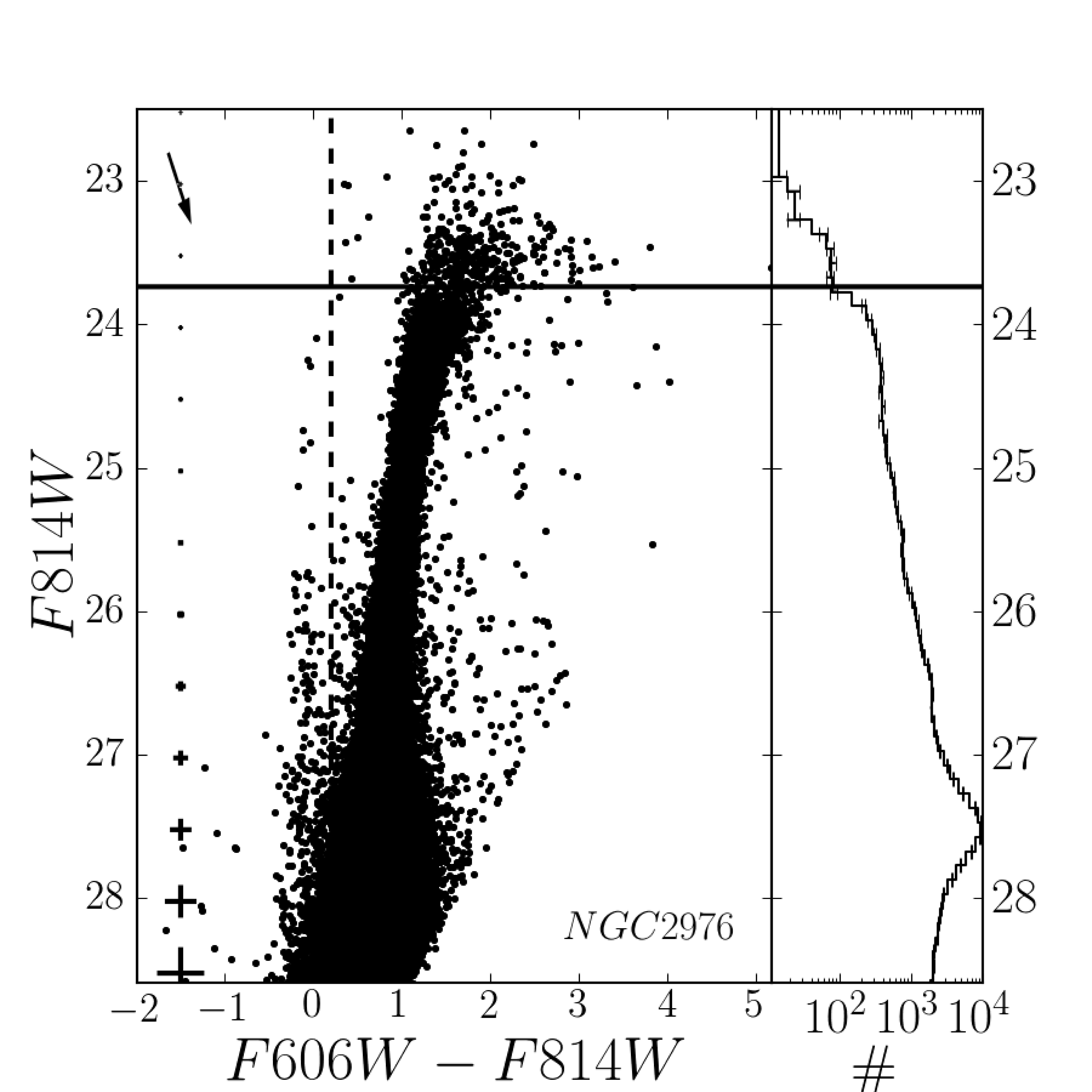}
\includegraphics[width=0.23\textwidth]{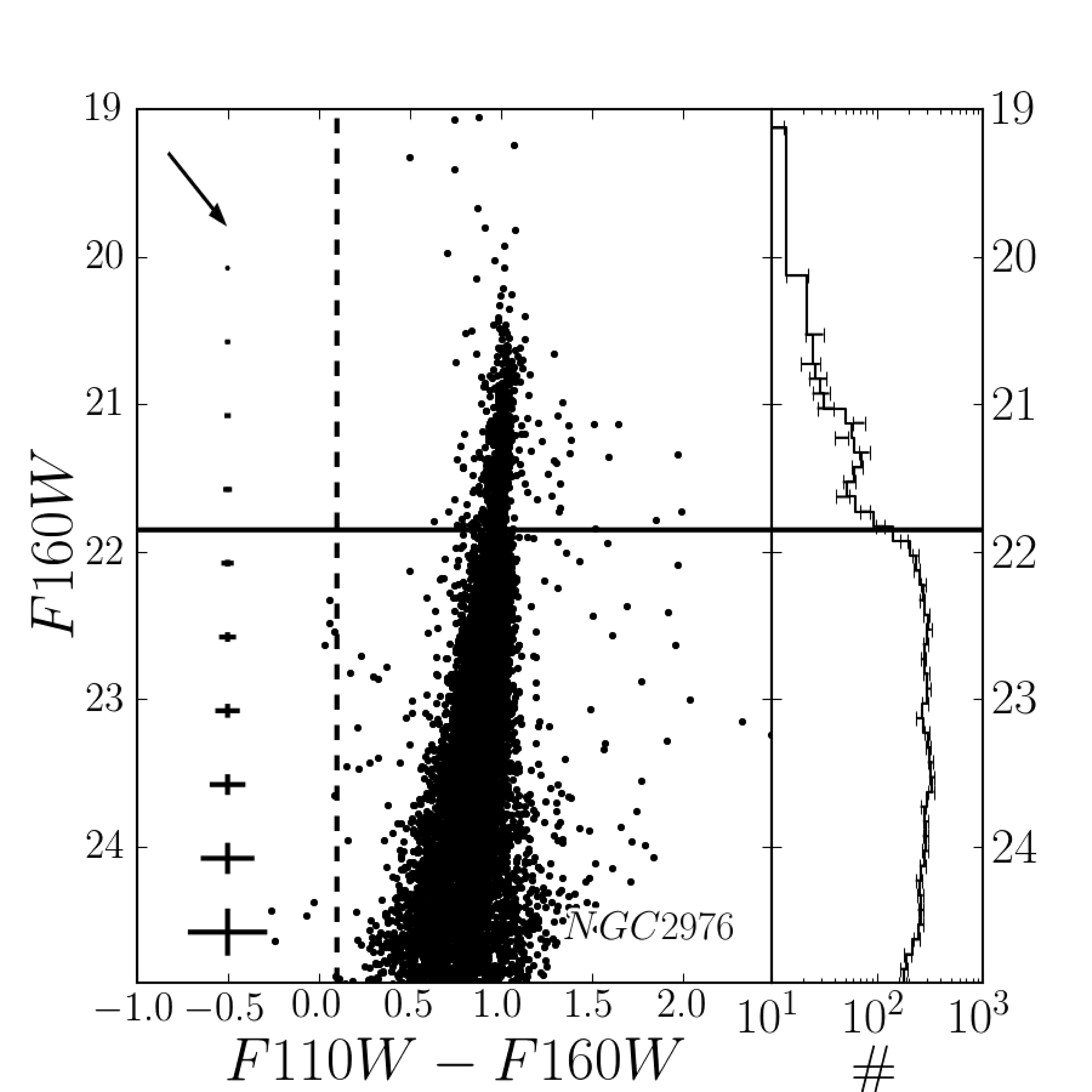}

\caption{Color magnitude diagrams and luminosity functions of the galaxy sample. Each CMD shows typical photometric uncertainties as a function of magnitude and each LF shows Poisson uncertainty in each bin. Reddening arrows for 0.5 mag extinction are shown in the upper left of each CMD. Horizontal lines mark the TRGB \citep{Dalcanton2009} and vertical lines mark the color cut discussed in Section \ref{sec_agb_rgb_selection}.}
\label{galaxy_sample}
\end{center}
\end{figure*}

We divided the AGB SNAP sample into galaxies with and without evidence of recent star formation, which effectively limits the masses of the AGB star populations to $\lesssim3 \Msun$ for the former and includes all possible AGB masses for the latter. For this paper, we focus on six galaxies, DDO~71, DDO~78, NGC~2976, SCL~DE1, HS~117, and KKH~37 shown in optical and NIR CMD and their LFs in Figure \ref{galaxy_sample}. These galaxies are suitable to be modeled with the isochrones derived from the Padova and TRieste Stellar Evolution Code \citep[][PARSEC]{Bressan2012}, which in its present version (v1.1) include masses up to $12\msun$ and typical metallicities ranging from $Z\sim 0.0001-0.06$ ([Fe/H]$\sim$ -2.18 $-$ -0.60. Notice that half of these galaxies (DDO~71, DDO~78, and SCL~DE1), where already analyzed in \citetalias{Girardi2010} but using the optical data only. In the next paper, we will extend the galaxy sample to those that show higher amounts of recent star formation with more massive stellar evolution models from PARSEC v2 (Bressan et al. in prep).

The paper is organized as follows. In Section \ref{sec_data}, we briefly summarize the data reduction and photometry of the AGB-SNAP sub-sample. Next, we describe the process of recovering the SFH from the deeper optical photometry and associated artificial star tests and our method of isolating TP-AGB and red giant branch (RGB) stars. We discuss the TP-AGB stellar evolutionary models and the pre-dust mass-loss prescriptions in Section \ref{sec_models}. Our method of using star counts to robustly model the TP-AGB stars in data is presented in Section \ref{sec_model_data}. We compare the observational constraints from the TP-AGB models to those of data in Section \ref{sec_analysis}, and conclude in Section \ref{sec_conc}.

\section{Data}
\label{sec_data}
\subsection{Reduction and Photometry}
We now briefly summarize the data reduction and photometry of the AGB-SNAP sample. For full details, we refer to the ANGST and AGB-SNAP survey papers \citep[][respectively]{Dalcanton2009, Dalcanton2012}.

Optical data from the STScI ACS pipeline data were photometered using DOLPHOT2.0 \citep{Dolphin2002} including the ACS module. Cosmic rays were rejected after combining all images into a single drizzled image using the {\tt multidrizzle} task within PyRAF \citep{Koekemoer2002}. We use the conservative ANGST photometric catalog ({\tt *gst}), which only includes objects with DOLPHOT parameters {\tt{ SNR} < 4}, (({\tt sharp$_1$} + {\tt sharp$_2$})$^2 \leq 0.075$), and crowding (({\tt crowd$_1$} + {\tt crowd$_2$}) $\leq 0.1$) in both filters. WFPC2 data from ANGRRR and ANGST (following the ACS camera failure) used the WFPC2 pipeline of \citet{Holtzman2006}, which processes STScI baseline output through HSTphot, a WFPC2 optimized predecessor of DOLPHOT but updated to July 2008 CTE corrections (derived by A. Dolphin). To assess the photometric uncertainties and completeness, at least $\sim10^5$ artificial star tests were calculated for each galaxy.

The NIR AGB-SNAP data were also reduced using the DOLPHOT package, using a significant update of the WFC3 module (among other enhancements). We again use the conservative photometry ({\tt *gst}), in this case with DOLPHOT parameters {\tt{ S/N} < 4, (({\tt sharp$_{F110W}$} + {\tt sharp}$_{F160W}$})$^2  \leq 0.12$) and (({\tt crowd}$_{F110W}$ +{\tt crowd}$_{F160W}$) $\geq 0.48$).

\subsection{Star Formation Histories of the Galaxy Sample}
\label{sec_sfh_z}

We use the optical CMDs to derive the SFH. ANGST observations are deep enough to robustly measure the color and magnitude of the red clump (RC) and the shape of the RGB. Both place strong constraints on the metallicity evolution and past star formation rate \citep[SFR; see][]{Dalcanton2009}. However, we exclude the AGB from contributing to the SFH fit by masking stars brighter than the TRGB (as was done in G10).

Star formation histories were derived using the CMD-fitting MATCH package \citep{Dolphin2000, Dolphin2012, Dolphin2013}. MATCH finds the most likely SFH and metallicity evolution that fits the observed CMD based on a given IMF, binary fraction, and stellar isochrones while taking into account photometric uncertainties and completeness. We adopt a \citet{Kroupa2001} IMF, a binary fraction of 0.35, and the PARSEC isochrones \citep{Bressan2012}. Photometric uncertainties and completeness were obtained from the artificial star tests described in \citet{Dalcanton2012} and \citet{Weisz2011}. In addition, we allow MATCH to search distance and $A_V$ parameter space, in finding the most likely SFH.

Following the discussion in G10, we used MATCH with the {\tt zinc} flag, which determines the best-fit CMD as a product of the most probable star formation history given an enrichment history in which the metallicity increases with galaxy age \citep[see discussions in][]{Dolphin2002, Dolphin2012}. MATCH derived SFR-weighted metallicities at ages older than 300 Myr range from Z=$0.0004-0.002$ ([Fe/H]=-1.54$-$-0.86; see Table \ref{tab_data_numbers}) . Random uncertainties in the SFH were determined with the Hybrid Monte Carlo tests described in \citet{Dolphin2013}. We use both the SFHs and their uncertainties to produce the model LFs (see Section \ref{sec_sfh_err}).

\begin{figure}
\includegraphics[width=\columnwidth]{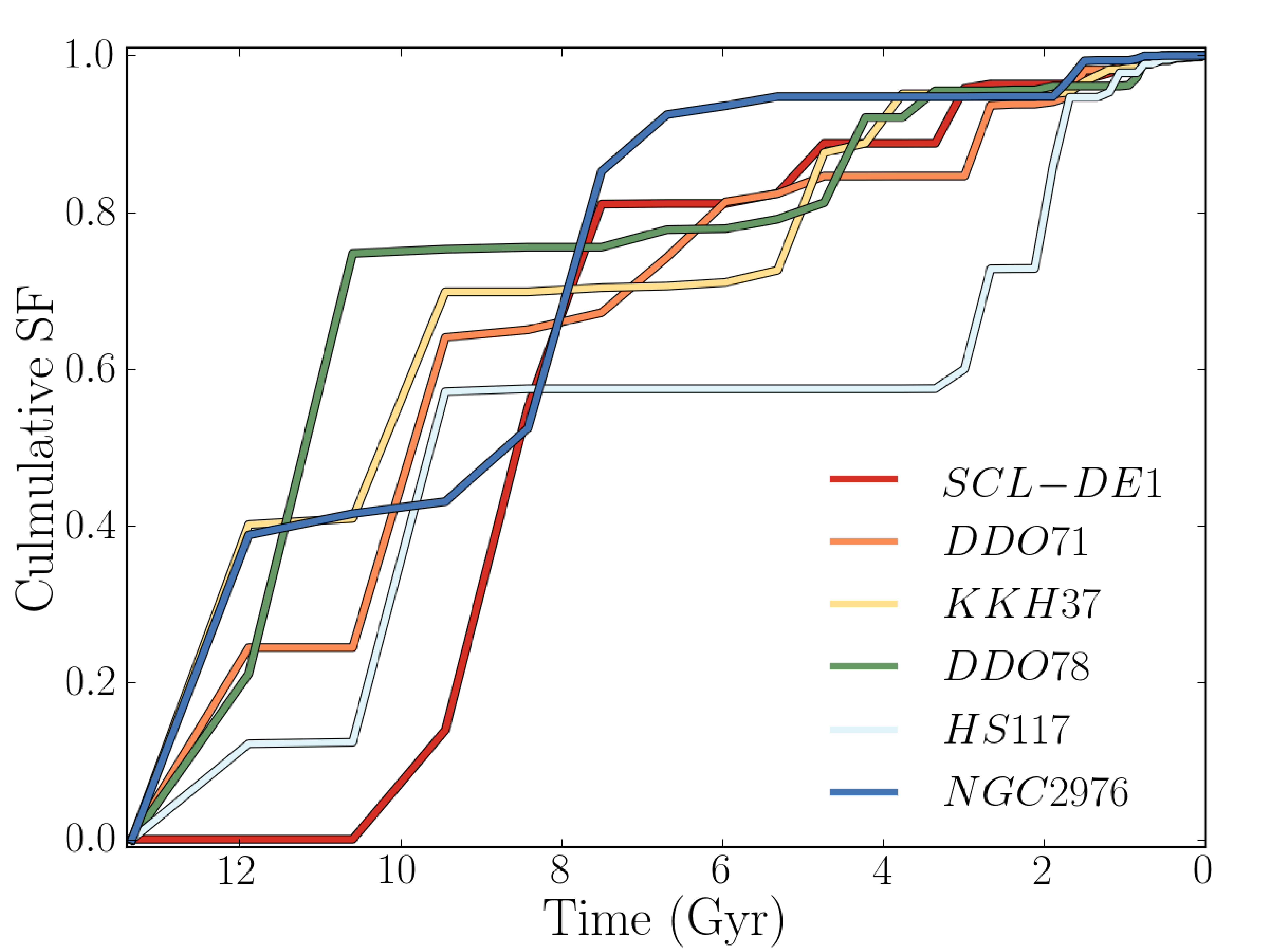}
\caption{The derived cumulative star formation histories in the galaxy sample.}
\label{fig_sfhs}
\end{figure}

A summary of the cumulative SFHs of the galaxy sample are shown in Figure \ref{fig_sfhs}. Derived SFHs all agree within uncertainties to those derived in \citet{Weisz2011} or for NGC2976 \citet{Williams2010} (both studies used \citet{Girardi2000} isochrones). In general, the CMD-fitting using PARSEC v1.1 compared to the previous Padova models produce better matches to the shape of the RGB found in the data.

Our method to constrain TP-AGB models is ultimately limited to how well we can model the rest of the stellar populations in the data. One measure of how well the SFH is recovered from the data is the effective $\chi^2$ \citep[for details see][]{Dolphin2013}. For the purposes of this study, we state the effective $\chi^2$ values derived from the SFH fitting as a means to compare one SFH fit to the next in Table \ref{tab_data_numbers}.

\subsection{Selecting AGB and RGB Stars}
\label{sec_agb_rgb_selection}

A successful TP-AGB model must be able to reproduce the lifetimes of TP-AGB stars. One classic observational comparison in stellar evolution modeling is to measure the relative lifetimes of an uncertain stellar evolution phase to a more certain stellar evolutionary phase by taking the ratio of stars found in different regions of a CMD. Therefore, a first test of the TP-AGB models is to calculate the number ratio of TP-AGB stars to RGB stars, \narratio. As we identify the TP-AGB and RGB stars, we must minimize the contamination from stars of other phases in the RGB region and the TP-AGB region.

\begin{figure}
\begin{center}
\includegraphics[width=\columnwidth]{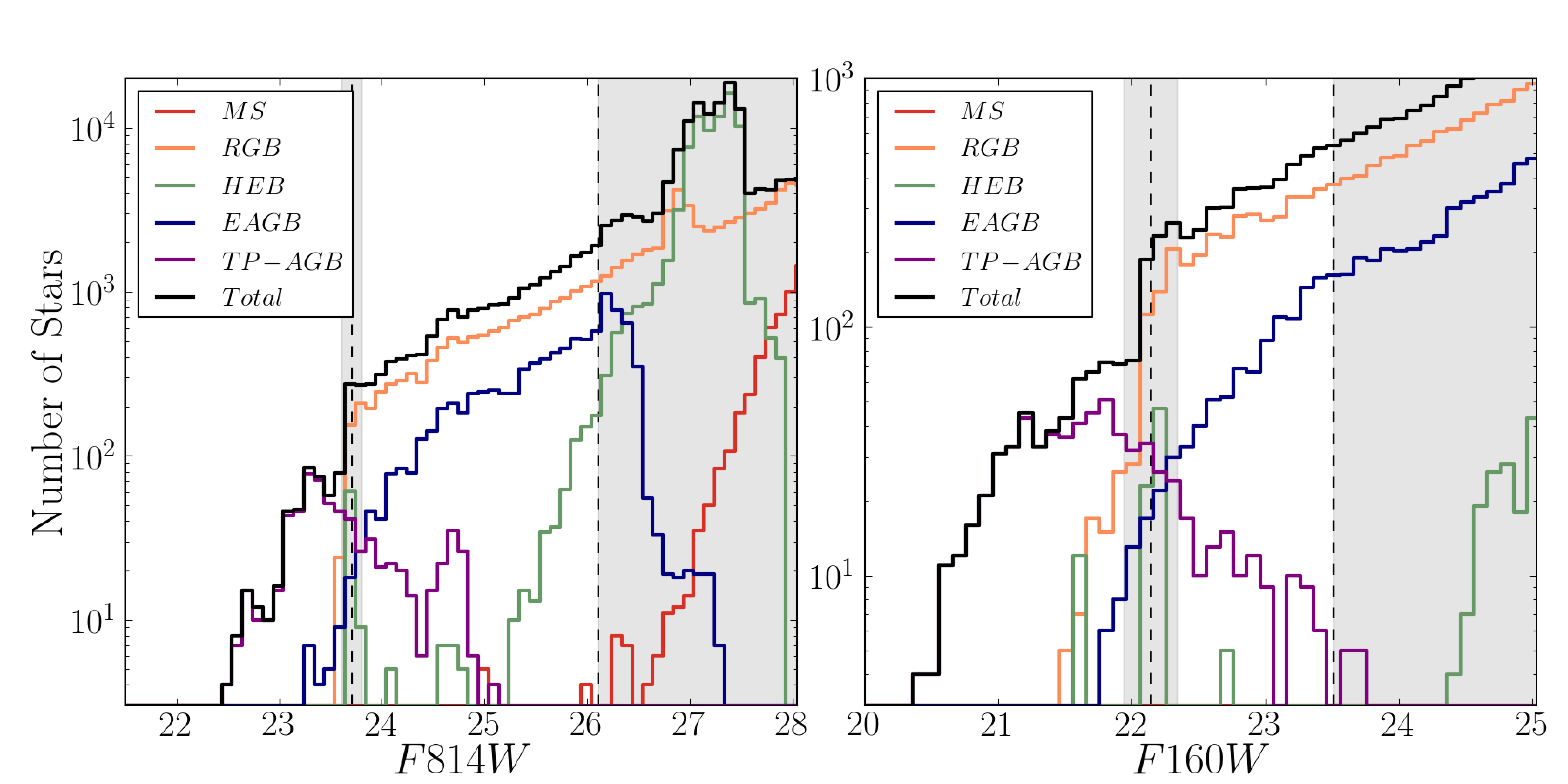}
\caption{Example simulated LFs (optical, left; NIR, right) of individual evolutionary stages (colored) and the total LF (black).  The red clump is seen in the optical (green, $F814W\sim27.5$) as is the RGB bump (blue, $F814W\sim27$). This simulation is from a M=$10^8$\msun TRILEGAL population synthesis model which used the MATCH-derived best fit SFH of DDO71. All stages from Pre-MS to Carbon burning or the first TP are modeled by PARSEC and the TP-AGB is modeled by COLIBRI using \NOV\ mass-loss. Regions of the LF that are excluded from the \narratio\ ratio counting are shaded grey (i.e., around the TRGB and faintward of the 90\% completeness limit). Stars with optical or NIR color bluer than 0.3  (see Figure \ref{galaxy_sample}) are considered a source of contamination for this study and are not included in any of the following LF figures or subsequent calculations (though are included here; see Section \ref{sec_agb_rgb_selection}).}
\label{fig_example_lf}
\end{center}
\end{figure}

We selected RGB and TP-AGB stars in two regions of the CMDs that were defined to minimize any possible contamination from stars in other evolutionary phases. Figure \ref{fig_example_lf} shows example optical and NIR LFs of several evolutionary stages of a simulation massive enough to ensure all phases were well populated and based on the SFH of DDO71.

We exclude any star within a magnitude offset around the TRGB to minimize the number of misclassified TP-AGB and/or RGB stars that scatter up or down in luminosity. Our simulations show that outside a region around the TRGB of $\pm0.1$ mag in $F814W$ or $\pm0.2$ mag in $F160W$ there is never more than 5\% contamination of TP-AGB stars in the RGB.  These ``excluded regions'' are shown in grey in all LF figures with the TRGB also indicated.

With the optical and NIR filters used in our observations, there is no way of completely separating RGB stars from Early AGB (EAGB) stars, Red Helium Burning stars (RHeB), or RC stars\footnote{Both RHeB and RC are red giants burning helium in their cores. The distinction between them is based on mass, and hence luminosity: RHeB have intermediate masses and hence cover a wide range of luminosities, while RC stars are low-mass stars occupying a very narrow region in the CMD.}. However, the RHeB and EAGB do not contribute more than $\sim 3\%$ and $\sim 20\%$ of the number of stars in the RGB region respectively, a percentage that greatly diminishes with decreasing brightness, based on our population synthesis of the galaxy sample. If we had included galaxies with more active recent SF, then the contamination from RHeB stars would have been higher.

Unlike the EAGB and RHeB, possible contamination from the RC increases with fainter magnitudes. RC stars are several magnitudes fainter than the TRGB, and have higher photometric uncertainties, potentially causing RC stars to blend with faint RGB stars. The possible contamination is only in the optical, since the NIR data is never deep enough to reach the RC. Regardless, we mitigate this contamination of the RGB by only including stars above the 90\% completeness limit in our analysis. In addition, high completeness assures precise counting of stars in each magnitude bin while being a bright enough limit to avoid stars from the RC (in the optical; the 90\% completeness magnitudes for each galaxy in $F814W$ and $F160W$ are listed in Table \ref{tab_sample}). RC stars never make up more than 10\% of stars in the RGB region included in our analysis. 

To exclude main sequence (MS) and blue He-burning (BHeB) stars, we restrict our analysis to stars redder than $F606W-F814W=0.2$, $F475W-F814W=0.3$ and $F110W-F160W=0.1$. Though the color cuts are easily drawn by eye, they were based on population synthesis combined with photometric uncertainties measured in the data. We take the color cut as the reddest MS star that is brighter than the 90\% completeness limit in a massive simulation calculated with constant SF ($\log({\rm age/yr})$ range from $7-10.13$) and constant, high metallicity ($Z=0.06$ for the reddest MS available in PARSEC). We then add to the color cut the typical 1$\sigma$ color uncertainty found in the data (typical uncertainties are shown in Figure~\ref{galaxy_sample}. From our simulations of each galaxy, the analysis regions contain on average 15\% MS stars and 1\% BHeB stars.

Finally, to robustly compare our models to data, we correct the data for completeness using the same artificial star tests derived for the SFH recovery. For our analysis, these corrections are minor as our faint limit is already at the 90\% completeness. In effect, due to the completeness corrections, the number of RGB stars increases at most by 10\% for optical data and 7\% for NIR data. 

In summary, the TP-AGB stars are defined as those (optical/NIR) stars (0.1/0.2) magnitudes brighter than the TRGB and the RGB stars are defined as the stars from the 90\% completeness limit to (0.1/0.2) magnitudes below the (optical/NIR) TRGB, excluding stars that are bluer in color than 0.2 for $F606W-F814W$, 0.3 for $F475W-F814W$, and 0.1 for $F110W-F160W$. The numbers of TP-AGB and RGB stars as well as their ratio and Poisson uncertainties are listed in Table \ref{tab_data_numbers}.

As eluded to in Section \ref{sec_wic}, $F160W$ observations can detect cooler TP-AGB stars than those using $F814W$ as the latter is more affected by circumstellar dust. However, with the exception of NGC~2976, we detect fewer TP-AGB stars in NIR than in the optical. The reason for the discrepancy is simply due to the observations. While the WFC3/IR fields overlap the ACS/WFC3 fields \citep[cf. Figure 1 of][]{Dalcanton2012a}, the ACS/WFC field of view (FOV; $202''\times202''$) is larger than the WFC3/IR FOV ($136''\times 123''$) by a factor of 2.4. Indeed, the number densities of NIR TP-AGB stars are between 1.9--2.2 times higher than that of the optical TP-AGB stars (excluding NGC~2976 which is 4 times higher).

The ratio of TP-AGB stars to RGB stars ranges from 0.017 to 0.253 and random errors in the measured ratio are never above 16\%. When all galaxies are combined, the measured optical and NIR \narratio\ ratio is $0.046\pm0.002$ and $0.116\pm0.005$, respectively. Therefore, random errors will have little effect when comparing the measured \narratio\ ratio to that predicted by a grid of TP-AGB models. 

\begin{deluxetable*}{lcccccc}
\tabletypesize{\scriptsize}
\tablecaption{Observational Data}
\tablehead{
    \colhead{Target} &
    \colhead{$A_V$} &
    \colhead{$(m-M)_0$} &
    \multicolumn{2}{c}{F814W} &
    \multicolumn{2}{c}{F160W}\\
    \colhead{} &
    \colhead{} &
    \colhead{} &
    \colhead{90\% Comp.} &
    \colhead{$m_{TRGB}$} &
    \colhead{90\% Comp.} &
    \colhead{$m_{TRGB}$} 
}
\startdata 
DDO71   & 0.30 & 27.74 & 26.11 & 23.71 & 23.51 & 22.14 \\ 
HS117   & 0.36 & 27.91 & 24.76 & 23.86 & 23.85 & 22.34 \\ 
KKH37   & 0.23 & 27.57 & 24.41 & 23.54 & 23.26 & 21.96 \\ 
NGC2976 & 0.22 & 27.76 & 25.52 & 23.74 & 22.67 & 21.85 \\ 
DDO78   & 0.07 & 27.82 & 24.76 & 23.80 & 23.63 & 22.05 \\ 
SCL-DE1 & 0.05 & 28.11 & 25.65 & 24.09 & 24.27 & 22.57 
\enddata
\tablecomments{Columns 2, 3, and 5 from \citet{Dalcanton2009} and column 7 is from \citet{Dalcanton2012},
columns 4 and 6 are the 90\% completeness magnitudes in $F814W$ and $F160W$, respectively.}
\label{tab_sample}
\end{deluxetable*}

\begin{deluxetable*}{lrrcrrccc}
\tabletypesize{\scriptsize}
\tablecaption{Star Counts and MATCH data}
\tablehead{
    \colhead{Target} &
    \multicolumn{3}{c}{F814W} &
    \multicolumn{3}{c}{F160W} &
    \multicolumn{2}{c}{Results from MATCH}\\
    \colhead{} &
    \colhead{$N_{\rm TP\!-\!AGB}$} &
    \colhead{$N_{\rm RGB}$} &
    \colhead{ $\frac{N_{\rm TP\!-\!AGB}}{N_{\rm RGB}}$} &
    \colhead{$N_{\rm TP\!-\!AGB}$} &
    \colhead{$N_{\rm RGB}$} &
    \colhead{ $\frac{N_{\rm TP\!-\!AGB}}{N_{\rm RGB}}$} &
    \colhead{SFR-weighted [Fe/H]} &
    \colhead{$\chi^2_{\rm eff}$ of CMD-fit}
}
\startdata 
DDO71   &  149 &  8739 & $0.017\pm0.002$ &  136 & 1730 & $0.079\pm0.009$ & -1.13 & 1.48 \\ 
HS117   &   70 &   498 & $0.141\pm0.023$ &   62 & 1003 & $0.062\pm0.010$ & -1.20 & 1.07 \\ 
KKH37   &  135 &   669 & $0.202\pm0.025$ &  122 &  923 & $0.132\pm0.016$ & -1.38 & 1.14 \\ 
NGC2976 &  290 &  7695 & $0.038\pm0.003$ &  490 & 1933 & $0.253\pm0.017$ & -0.86 & 1.55 \\ 
DDO78   &  273 &  2783 & $0.098\pm0.008$ &  215 & 2987 & $0.072\pm0.006$ & -0.86 & 1.27 \\ 
SCL-DE1 &   83 &  1144 & $0.073\pm0.010$ &   66 &  867 & $0.076\pm0.012$ & -1.54 & 0.89 \\ 
Total   & 1000 & 21528 & $0.046\pm0.002$ & 1091 & 9443 & $0.116\pm0.005$ & ... & ... 
\enddata
\tablecomments{The number of stars in the TP-AGB and RGB regions defined in Section \ref{sec_agb_rgb_selection} and their ratio in $F814W$ (columns 2-4) and $F160W$ (columns 5-7). The final two columns come from the MATCH derived star formation histories, the sfr-weighted averaged metallicity older than 300 Myr (column 8) and the effective $\chi^2$ of the MATCH solution \citep[see][]{Dolphin2002}.}
\label{tab_data_numbers}
\end{deluxetable*}

\section{TP-AGB evolutionary models}
\label{sec_models}
PARSEC \citep{Bressan2012} 
is a thoroughly revised version of the popular Padova stellar evolution code used to compute stellar evolution tracks. We use its v1.1 release \citep[][see also http://stev.oapd.inaf.it]{Bressan2013}, which offers stellar tracks spanning the ranges $0.0001 \leq Z \leq 0.06$, $0.1\msun \leq M \leq 12\msun$, and from the Pre-MS phase to the beginning of either the TP-AGB or the core Carbon ignition phase. PARSEC is used in this work both to derive the SFH through the MATCH package (see Section \ref{sec_sfh_z}) as well as the input isochrones to the population synthesis code TRILEGAL (see Section \ref{sec_trilegal}), which simulates the LF under different AGB model assumptions.

Following the first thermal pulse on the AGB, COLIBRI \citep{Marigo2013} takes over the stellar evolution calculations from PARSEC. COLIBRI is an ``almost-full'' TP-AGB modeling code, that is, it relaxes many of the analytic forms of other synthetic TP-AGB models. All details are found in \citet{Marigo2013}. The most relevant to this study include: 1) a complete envelope model from the bottom of the quiescent H-burning shell up to the atmosphere; 2) Rosseland mean opacities calculated on-the-fly with the Opacity Project tools in the high temperature regime for $T > 20,000$ K \citep{Seaton05} and with the \AE SOPUS code \citep{Marigo2009} at lower temperatures for $1,500$ K $< T < 20,000$ K, including the equation of state of $\simeq 800$ atomic and molecular species, assuring complete consistency with current chemical abundances; 3) complete nuclear network to follow the nucleosynthesis occurring at the base of the convective envelope (hot-bottom burning; HBB) in more massive AGB stars ($M\gtrsim4\msun$) and in the pulse-driven convective zone at thermal pulses; and 
4) efficiency of mass-loss treated as a free parameter to be calibrated by observations. 

In this paper, we focus on mass-loss and test three prescriptions by how well they reproduce the \narratio\ ratio and the shape of the observed LFs. Although we also wish to constrain the efficiency of the third dredge up, such a test would require measuring the C/O ratio of the stellar population, and the NIR filters available are not red enough to robustly separate C-rich and O-rich stars \citep{Dalcanton2012, Boyer2013}. As we show below, the variation of the magnitude of the efficiency of the third dredge up does not change the TP-AGB lifetimes in mass-loss prescriptions with a high \teff\ dependence. 

\subsection{Mass-Loss Prescriptions on the TP-AGB}
For a star to have a wind, there must be an outward force that provides momentum and energy input, accelerating the surface layers to velocities larger than the escape velocity. This may be realized in various ways, including the scattering of UV radiation by resonance line opacity in hot stars, the generation of magneto-acoustic waves above the photosphere in red giants, or the absorption of photons by dust grains in the outer atmospheres of the coolest and most luminous stars \citep[e.g.,][]{Lamers1999}. 

Mass-loss dominates an AGB star's evolution and fate. It is clear from observations of Mira and OH/IR stars  that mass-loss rates increase exponentially along the AGB until they reach super-wind values of $\sim 10^{-5}\!-\!10^{-4}\,\msun$ {\rm yr}$^{-1}$  \citep[e.g., reviews by][]{Willson2000, Olofsson03}. Despite the recent progress in the theory of AGB mass-loss \citep[e.g.,][]{Wachter08, Mattsson10, Bladh12}, we still lack complete understanding of all the factors and their complex interplay which control the stellar winds on the AGB \citep[e.g.,][]{Woitke06}.

Combining theoretical efforts and empirical evidence, a reasonable scenario takes form in which mass-loss on the AGB can be divided into three regimes: an initial period before the onset of the dust-driven wind (designated as ``pre-dust mass-loss"); a subsequent phase characterised by an exponential increase of mass-loss driven by the combined action of dust and pulsation (designated as ``dust-driven mass-loss"); and a final brief regime with high mass-loss (designated as ``super-wind mass-loss").

In our scheme, the phase of {\em pre-dust mass-loss} (with rate $\dot M_{\rm pre\mhyphen dust}$) is thought to apply to the early stages on the AGB in which either dust has not yet formed in the outermost atmospheric layers, or if present in some small amount, is unable to generate an outflow. In these conditions a likely wind mechanism could be related to a strong flux of pressure waves or Alfv\'en waves able to cause the spillover of the extended and highly turbulent chromospheres typical of red giants. The same mechanism might be at work during both the ascent along the RGB and the early stages of the AGB \citep[e.g., ][]{Schroeder2005, Cranmer2011}. 

In stellar evolutionary calculations a frequent choice to describe mass-loss during the early phases is the classical \citet{Reimers1975} law, a simple scaling relation of stellar parameters based on observations of few red giants and supergiants. The Reimers relation is commonly multiplied by an efficiency parameter $\eta_{\rm R}$, whose value is calibrated such that it recovers the observed morphology of horizontal branch stars in Galactic Globular clusters. The calibration however, still depends on the residual envelope mass left over from the RGB \citep{Renzini1988}.

More recently \citet[][]{Schroeder2005} proposed a modified version of the \citet{Reimers1975} law, in which additional dependencies on the effective temperature and surface gravity follow from a physically-motivated consideration of the mechanical flux responsible for the wind. The role of the chromosphere in driving mass-loss in late-K to early-M giants is supported by the analysis \citet[][]{MCDonald2007} of the H$\alpha$ and infrared calcium triplet lines in a sample of red giant stars hosted in Galactic globular clusters. Similarly to the Reimers relation, the \citet{Schroeder2005} formula also needs an efficiency parameter $\eta_{\rm SC}$ to be specified. 

Novel efforts to model stellar winds from red giants were carried out by \citet{Cranmer2011}. A self-consistent and more detailed theoretical approach is developed to follow the generation of energy flux due to magnetohydrodynamic turbulence from subsurface convection zones to its eventual dissipation and escape through the stellar wind. One major difference is that, while in \citet{Schroeder2005} the mass-loss rate is assumed to scale linearly with the photospheric mechanical energy flux ($F_{\rm M}$) of Alfv\'en waves ($\dot M_{\rm pre\mhyphen dust} \propto F_{\rm M}$), the analysis of \citet{Cranmer2011} yields a higher dependence ($\dot M_{\rm pre\mhyphen dust} \propto F_{\rm M}^{12/7}$). Analytic models for magnetic wave generation indicate that the mechanical energy flux scales as $F_{\rm M} \propto T_{\rm eff}^{7.5}$ \citep{Musielak88}. Hence, considering that the mass-loss rate is proportional to the surface-integrated mechanical energy flux, $L_{\rm M} = 4 \pi R^2 F_{\rm M}$, and expressing the stellar radius $R$ with the Stefan-Boltzmann law for a black body, we eventually obtain a significantly steeper dependence of the mass-loss rate on the effective temperature, i.e.  $\dot M_{\rm pre\mhyphen dust} \propto T_{\rm eff}^{3.5}$ for \citet{Schroeder2005} and $\dot M_{\rm pre\mhyphen dust} \propto T_{\rm eff}^{8.86}$  following the results of \citet{Cranmer2011}.

Following the pre-dust phase of mass-loss, as the star climbs the AGB at increasing luminosity, suitable conditions can be met in the cool atmosphere for stellar winds to be generated through a different intervening mechanism. The most plausible hypothesis resides in the momentum input when the stellar radiation field is absorbed (or scattered) by dust grains and transferred to the gas via collisions.  This wind is enhanced by pulsations that shock the envelope and periodically levitate matter up to regions where dust can more efficiently condense \citep{GustafssonHoefner03}. Observationally there is a clear correlation, though with a large scatter, between  the mass-loss rate (here designated with $\dot M_{\rm dust}$) and the pulsation period $P$ of AGB variables, such that $\dot M_{\rm dust}$ is seen to increase exponentially with the period \citep{Vassiliadis1993}.

Finally, close to tip of the TP-AGB, the mass-loss rates almost level out to $10^{-5}-10^{-4}\,M_{\odot}$ yr$^{-1}$ of the so-called super-wind phase ($\dot M_{\rm SW}$), corresponding to the condition in which the maximum momentum of the radiation field is transferred  to the stellar atmosphere.

Within this framework, the mass-loss prescriptions adopted in the TP-AGB stellar models computed for this study are as follows. For the dust driven wind phase we adopt a formula similar to \citet{Bedijn1988}, which predicts an exponential increase of mass-loss $\dot{M}_{\rm dust} \propto \exp(R^a M^b)$ dependent on stellar parameters derived from models of periodic shocked atmospheres. Coefficients $a$ and $b$ are calibrated on a sample of Galactic Mira stars.  This prescription is also discussed in \citet{Marigo2013}, \citet{Nanni13}, and \citet{Kalirai14}.

For the super-wind phase we adopt the formalism of \citet[][their equations 1 and 2]{Vassiliadis1993}, in which the mass-loss rate, $\dot M_{\rm SW}$, is proportional to the ratio of the stellar luminosity to the terminal velocity of the gas, which itself scales linearly with the 
pulsation period.  In practice as soon as $P > 500 - 600$ days the super-wind regime is expected to set in. 

We keep the same prescriptions for $\dot M_{\rm dust}$ and  $\dot M_{\rm SW}$ and vary only the $\dot M_{\rm pre\mhyphen dust}$. For the mass-loss rates $\dot M_{\rm pre\mhyphen dust}$ before the onset of dust-driven winds we consider four options: 

\begin{description}
\item[$\bullet$ \NOVeta] no mass-loss before the possible onset of the dust-driven wind, $\dot M_{\rm pre\mhyphen dust}=0$; 
\end{description}

\begin{description}
\item[$\bullet$ \OCT] the traditional \citet{Reimers1975} mass-loss
\end{description}
\begin{equation*}
\dot M_{\rm pre\mhyphen dust}= \displaystyle 4 \times 10^{-13} \eta_{\rm R} \frac{L}{g R}
\end{equation*}

\noindent with the efficiency parameter $\eta_{R}=0.4$;
\begin{description}
\item[$\bullet$ \SC] the original  \citet{Schroeder2005} law
\end{description}
\begin{equation*}
\dot M_{\rm pre\mhyphen dust} =  \displaystyle 10^{-14}\,\eta_{\rm SC} \frac{L R}{M} \left(\frac{T_{\rm eff}}{4000\, {\rm K}}\right)^{3.5} \left(1+\frac{1}{4300 g}\right)
\end{equation*}

\noindent with the efficiency parameter $\eta_{\rm SC}=8.0$;

\begin{description}
\item[$\bullet$ \NOV] a modified version of the 
\citet{Schroeder2005} scaling relation 
\end{description}
\begin{equation*}
\dot M_{\rm pre\mhyphen dust} =  \displaystyle 10^{-12}\,\eta_{\rm mSC} \frac{L R}{M} \left(\frac{T_{\rm eff}}{4000\, {\rm K}}\right)^{8.9} \left(1+\frac{1}{4300 g}\right)
\end{equation*}
in which, for the reasons explained above, the power-law dependence on the effective temperature is steepened; here the efficiency parameter is set to $\eta_{\rm mSC}=0.4$.

In all formulas the mass-loss rate is given in $M_{\odot}{\rm yr}^{-1}$, the effective temperature $T_{\rm eff}$ is in Kelvin,  the stellar radius $R$, luminosity $L$, the mass $M$, and surface gravity $g$ are expressed in solar units. Following \citet{Marigo2013}, at each time during the TP-AGB evolution, the current mass-loss rate is taken as  $\dot M = {\rm max}[\dot M_{\rm pre\mhyphen dust}, {\rm min}(\dot M_{\rm dust}, \dot M_{\rm SW})]$.

Finally, we caution the reader that our modified Schr\"oder \& Cuntz  relation, with $\eta_{\rm mSC}=0.4$, set for the early stages of the TP-AGB, may be too efficient to be extended to lower luminosities of RGB stars based on a quick comparison to the measured mass-loss rates of the sample, 
collected by \citet{Cranmer2011}, that includes metal poor RGB stars with and effective temperatures in the range 3800 - 5800 K, as a function of the luminosity. 
In this context, the role of RGB mass-loss and its possible influence on the subsequent AGB evolution of low-mass stars is postponed to a future work.

\subsection{Main processes affecting the TP-AGB lifetimes}

Many physical processes and events are at work during the TP-AGB phase. However, there is no doubt that mass-loss is the principal mechanism that controls the duration of this phase, which ends when almost all the stellar mantle is ejected into the interstellar medium.

In this study  we opt to analyze the significance of the early stages of AGB mass-loss, since 
this regime may be particularly important for low-mass stars whose small envelopes may already be removed before the onset of the dust-driven wind (c.f., G10). This choice seems appropriate given the sample of galaxies under consideration, which are all characterized by a significant fraction of old stellar populations and thus will have TP-AGB populations dominated by lower mass stars (Figure \ref{fig_sfhs}). While analysing the impact of different laws for $\dot M_{\rm dust}$ and $\dot M_{\rm SW}$ is postponed to future works, it is worth mentioning that the prescriptions adopted here have already successfully passed a few observational tests, including the recovery of the expansion velocities of AGB circumstellar envelopes \citep{Nanni13}, and the Galactic initial-final mass relation \citep{Kalirai14}. 

\begin{figure*}
\begin{center}

\includegraphics[width=.48\textwidth]{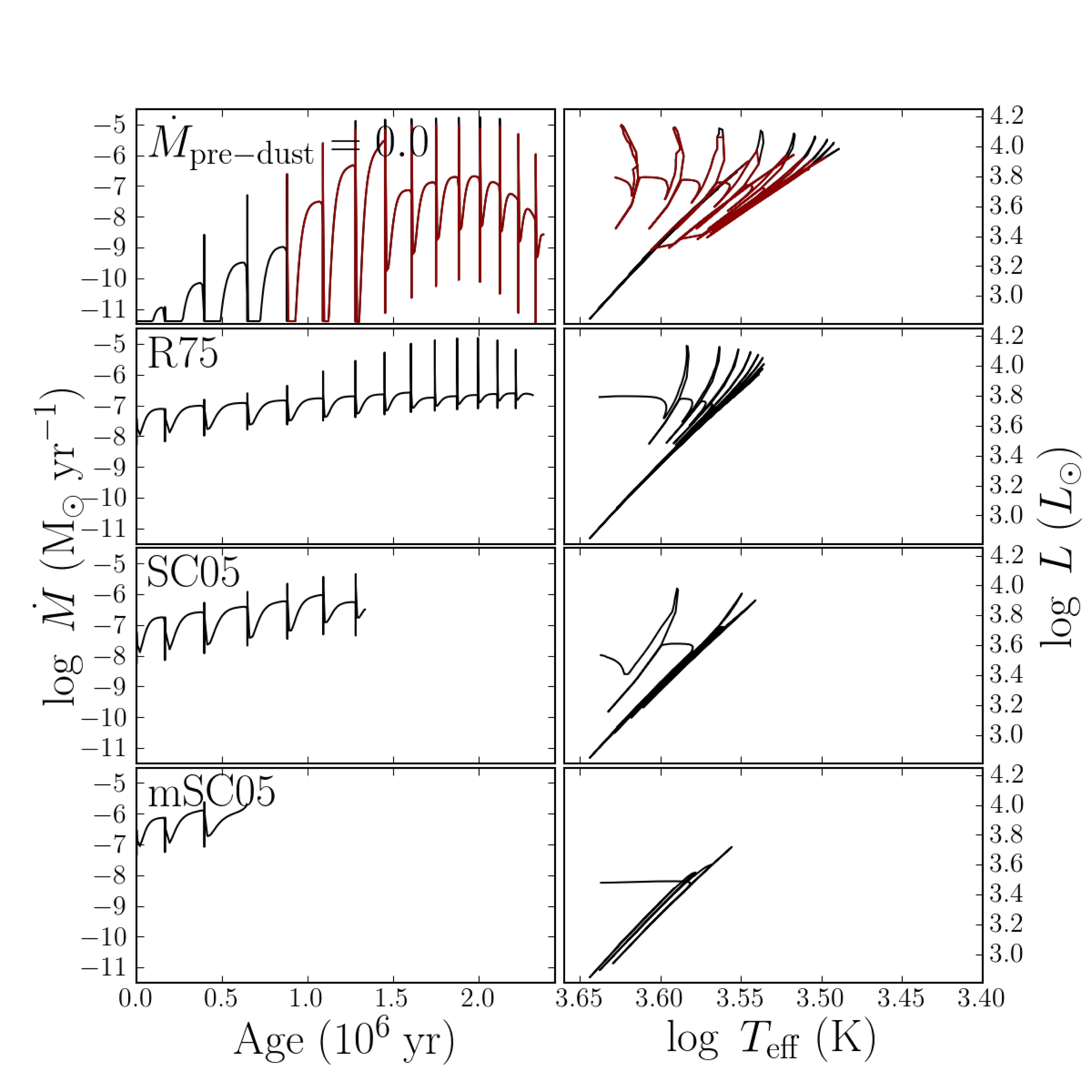}
\includegraphics[width=.48\textwidth]{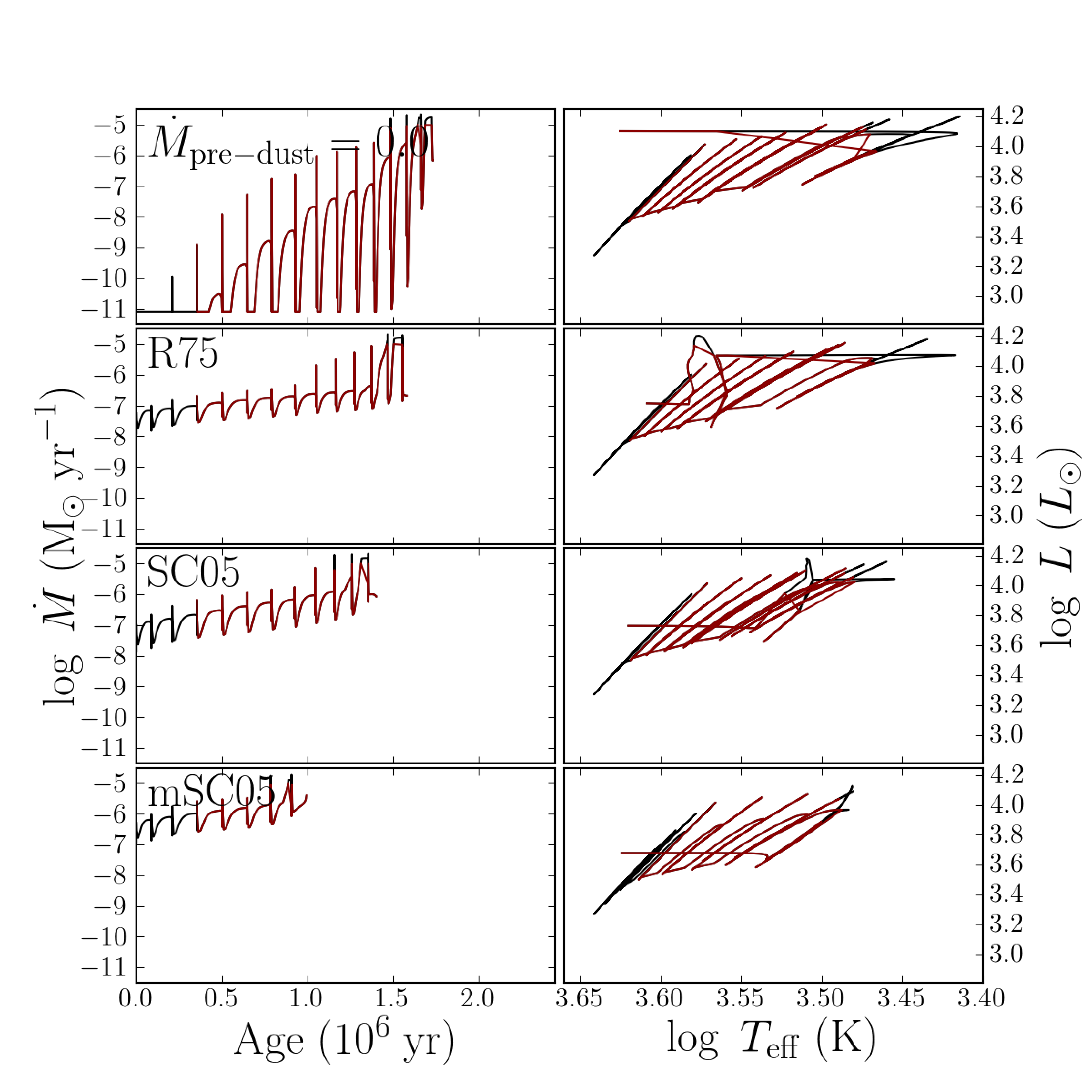}

\caption{Sample TP-AGB evolutionary tracks for initial masses $M=1.00$, (left) and $M=2.00 \msun$ (right) calculated 
with COLIBRI for TP-AGB stars under four mass-loss prescriptions. Each TP-AGB track has initial composition 
$Z=0.001$, $Y=0.25$. Left panels: mass-loss rate over time, right panels: example HR diagrams. Red lines mark when the star becomes C-rich.} 
\label{fig_compare_mass_loss}
\end{center}
\end{figure*}

It turns out that, indeed, the efficiency of $\dot M_{\rm pre\mhyphen dust}$ plays a major role in determining the lifetimes of TP-AGB stars of lower TP-AGB mass. In Figure \ref{fig_compare_mass_loss} we compare the four mass-loss options applied to compute the TP-AGB evolution of two stars with initial masses of 1\,$M_{\odot}$ and 2\,$M_{\odot}$.
The panels are organized from top to bottom following a sequence of progressively more efficient mass-loss.

As expected, the larger the mass-loss rates, the shorter the TP-AGB lifetimes. 
From the case of no mass-loss \NOVeta\ to the \NOV\ case, the TP-AGB lifetimes are reduced by a factor that depends on the stellar mass, being roughly a factor of 4 for the 1\,$M_{\odot}$ star and a factor of 2 for the 2\,$M_{\odot}$ star.

The \NOVeta\ case corresponds to the longest duration of the TP-AGB phase, and also to the reddest excursion on the HR diagram, while the TP-AGB tracks computed with \NOV\  have the shortest lifetimes  and exhibit a smaller displacement towards the coolest \Teff\ region. In general, comparing the panels from top to bottom we obtain a sequence of decreasing lifetimes and cooler HR tracks, a trend which is more pronounced at lower masses.  The differences among models for the predicted TP-AGB lifetimes and effective temperatures become less pronounced at larger masses, as shown by the results for the 2\,$M_{\odot}$ star (right panels of Figure \ref{fig_compare_mass_loss}).

The correlation between the TP-AGB lifetime and the \Teff\ redward excursion is explained as an effect related to C-star formation (when the surface C/O ratio increases from below to above unity): as more third dredge-up events are allowed to occur during the TP-AGB, a higher C-O excess is built in the atmosphere, leading to a stronger C-bearing molecular opacity, hence to a cooler Hayashi line.  This latter aspect is particularly evident for the 1\,$M_{\odot}$ star, which is able to become a  carbon star with \NOVeta\ (marked in red in Figure \ref{fig_compare_mass_loss}),  whereas it remains oxygen-rich with the other mass-loss prescriptions. 

The third dredge-up is expected to affect the TP-AGB lifetimes, essentially due to its impact on the surface chemical composition (mainly  in terms of the C/O ratio), which in turn controls both the atmospheric molecular opacity \citep{Marigo2009}, and the mineralogy of the dust grains that grow in the expanding circumstellar envelope \citep{Nanni13}. Both factors combine to influence the mass-loss rates \citep[e.g.][]{Marigo2002, Mattsson10}. In general, at the transition to the C-star regime, TP-AGB models with variable molecular opacities predict a sudden cooling of the track that makes the mass-loss rates increase (provided the adopted prescription is a sensitive function of \teff) with consequent reduction of the lifetimes. This point is fully discussed in \citet{Marigo2002}.

\begin{figure}
\begin{center}
\includegraphics[width=\columnwidth]{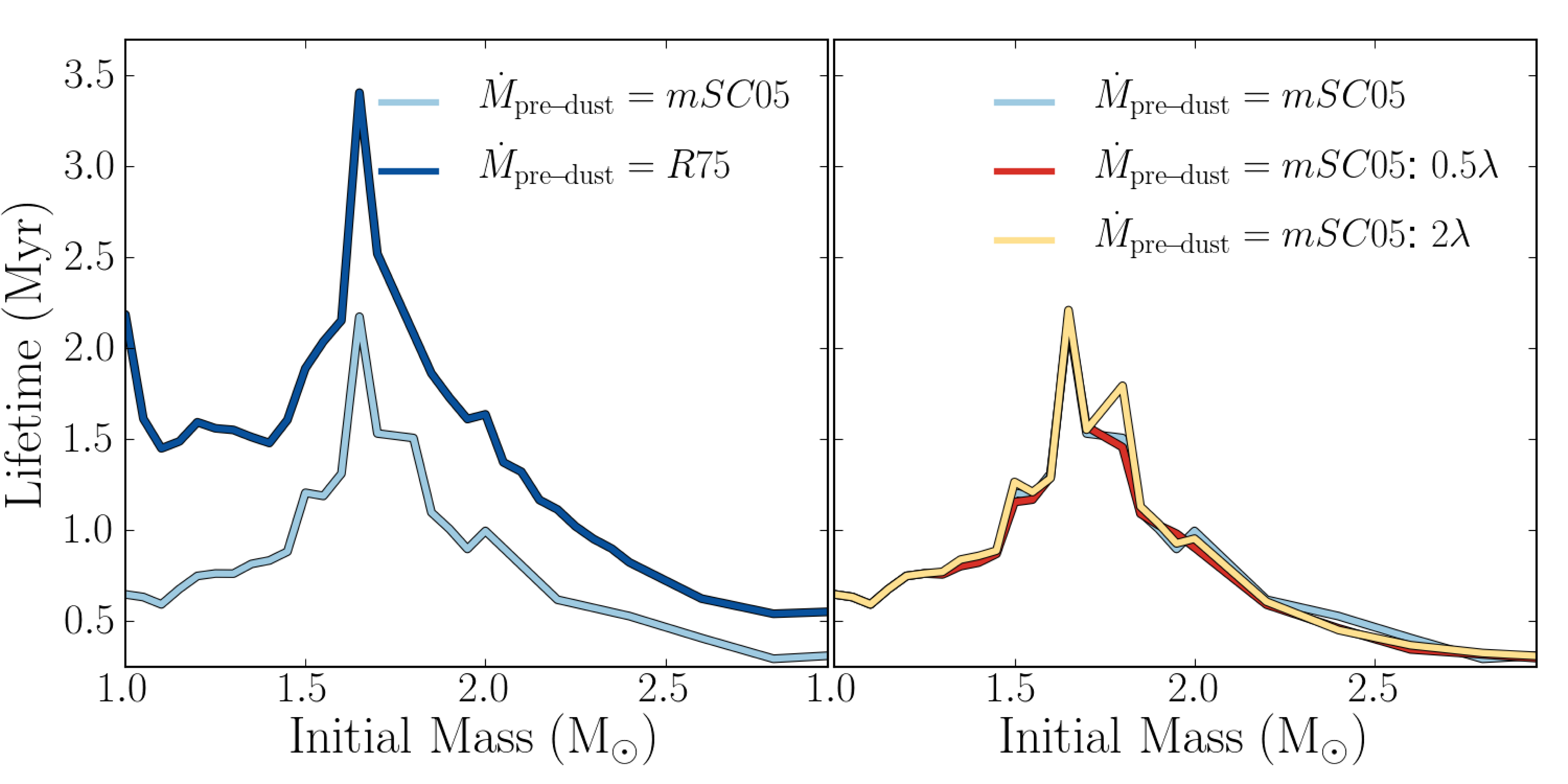}
\caption{Comparisons of expected TP-AGB lifetimes with the same dredge up efficiencies and differing mass-loss prescription (left panel) and differing dredge up efficiencies and the same mass-loss prescription (\NOV\ model, right panel), shown for one metallicity, $Z=0.001$. In the low mass, low metallicity regime, TP-AGB lifetimes are more affected the pre-dust mass-loss prescription than the efficiency of the third dredge up.}
\label{fig_lambda}
\end{center}
\end{figure}

The TP-AGB lifetime may also be sensitive to changes of the third dredge-up efficiency {\em after} the transition C-star phase. To test this possibility, we computed additional sets of TP-AGB tracks by varying the efficiency of the third dredge-up, while keeping the same \NOV\ mass-loss formalism. Expressing the efficiency of the third dredge-up with the classical parameter $\lambda$\footnote{$\lambda=\Delta M_{\rm dup}/\Delta M_{\rm c}$ is defined as the fraction of the core mass growth $\Delta M_{\rm c}$ over an interpulse period which is dredged-up up to the surface in the next TP.}, we adopt the original formalism for $\lambda$ proposed by \citet{Karakas2002}  ($\lambda_{\rm K02}$), and test the two additional choices of doubling  (2 $\lambda_{\rm K02}$) and halving (0.5 $\lambda_{\rm K02}$) the reference efficiency. We find that the predicted TP-AGB lifetimes barely change (see Fig.~\ref{fig_lambda}, right panel), even when the total amount of dredged-up material varies by a factor of $\simeq 3$. The limited effect of the third dredge-up can be explained as a combination of two main factors. The first is due to the efficiency of the mass-loss prescription adopted here. The \NOV\ model is quite efficient and the TP-AGB phase terminates quickly (for instance, at $Z=0.001$, the total number of thermal pulses is $3-15$ in the relevant initial mass range). Therefore, there simply is not enough time for the third dredge-up to produce dramatic effects, no matter how it is varied. In addition, as the star reaches the super-wind phase, the mass-loss rates settle to typical values that are little affected by variations of other stellar parameters.

The second factor is related to the sensitivity of the effective temperature as a function of the carbon excess in the atmosphere of carbon stars. In general, more carbon excess corresponds to lower effective temperature which causes higher rates of mass-loss. This response of the effective temperature to the increase of carbon can be described by the cooling rate (the derivative  $|d(\log T_{\rm eff})/d({\rm C/O})|$) which is expected to progressively decrease at increasing C/O ratio \citep[see Section 7.3 and Figure 17 of][]{Marigo2013}. In other words, after the initial sizeable drop of \teff\ once C/O exceeds unity, as more third dredge-up events continue to take place, the atmospheric structure becomes less and less sensitive to further increase of carbon. As a consequence, the impact of the third dredge-up on the effective temperature becomes progressively weaker as more mixing events occur.

In short, the duration of the TP-AGB phase is mainly controlled by mass-loss--at least for the mass/metallicity interval being considered here--while the effect of the third dredge-up is limited (see Fig.~\ref{fig_lambda}, right panel).  This result is important, as it strengthens the robustness of the analysis described next, whose primary aim is to obtain a quantitative estimation of the TP-AGB lifetimes as a function of the initial stellar mass, in the low-metallicity regime.

However, we must also emphasize that although the TP-AGB lifetimes are found to be little influenced by the third dredge-up, the chemical composition of the ejecta is much affected by the properties of the mixing events. In particular, for the same amount of mass lost, the quantity of primary carbon, and hence of carbonaceous dust, that is injected in the interstellar medium does depend strongly on the efficiency of the third dredge-up. Therefore, although it is outside the scope of this paper, future calibration of the third dredge-up process is an essential step towards a comprehensive description of the TP-AGB phase that includes not only the spectro-photometric but also the chemical role of TP-AGB stars in the context of galaxy evolution.

On a final note, we exclude the original \citet{Schroeder2005} from the subsequent discussion and analysis. As is shown in Figure \ref{fig_compare_mass_loss}, \citet{Schroeder2005} mass-loss rate predicts a longer lifetime than the \NOV\ model. 
There are two main reasons that the \citet{Schroeder2005} prescription is no longer satisfactory for the present work. First, the computations discussed here are based on a completely new release of stellar evolution models \citep{Bressan2012, Marigo2013}  in which major modification and update of the input physics were introduced. This leads, for instance, to produce Hayashi lines that are on average somewhat cooler than in G10. Second, we found that if we vary the mass-loss efficiency ($\eta_{\rm SC05}$) enough to account for the number of AGB stars in our predominantly  old and metal-poor galaxies  (increasing the efficiency  parameter $\eta_{\rm SC}$), the effect on higher mass AGB stars (i.e. $M_{\rm i} \ga 2 M_{\odot}$)  would be dramatic. In fact, that would lead to an extreme shortening of their lifetimes producing a deficit in contrast with the observations of AGB stars at typical metallities of the SMC  and LMC ($Z\approx 0.004 - 0.008$). Finally, we underline that the modification to the original \citet{Schroeder2005} arises from replacing a simple assumption with a more physically-sound consideration of the dependence of the mass-loss rate on the mechanical magnetic flux  \citep{Cranmer2011}. For these reasons, as discussed above, a new \teff\ scaling was in order, so we proceed with the \NOV\ model in lieu of $\dot{M}_{\rm pre\mhyphen dust}^{\rm SC05}$.

\section{Modeling the Data}
\label{sec_model_data}

With optically derived SFHs of each galaxy and three sets of TP-AGB models, we now turn to our method of using star counts to robustly constrain TP-AGB lifetimes. We now discuss how we apply the derived SFHs and use TRILEGAL to create model LFs. 

\subsection{Population Synthesis with TRILEGAL}
\label{sec_trilegal}
Following \citetalias{Girardi2010}, we use TRILEGAL \citep{Girardi2002} to synthesize the stellar populations for direct comparison to observations. TRILEGAL takes as input the PARSEC and COLIBRI stellar evolution libraries, a specified initial mass function (IMF), binary fraction, and the SFH.  Importantly, TRILEGAL also simulates the $L$--\Teff\ variations due to the thermal pulse cycles on the TP-AGB \citep{Girardi2007}, and the reprocessing of radiation by their circumstellar dust-shells \citep[as in][]{Marigo2008}. For previous evolutionary phases, TRILEGAL provides simulations which are essentially identical to those performed by MATCH.

The TRILEGAL input parameters are set to remain consistent with the parameters used in the SFH recovery (see Section \ref{sec_sfh_z}).  The stars produced by TRILEGAL are converted into absolute magnitudes in $HST$ filters using the set of bolometric corrections and extinction coefficients described in \citet{Girardi2008}, which are mostly based on ATLAS9 \citep{Castelli2003} synthetic spectra, but with two important updates for cool giants: M giants now come from an extended database from Aringer et al. (in prep.) that covers the all relevant space of parameters (\Teff, \logg, and \feh). For C-type stars, we adopt the \citet{Aringer2009} library of C star models, interpolating inside the grids as a function of \Teff, \logg, \feh\ and C/O ratio. Radiation reprocessing by circumstellar dust shells in mass-losing stars are taken into account as in \citet{Marigo2008}, using the results of \citet{Groenewegen2006}'s radiation transfer models for mixtures of 60\% AlOx and 40\% silicate, and of 85\% amorphous Carbon and 15\% silicon carbide (for M and C stars, respectively). Finally, the synthetic CMDs are corrected for distance and extinction, $A_V$ using extinction coefficients from \citet{Girardi2008}, as listed in Table \ref{tab_sample}.

\subsection{Accounting for Uncertainties in SFH}
\label{sec_sfh_err}
To obtain a robust number ratio of TP-AGB stars to RGB stars as well as the range of probable LFs expected from a given TP-AGB model, one must account for the random uncertainties in SFH. We synthesize at least 50 stellar populations with SFHs that are randomly sampled within the uncertainties of the best fit SFH. Specifically, for each SFH sample, we take the value of SF in each time bin as a random draw of a Gaussian distribution whose mean is the best fit SFR in that time bin and whose $\sigma$ is the uncertainty associated with that time bin. If the SFR is zero in the time bin, we adopt only positive uncertainties.

\begin{figure}
\begin{center}
\includegraphics[width=\columnwidth]{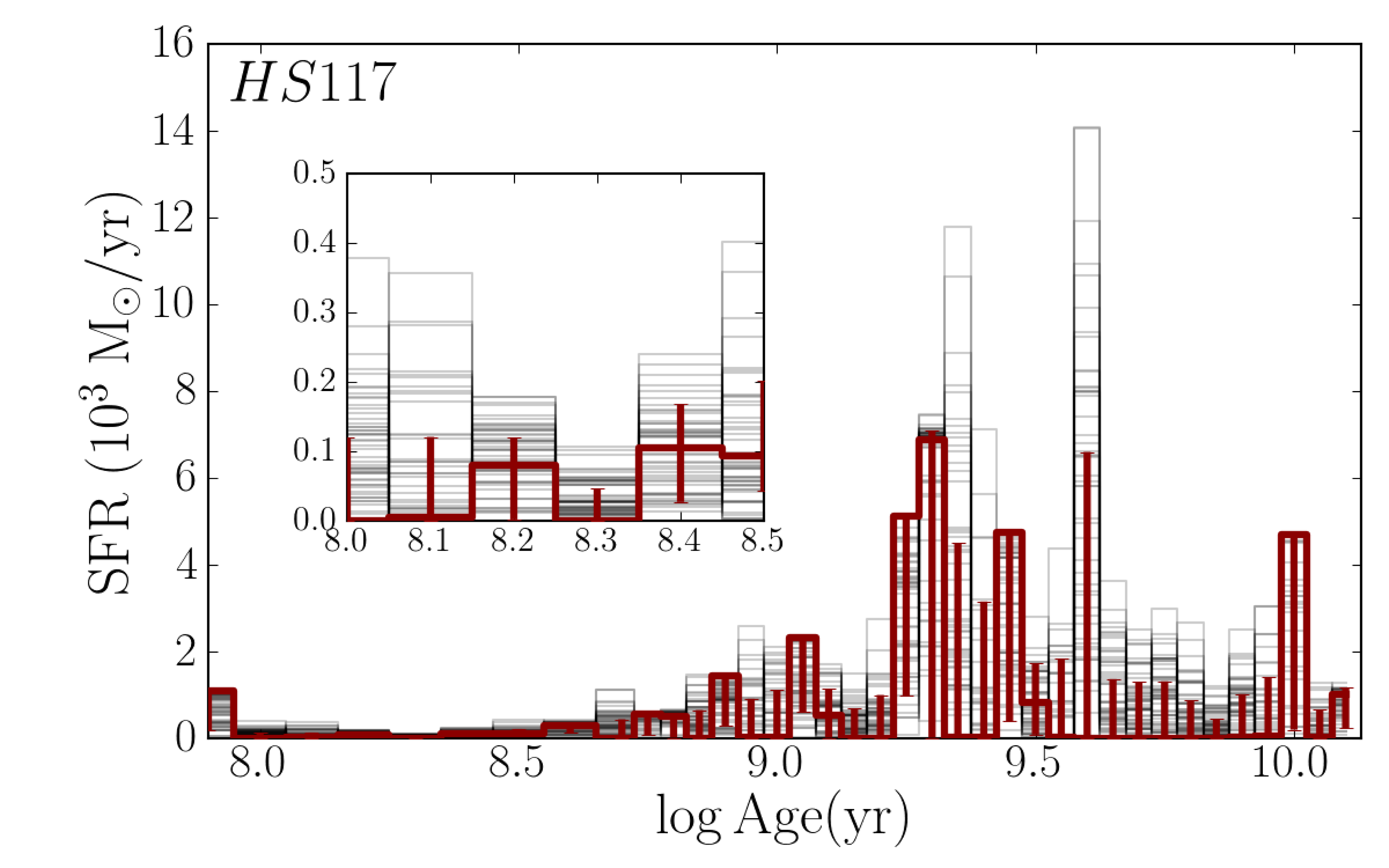}
\caption{Example of 50 SFHs (black) sampled from the best-fit SFH (red) and random and systematic uncertainties derived in MATCH from HS117 optical photometry and PARSEC isochrones.}
\label{fig_random_sfr}
\end{center}
\end{figure}

Figure \ref{fig_random_sfr} shows an example of 50 SFH realizations based on the MATCH-derived SFH and hybrid Monte Carlo uncertainties \citep[see][]{Dolphin2013}. The effect of the randomly sampled SFH can be seen in the Figures \ref{fig_oct13_lfs}-\ref{fig_nov13_lfs} as a spread in LF at bright magnitudes. Accounting for uncertainties in the SFH shows a clearer picture of the model predictions on the LF by introducing a spread in the number of stars that are expected to be found in each magnitude bin on a LF. The effect on the mean \narratio\ ratio is to consistently produce a standard deviation of $\sim 10\%$, independent of the TP-AGB model. 

\subsection{Creating LFs of the Galaxy Sample}
\label{sec_model_lfs}
For each SFH of each galaxy and each TP-AGB model, we generate a model stellar population with TRILEGAL of sufficient size to both to completely sample the IMF and to have at least twice the number of RGB stars in the sample as there are in the data. We then correct for the discrepancy in total stellar mass by scaling the model LF by the number of stars in the RGB region of the data (defined in Section \ref{sec_agb_rgb_selection}). In other words, we multiply the simulated LF by a factor $x$, such that,

\begin{equation}
x = N_{\rm RGB, data}/N_{\rm RGB, simulation} \ .
\end{equation}

\noindent Example LFs of an un-scaled simulation are shown in Figure \ref{fig_example_lf}, all simulated LFs in the figures following are scaled.

To test the adequacy of the TP-AGB model, we compare the amplitude and the shape of the model LF to the observations. The number of stars in the model LF are compared to that in the data by calculating the \narratio\ ratio as described in Section \ref{sec_agb_rgb_selection}.  The \narratio\ ratio is related  to the average TP-AGB lifetime of the observed population, as it combines all the star counts into one data point.

The \narratio\ ratio is a useful first comparison to make between models, as well as comparisons to other studies (e.g., G10). A successful TP-AGB model must also match the shape of the observed LF. Therefore, we also compare the predicted LF with those of the observations.

We calculate the Poisson-equivalent of the Gaussian $\chi^2$ statistic \citep[see e.g., discussion in][]{Dolphin2002} in two regions of the LF to compare the model LF shape to that in the data. The first region is the ``full LF'' that is, from the 90\% completeness magnitude and brighter (including the regions excluded from the \narratio\ ratio calculation). The second region is only the TP-AGB region, defined as brighter than a small offset above the TRGB (0.1 mag in $F814W$ and 0.2 mag in $F160W$; see Section \ref{sec_agb_rgb_selection}). 

\section{Analysis}
\label{sec_analysis}

\subsection{Ratio of TP-AGB to RGB stars}
\label{sec_res_narratio}

Tables \ref{tab_opt_narratio} and \ref{tab_ir_narratio} list the mean \narratio\ and standard deviation of their Poisson uncertainties calculated from 50 simulations for each galaxy for each TP-AGB model (columns 2, 4, and 6). Next to each mean \narratio\ ratio are the fractional difference between the mean model ratio and that found in the data (columns 3, 5, and 7). We define the fractional difference to be $(\narratio)_{\rm data} = f (\narratio)_{\rm model}$. Therefore, a value of $f=1$ would be perfect agreement between data and model while $f>1$ would mean the model is overpredicting the number, and thus, the lifetimes of TP-AGB stars.

For each individual galaxy, the \OCT\ and \NOVeta\ models consistently overpredict the \narratio\ ratio. On average, the \NOVeta\ mass-loss prescription overpredicts the number of optical TP-AGB stars by nearly a factor of 3 and the \OCT\ mass-loss law overpredicts them by more than a factor of 2. In the NIR, the overpredictions of TP-AGB stars are similar or higher, nearly a factor of 3 when not accounting for pre-dust mass-loss, and a factor of 2.5 when only including Reimers' mass-loss. The \NOV\ model, however, is on average consistent or slightly lower within uncertainties to the observed \narratio\ ratio in both filters.

The conclusion from comparing the \narratio\ ratio from model to model, is that there are too many TP-AGB stars predicted by the mass-loss prescriptions that neglect pre-dust mass-loss, or assume only Reimers' relation. In contrast, pre-dust mass-loss in the \NOV\ model is most consistent with the average lifetimes of observed TP-AGB stars.

\subsection{Luminosity Functions in the Optical and NIR}
\begin{figure*}
\includegraphics[width=0.48\textwidth]{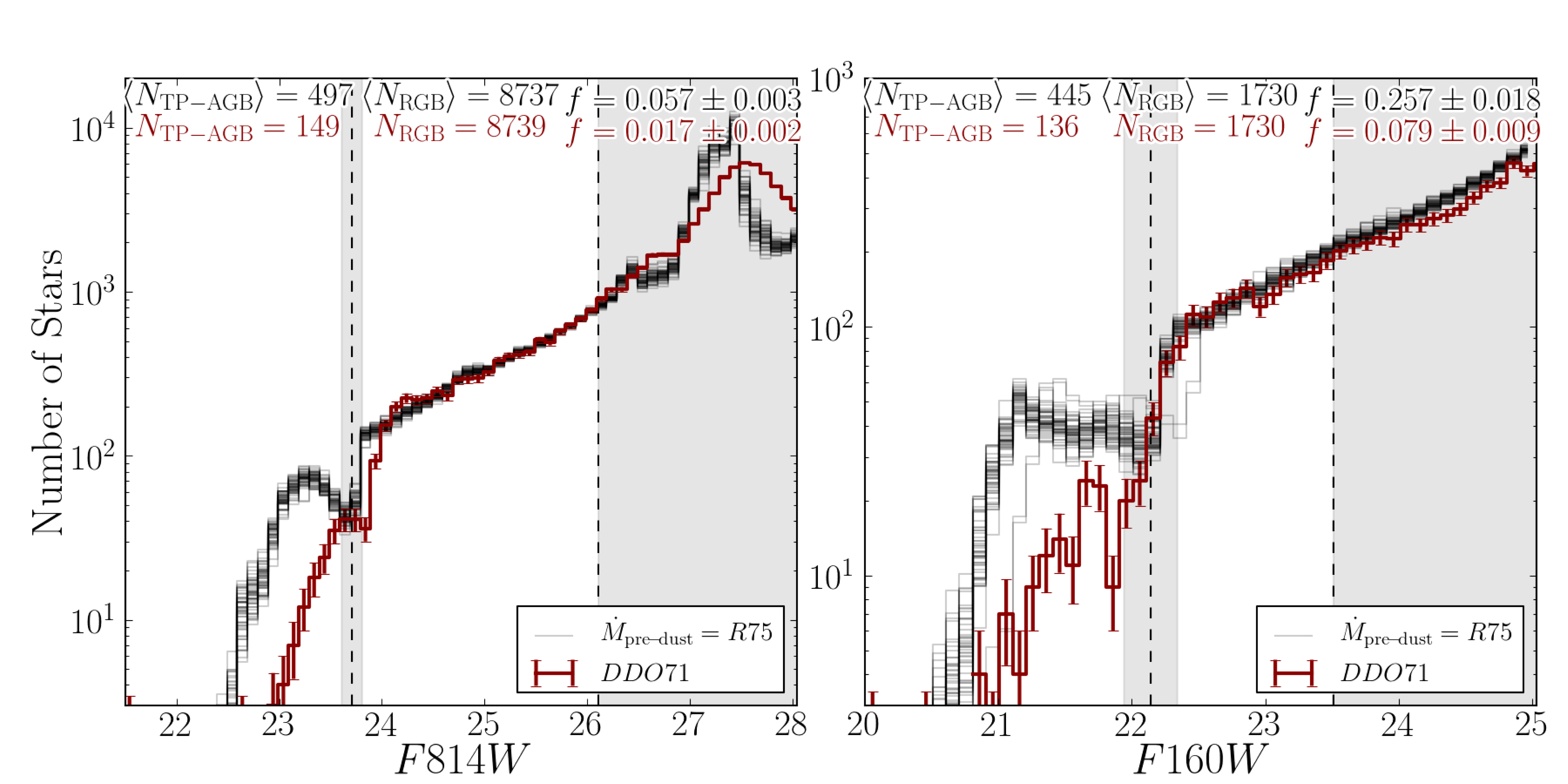}
\includegraphics[width=0.48\textwidth]{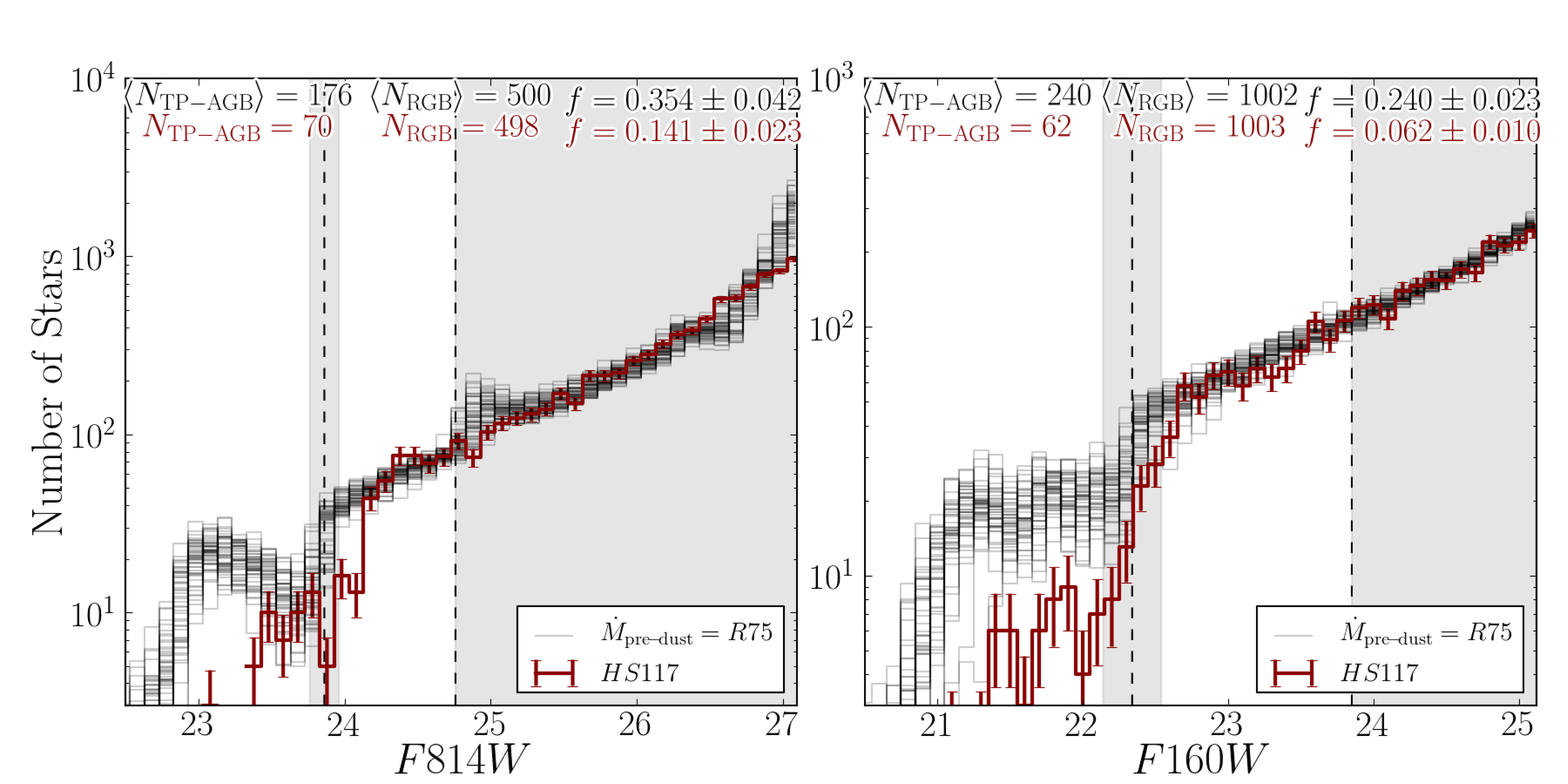}
\\
\includegraphics[width=0.48\textwidth]{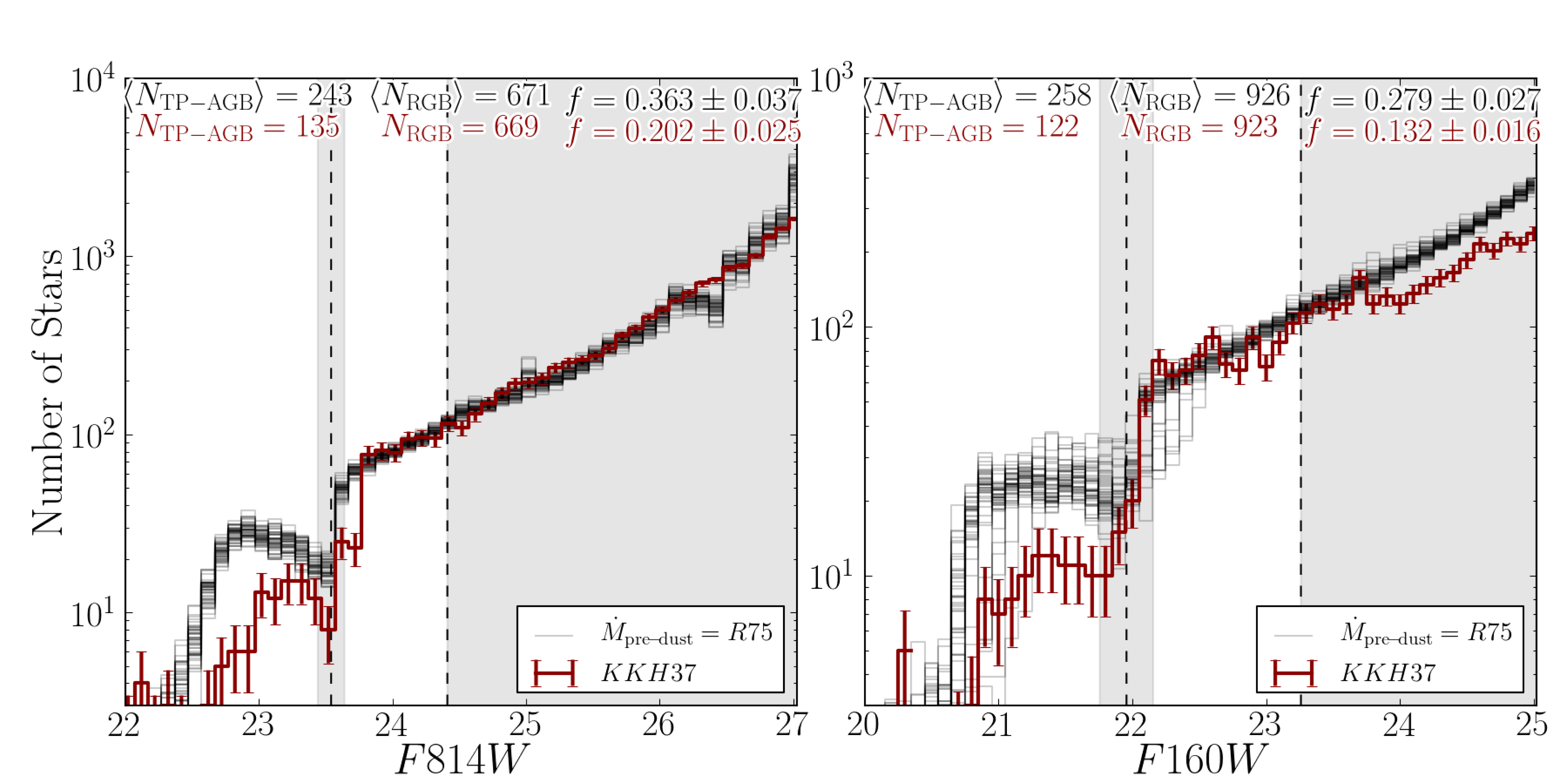} 
\includegraphics[width=0.48\textwidth]{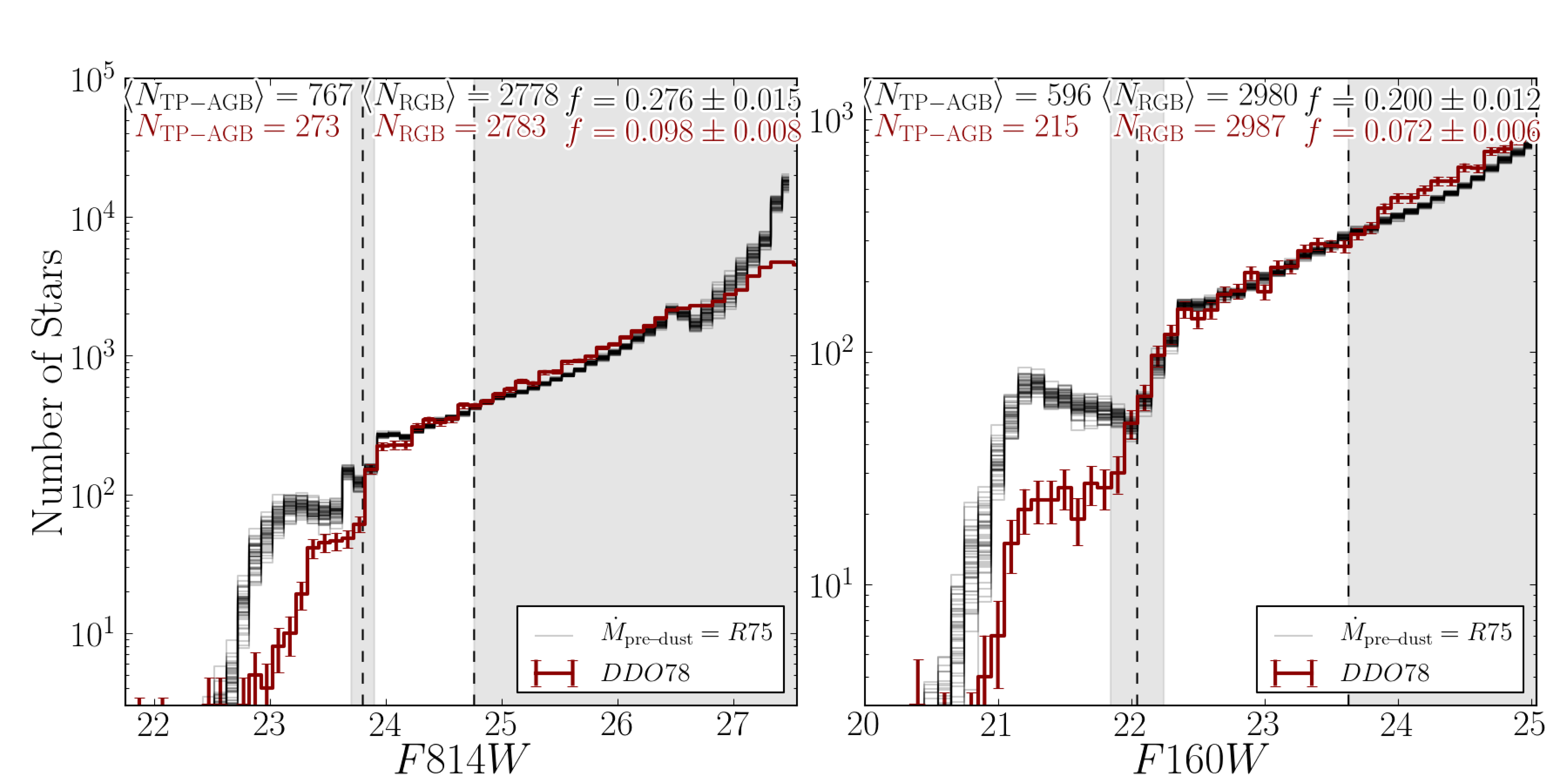}
\\
\includegraphics[width=0.48\textwidth]{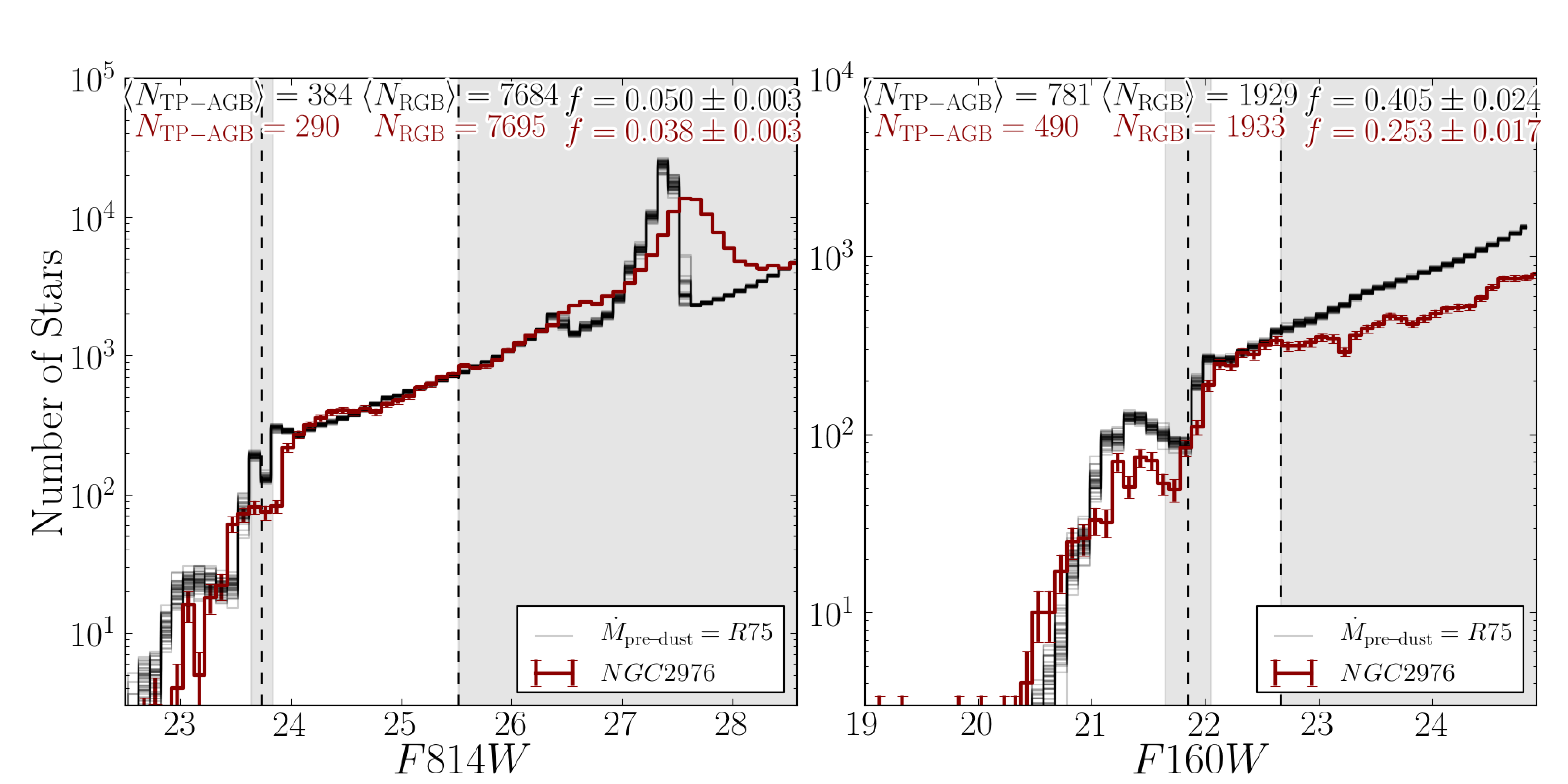}
\includegraphics[width=0.48\textwidth]{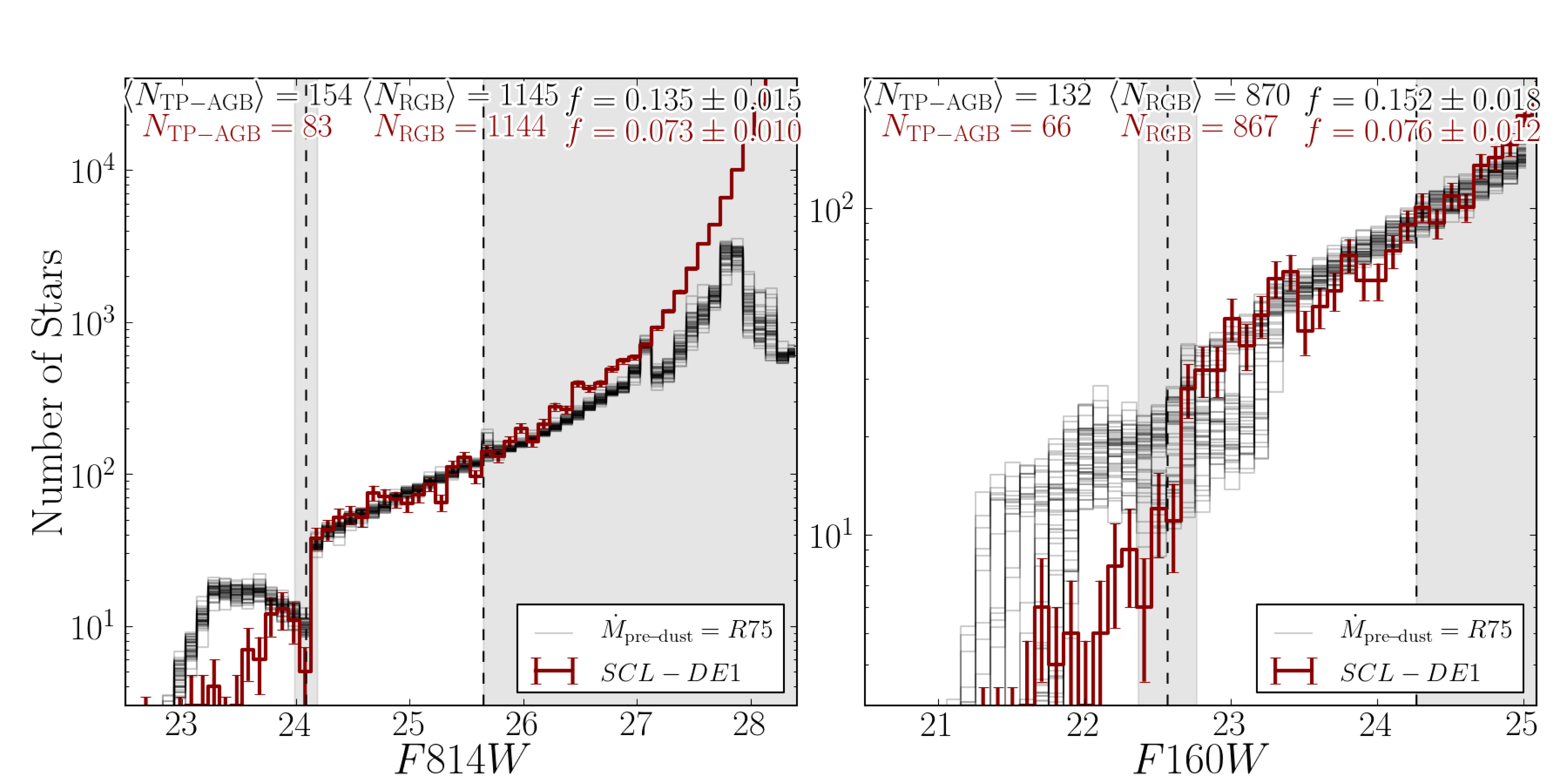}
\caption{Data and \OCT\ model LFs. Each panel has optical (left) and IR (right) LFs. Observations are plotted in red with Poission uncertainties. Each SFH realization is shown in grey. Data and mean model values of the \narratio\ ratio (Tables \ref{tab_opt_narratio} and \ref{tab_ir_narratio}) are written at the top of each panel. Faint axes limits are the 50\% completeness in the data, faintest dashed line denotes the 90\% completeness limit. Shaded regions around the TRGB are excluded in \narratio\ ratio calculations. All model LFs are scaled to match the number of stars in the data in the RGB region. Deviations between data and model faintward of the 90\% completeness limit is likely caused by photometric uncertainties. Deviations between data and model above the TRGB are due to inadequacies in the TP-AGB model.}
\label{fig_oct13_lfs}
\end{figure*}

\begin{figure*}
\includegraphics[width=0.45\textwidth]{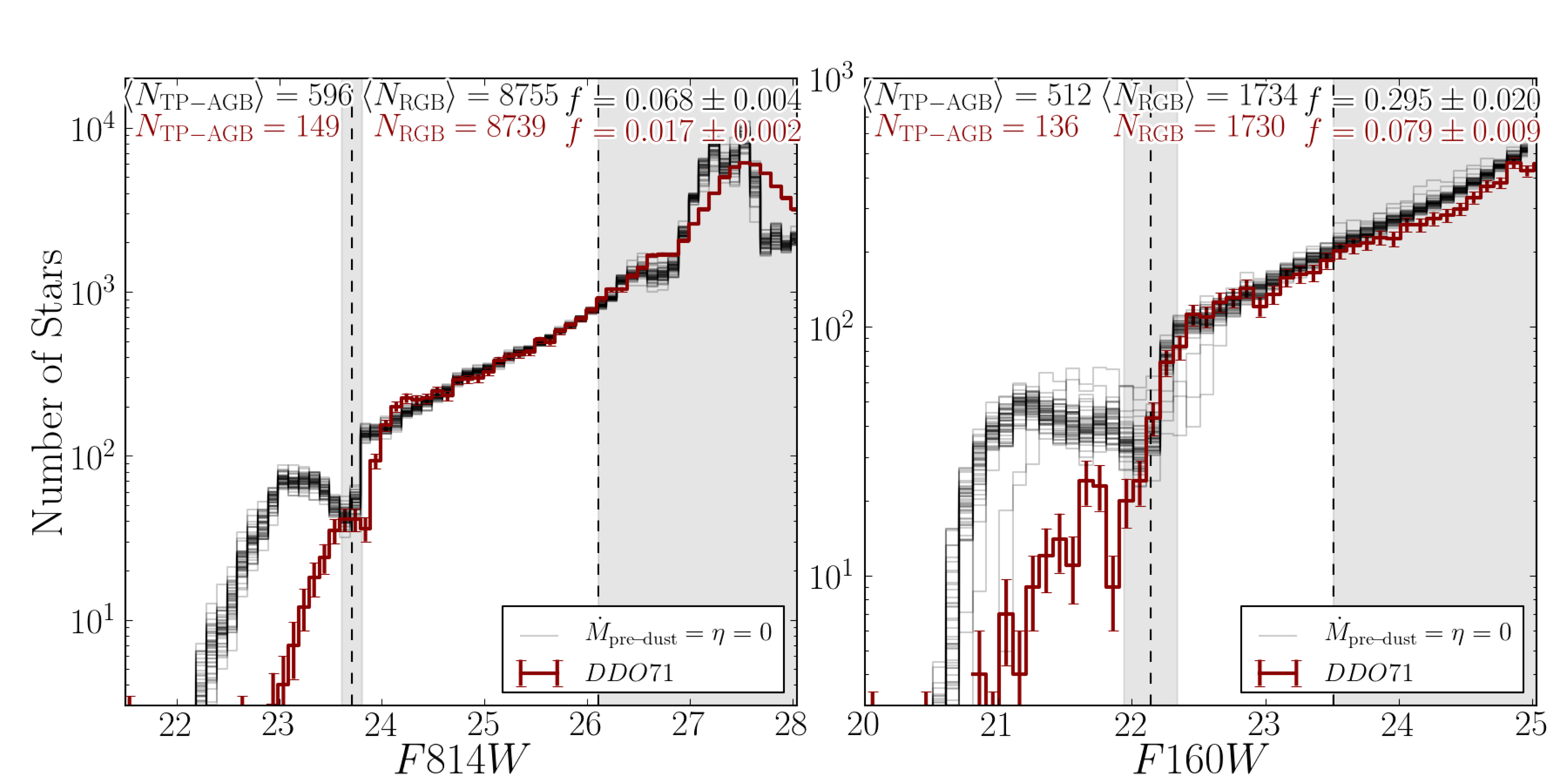}
\includegraphics[width=0.45\textwidth]{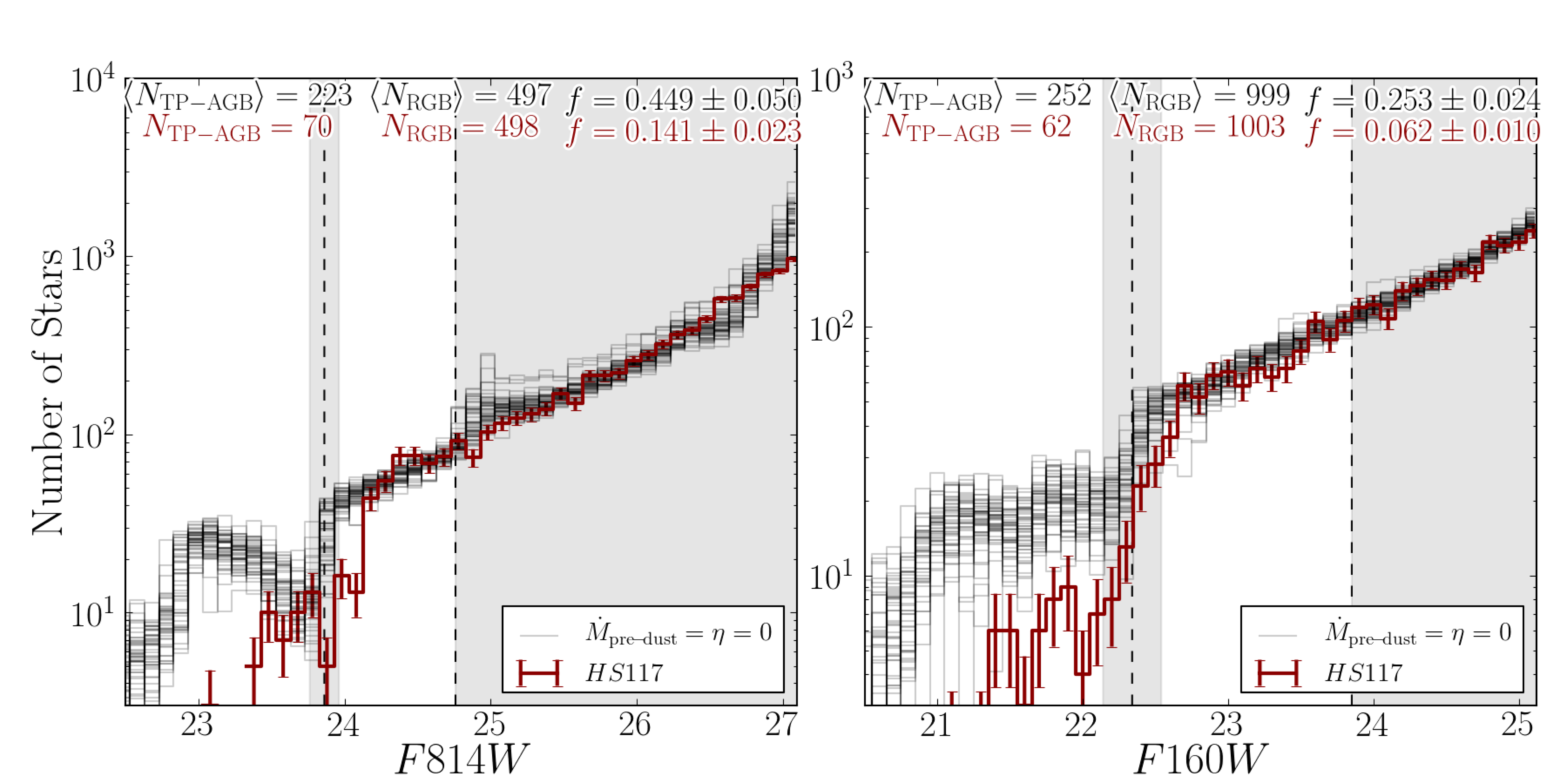}
\\
\includegraphics[width=0.45\textwidth]{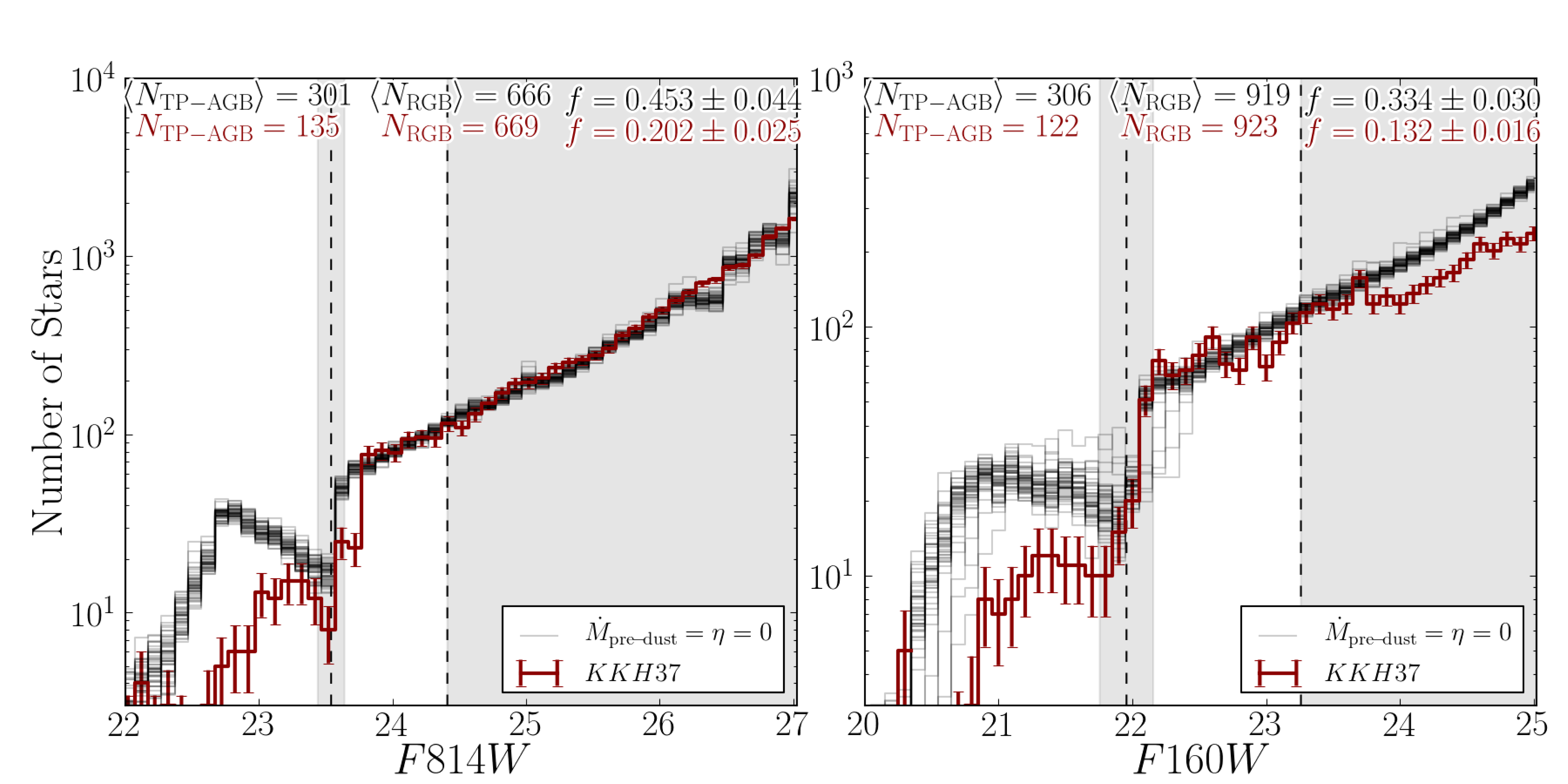}
\includegraphics[width=0.45\textwidth]{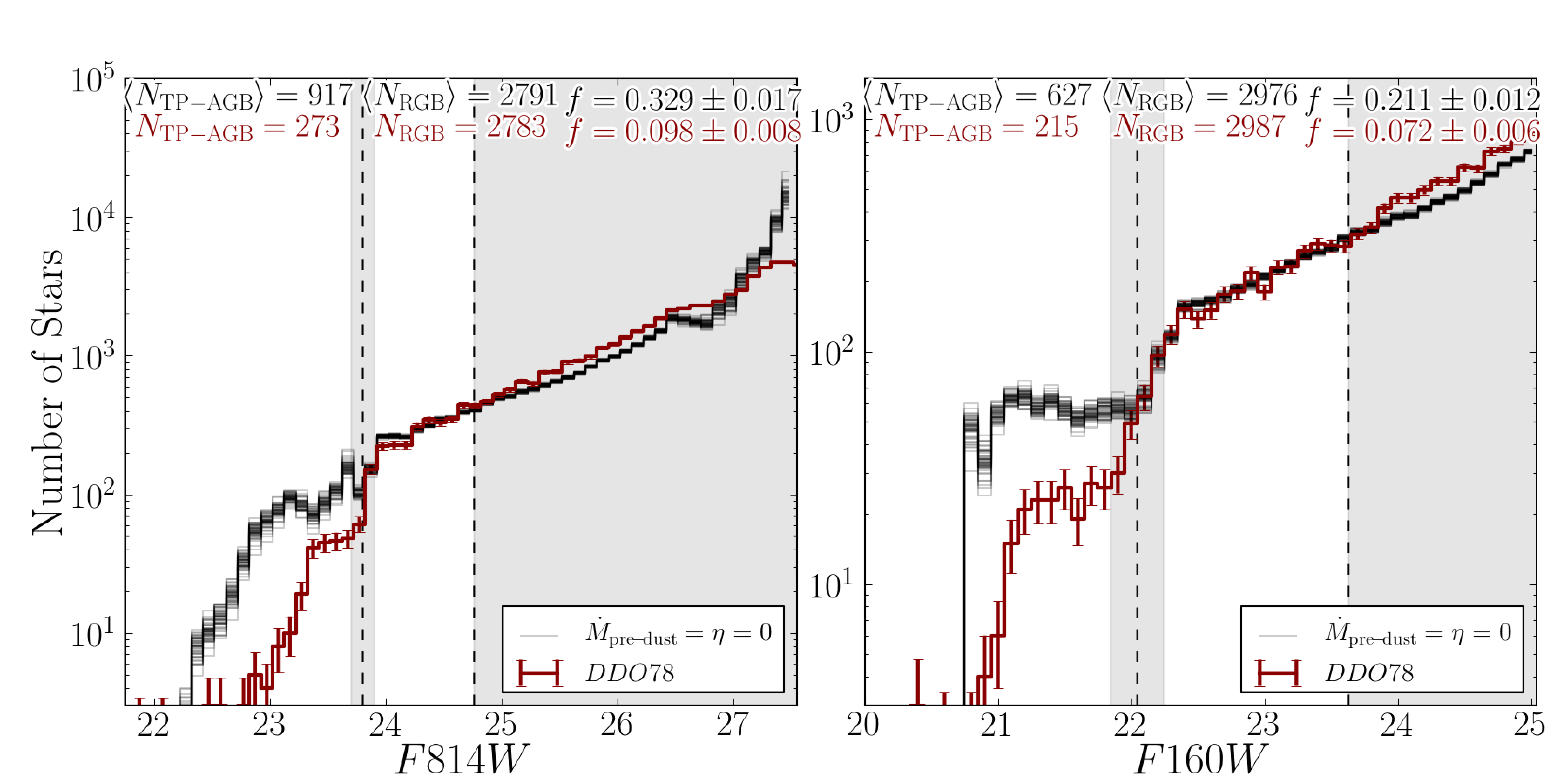}
\\
\includegraphics[width=0.45\textwidth]{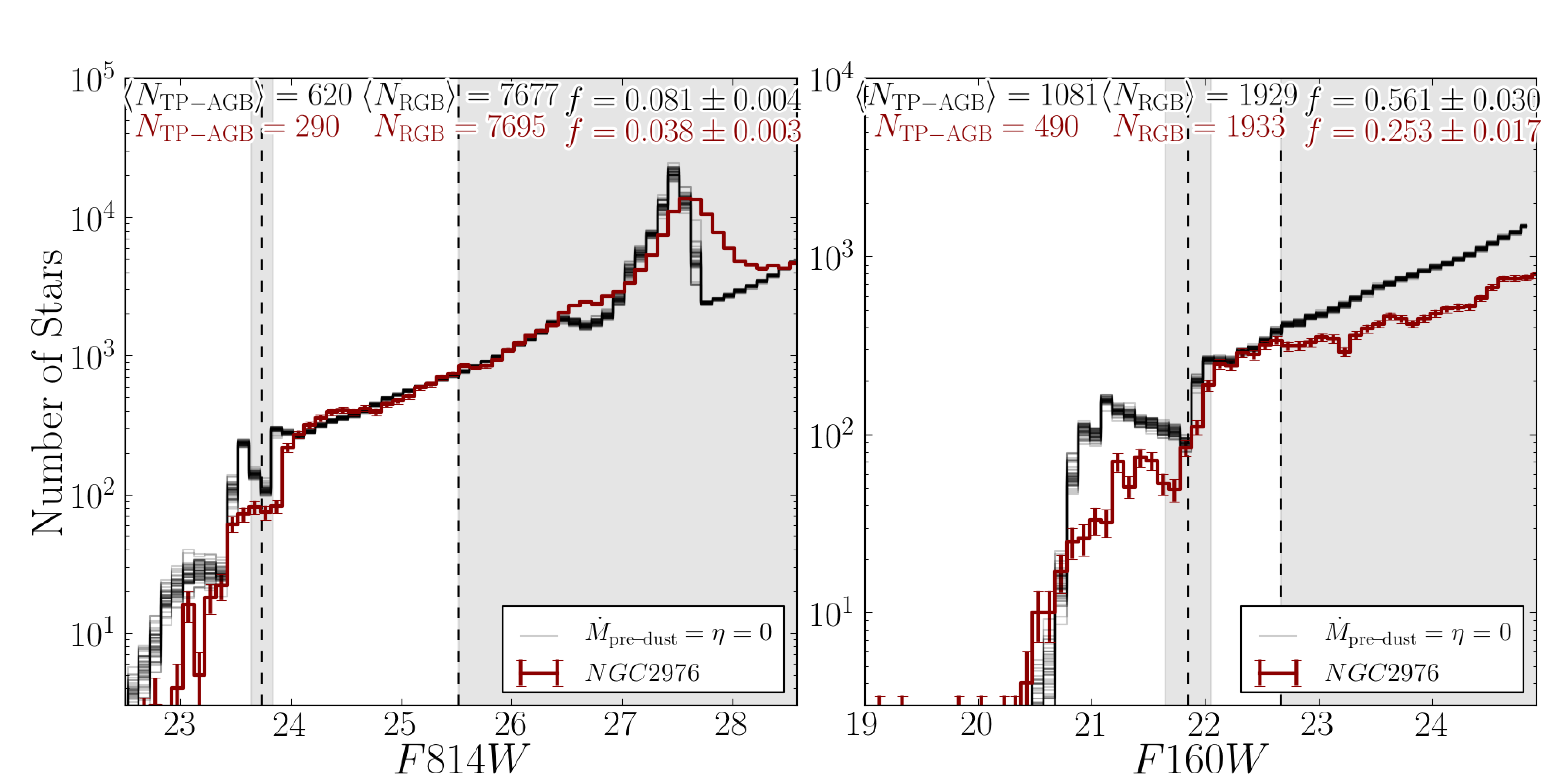}
\includegraphics[width=0.45\textwidth]{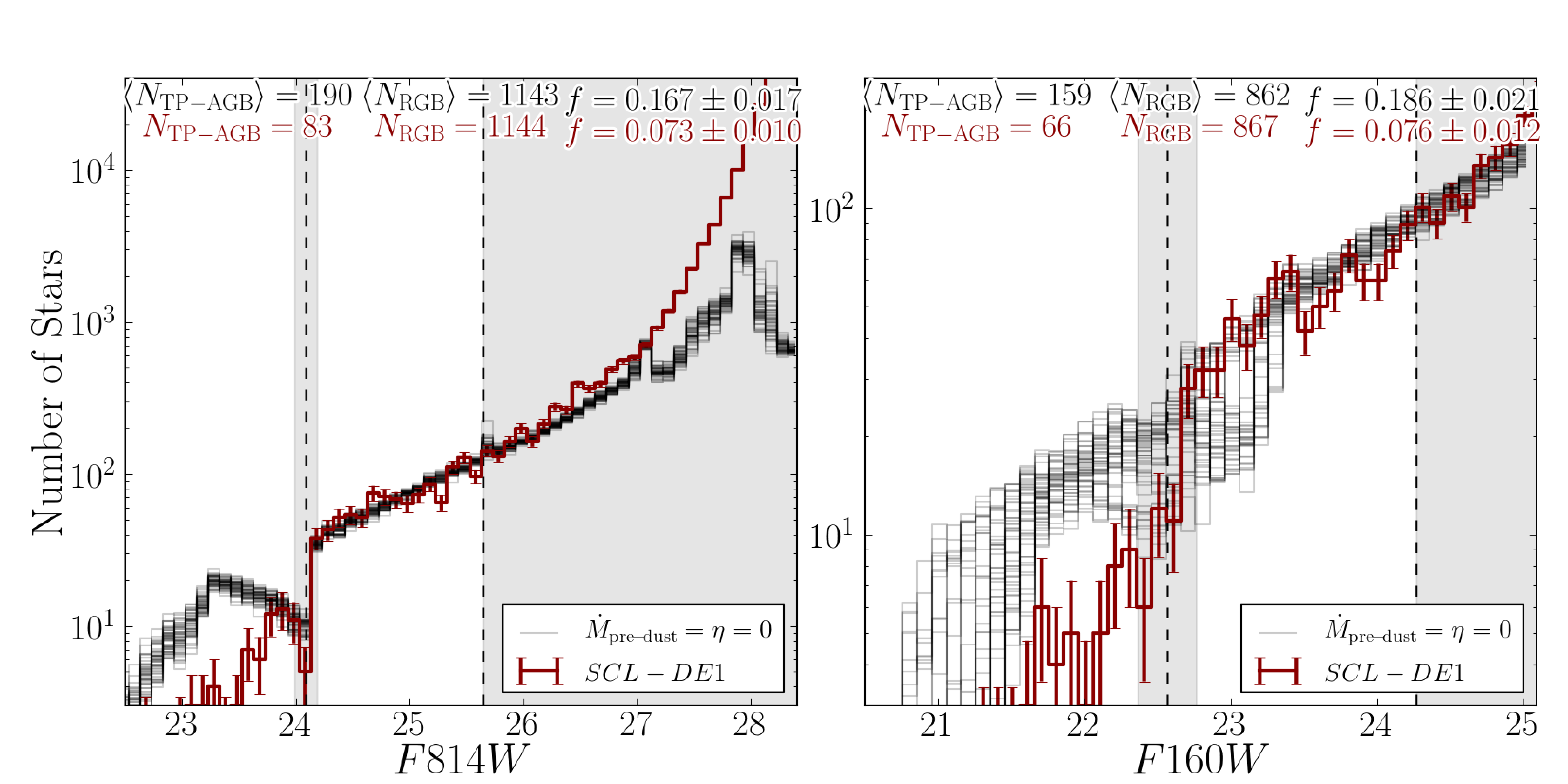}
\caption{Same as Figure \ref{fig_oct13_lfs} but with \NOVeta.}
\label{fig_nov13eta0_lfs}
\end{figure*}

\begin{figure*}
\includegraphics[width=0.45\textwidth]{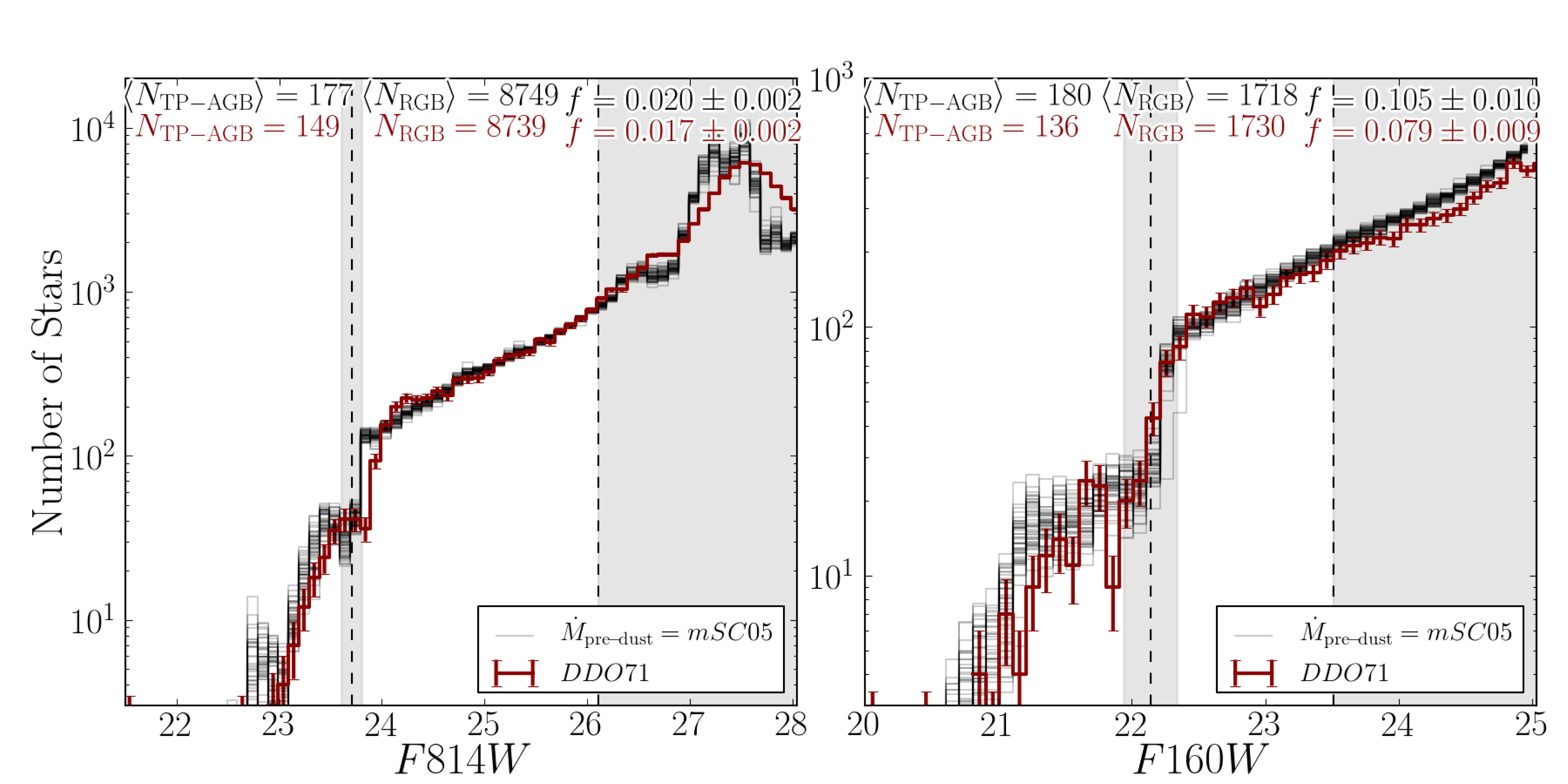}
\includegraphics[width=0.45\textwidth]{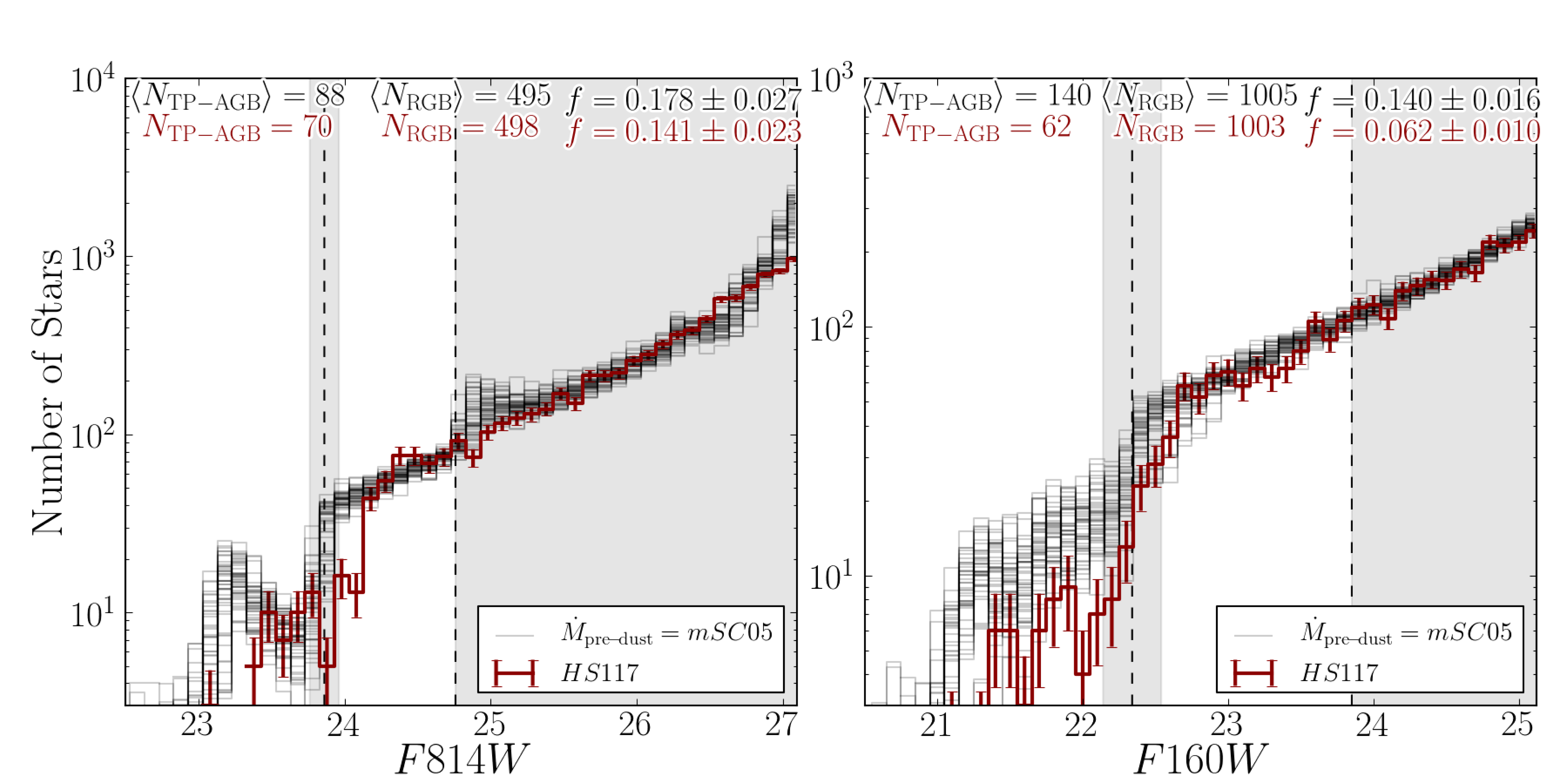}
\\
\includegraphics[width=0.45\textwidth]{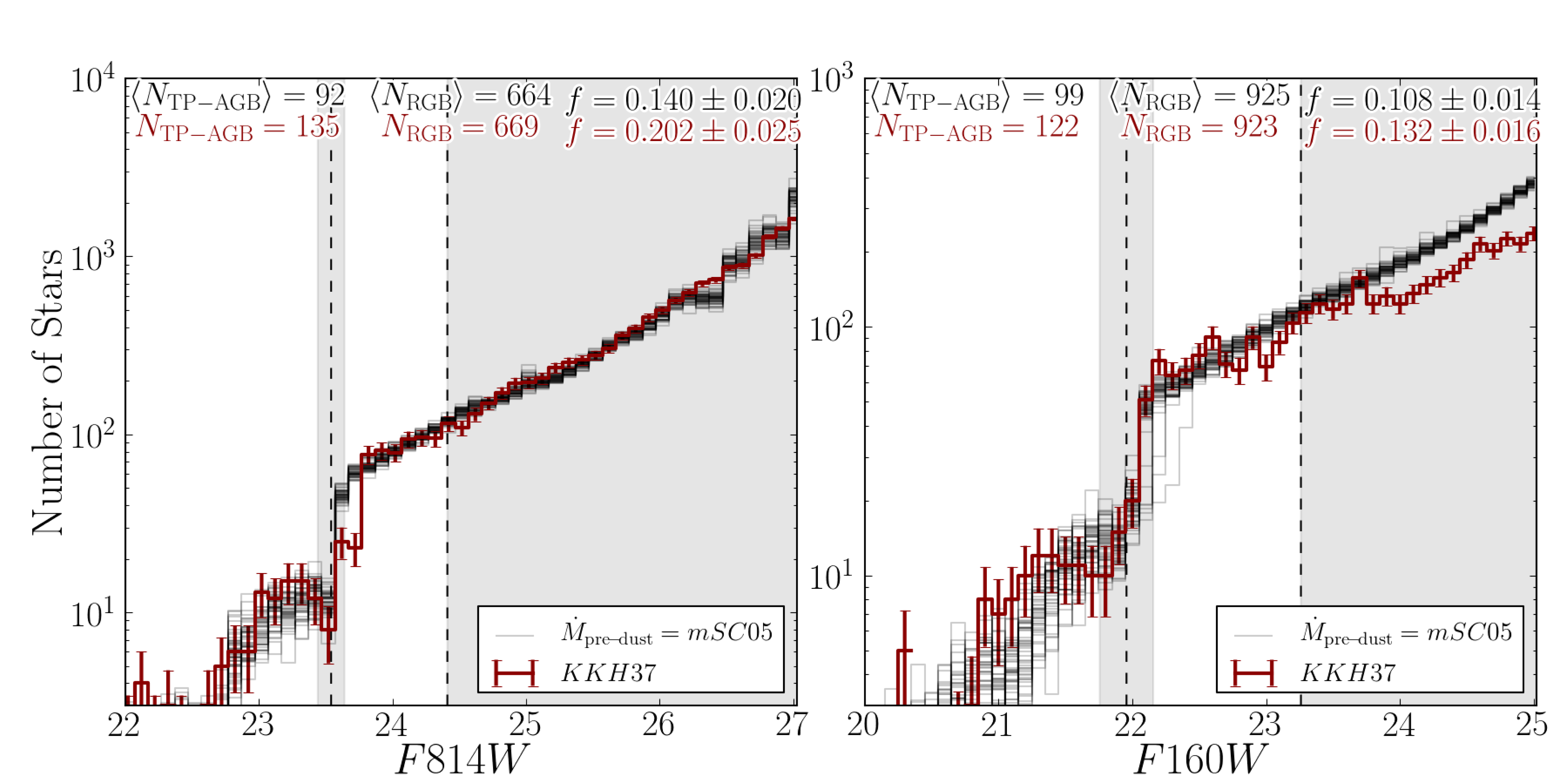}
\includegraphics[width=0.45\textwidth]{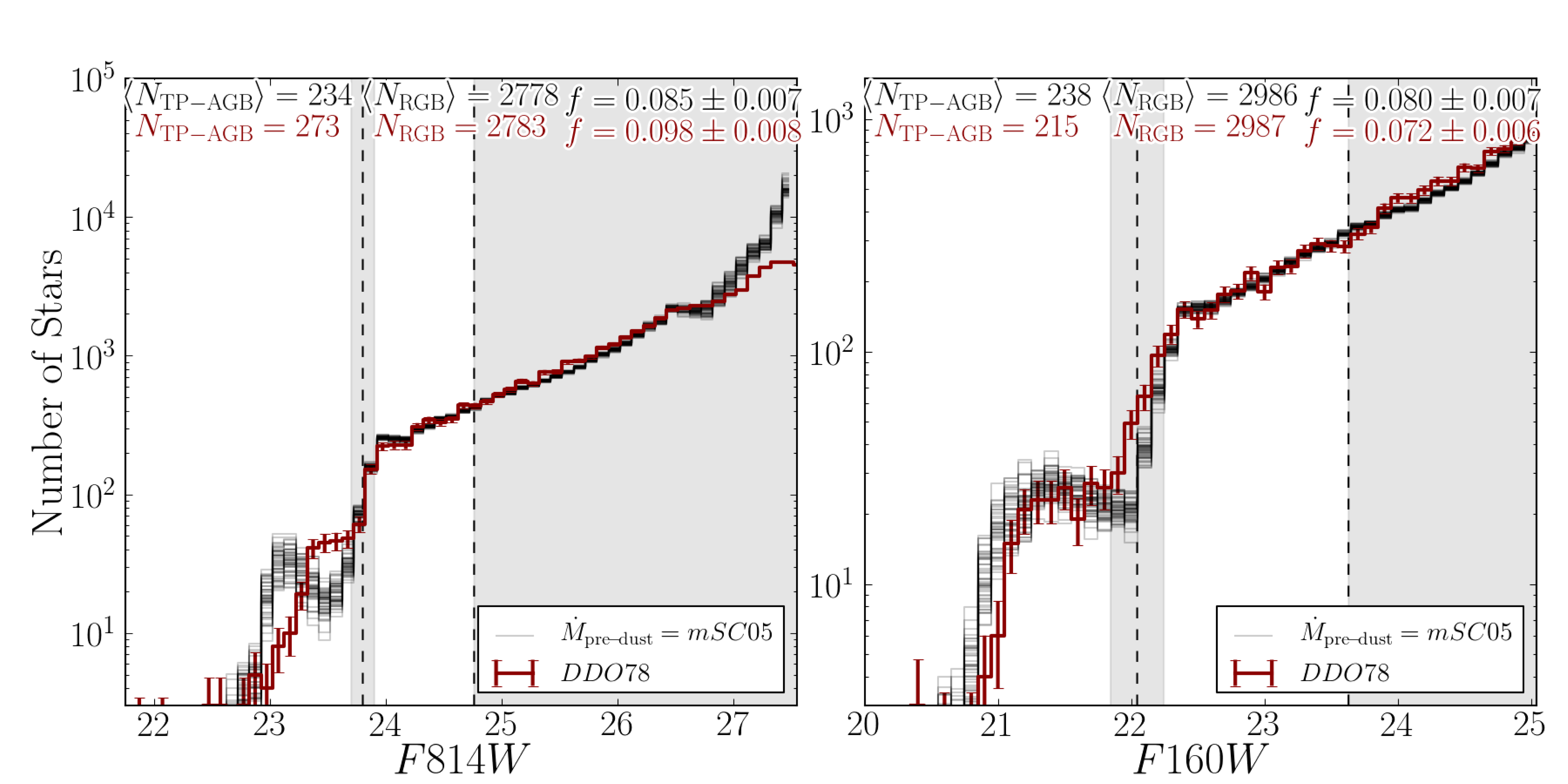}
\\
\includegraphics[width=0.45\textwidth]{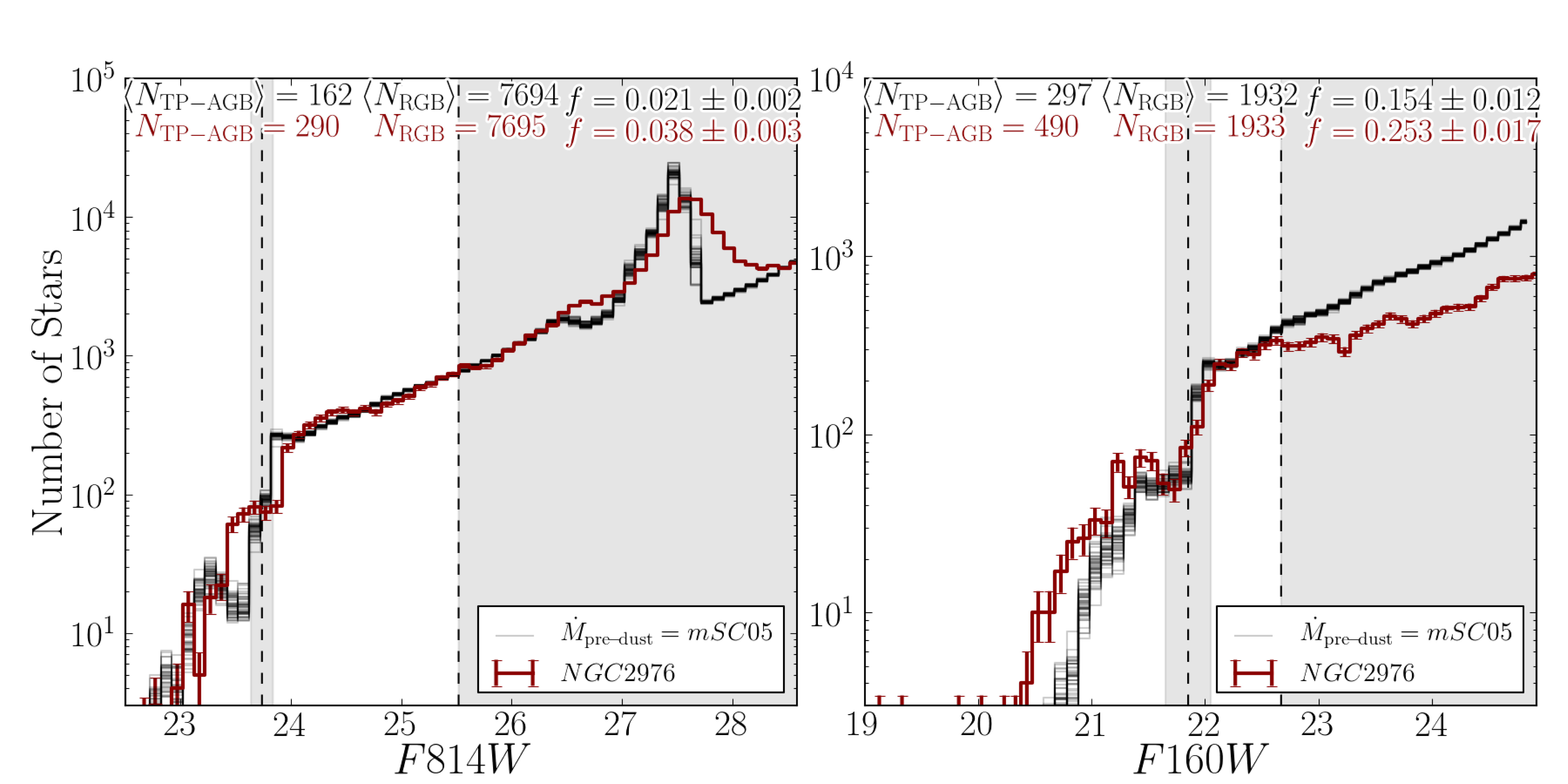}
\includegraphics[width=0.45\textwidth]{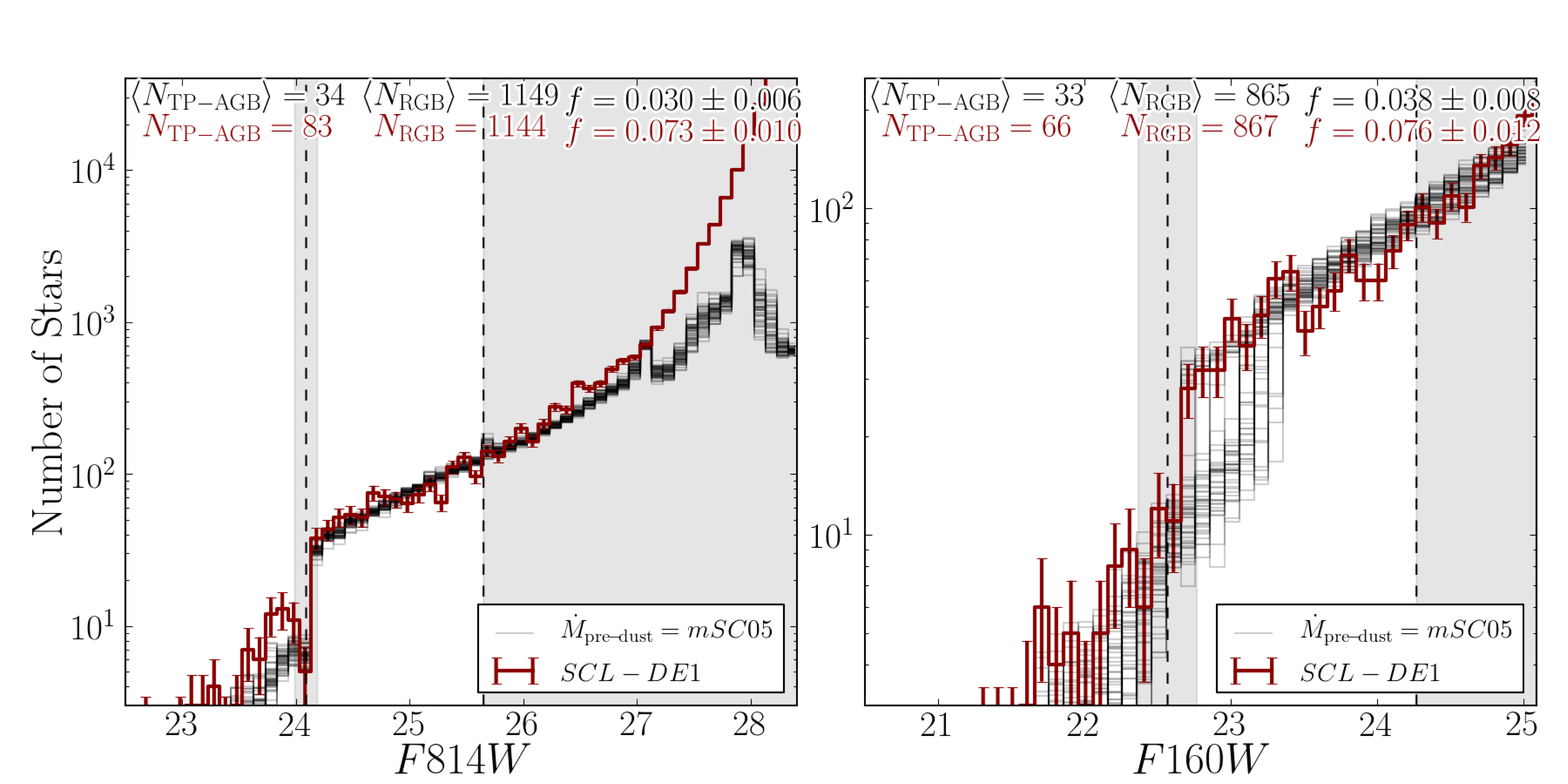}
\caption{Same as \ref{fig_oct13_lfs} but with \NOV.}
\label{fig_nov13_lfs}
\end{figure*}

Figures \ref{fig_oct13_lfs}, \ref{fig_nov13eta0_lfs}, and \ref{fig_nov13_lfs} show a set of panels with optical (left) and IR (right) LFs for each galaxy in the sample. Observations (corrected for completeness; red) are shown with Poisson uncertainties for each of the 50 model LFs per panel overplotted (grey). The regions around the TRGB which were excluded in calculating the \narratio\ ratio are shaded (see Section \ref{sec_agb_rgb_selection}). The fainter magnitudes between 90\%-50\% completeness are also shaded; they are never included in the analysis, though the RGB and RC are important constraints used to derive the SFH and best fit metallicity enrichment law.  

To propagate the completeness and photometric uncertainties from the observations to the model LFs, we make use of the uncertainties reported by MATCH in the derived SFHs (which include uncertainties from the artificial star tests). An example set of SFHs is shown in Figure \ref{fig_random_sfr}.

Qualitatively, the model and data LF agree in the RGB region (between the shaded regions) and diverge at fainter magnitudes. The agreement around the RGB is by design, as discussed in Section \ref{sec_agb_rgb_selection}, because the model LFs are scaled to match the total number of observed RGB stars in this region. However, there are two observational sources that give rise to the disagreement between data and scaled models fainter than $\sim 90\%$ completeness. The first observational source is decreasing completeness with increasing magnitude, and the second cause is due to photometric uncertainties (median uncertainties in each CMD are shown in each panel of Figure \ref{galaxy_sample}).

In general, the model LFs follow the results in Section \ref{sec_res_narratio} for the \narratio. The \NOVeta\ and \OCT\ LFs predict far more TP-AGB stars than are observed in each magnitude bin (the only exception is for bright TP-AGB stars in NGC2976 in the NIR). The \NOV\ model however, shows excellent agreement in the TP-AGB region of LF compared to the observed LF in many cases.

\begin{figure}
\includegraphics[width=\columnwidth]{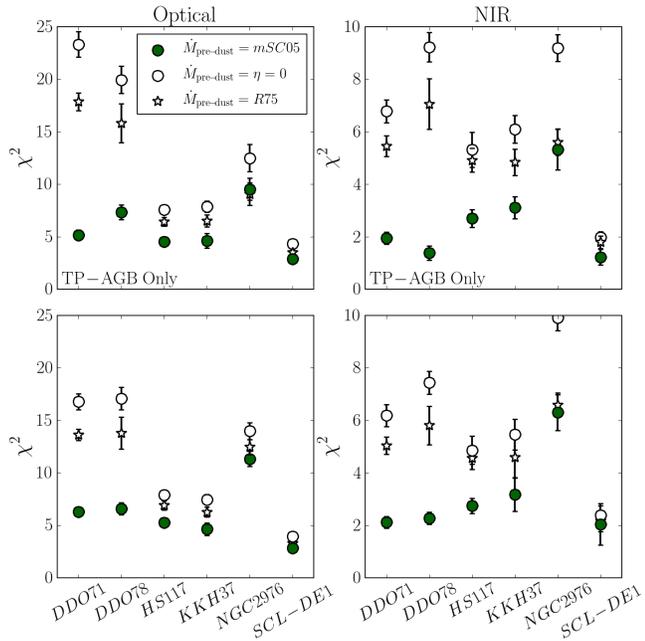}
\caption{The Poission-like $\chi^2$ test for each galaxy LF with each TP-AGB model (left, optical; right, NIR). Bottom panels show the $\chi^2$ value using the LF down to the 90\% completeness limit and top panels show the $\chi^2$ value of only the TP-AGB (brighter than $F814W=0.1$ or $F160W=0.2$ mags above the TRGB.}
\label{fig_chi2}
\end{figure}

To robustly compare one TP-AGB model to another, we calculate the Poission-like $\chi^2$ likelihood that the observational LF is randomly drawn from the model LF. Figure \ref{fig_chi2} shows the results of this calculation for only the TP-AGB region compared to the same region in the data (top panels) and the LF brighter than the 90\% completeness limit (bottom panels) for the optical (left panels) and the NIR (right panels). Each point of Figure \ref{fig_chi2} represents the mean value of the $\chi^2$ calculated individually for each of the 50 model LFs, with uncertainties corresponding to the standard deviation of the mean.

The absolute placement of the $\chi^2$ values for each galaxy are likely dominated by the goodness of the SFH recovery mentioned in Section \ref{sec_sfh_z}. For example, NGC~2976 has the largest effective $\chi^2$, corresponding to the least-good SFH fit, while SCL-DE1 is the opposite. The same trend is seen in Figure \ref{fig_chi2} but using TRILEGAL simulated LFs. Therefore, it is more meaningful to consider the relative $\chi^2$ values in Figure \ref{fig_chi2} to compare TP-AGB models. The $\chi^2$ values follow the qualitative picture from the LF agreement described above. The \NOV\ mass-loss produces the most consistent LF compared to that observed.

\begin{figure*}
\includegraphics[width=.45\textwidth]{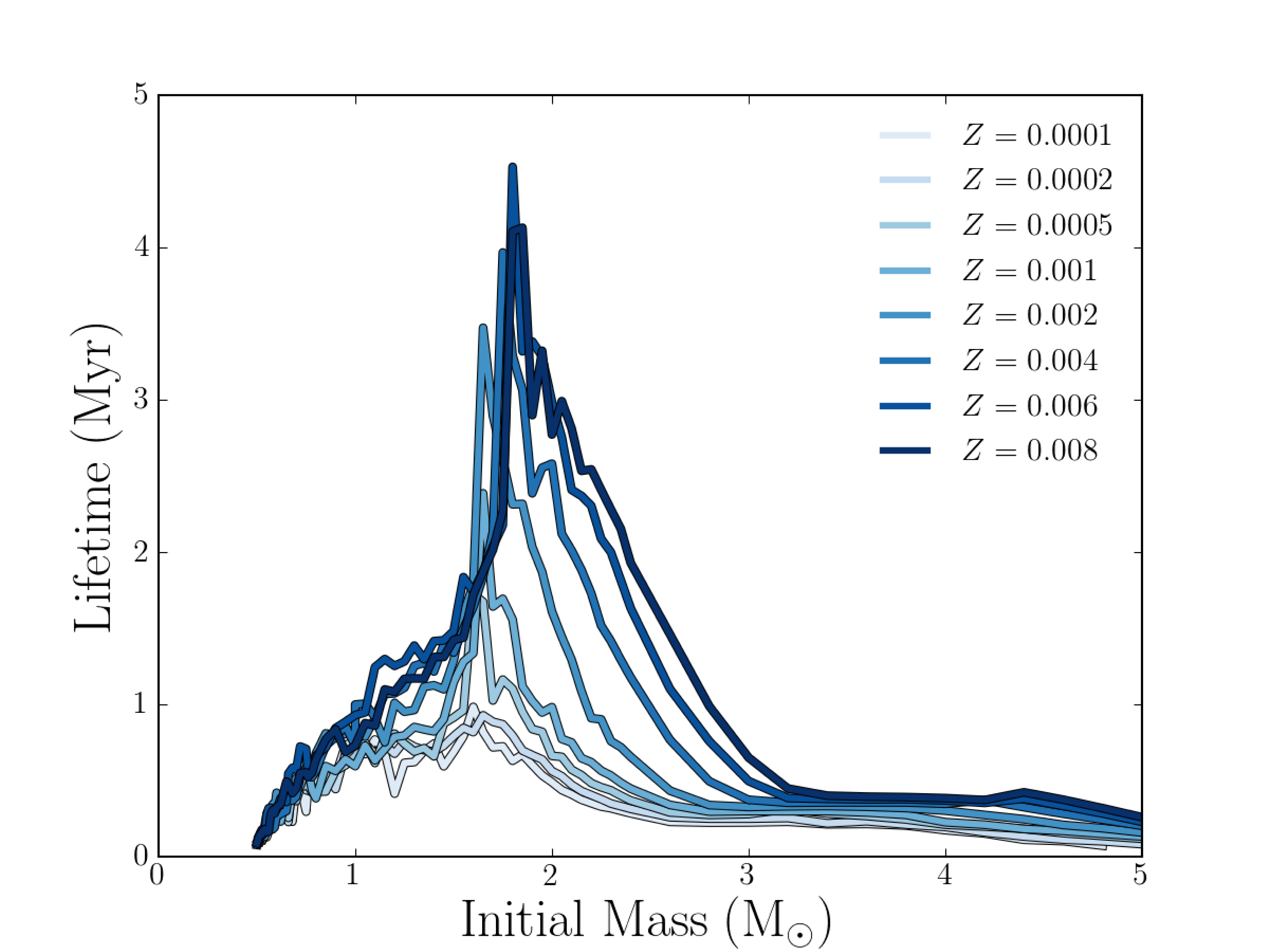}
\includegraphics[width=.45\textwidth]{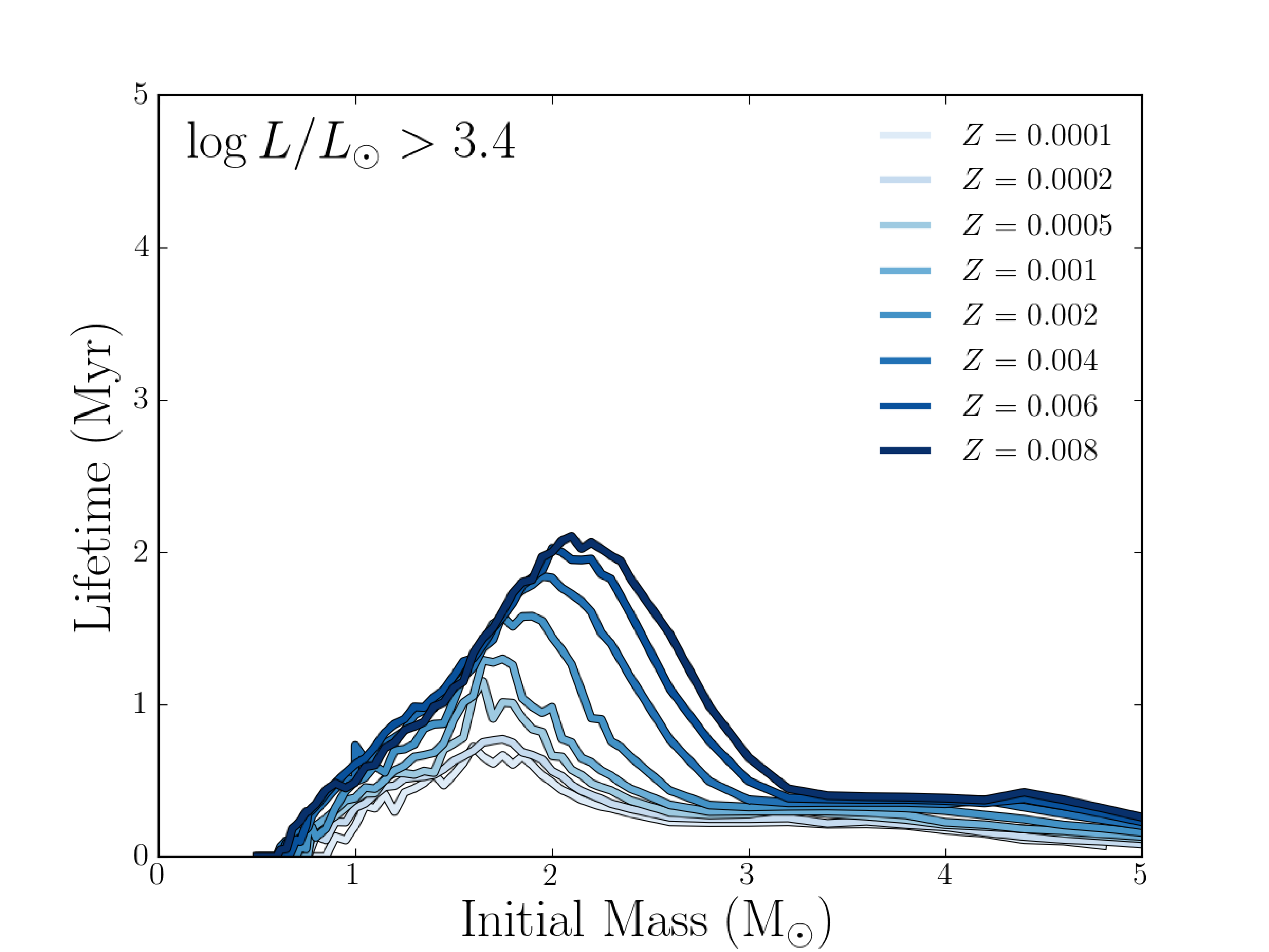}
\caption{The lifetime of TP-AGB stars as a function of initial mass for several metallicities predicted by the \NOV\ model. Left panel shows the total lifetime of the TP-AGB phase. Right panel shows the lifetime of the TP-AGB phase that is brighter than the TRGB, and thus measurable from observations.}
\label{fig_tpagb_lifetimes}
\end{figure*}

\begin{figure*}
\includegraphics[width=0.45\textwidth]{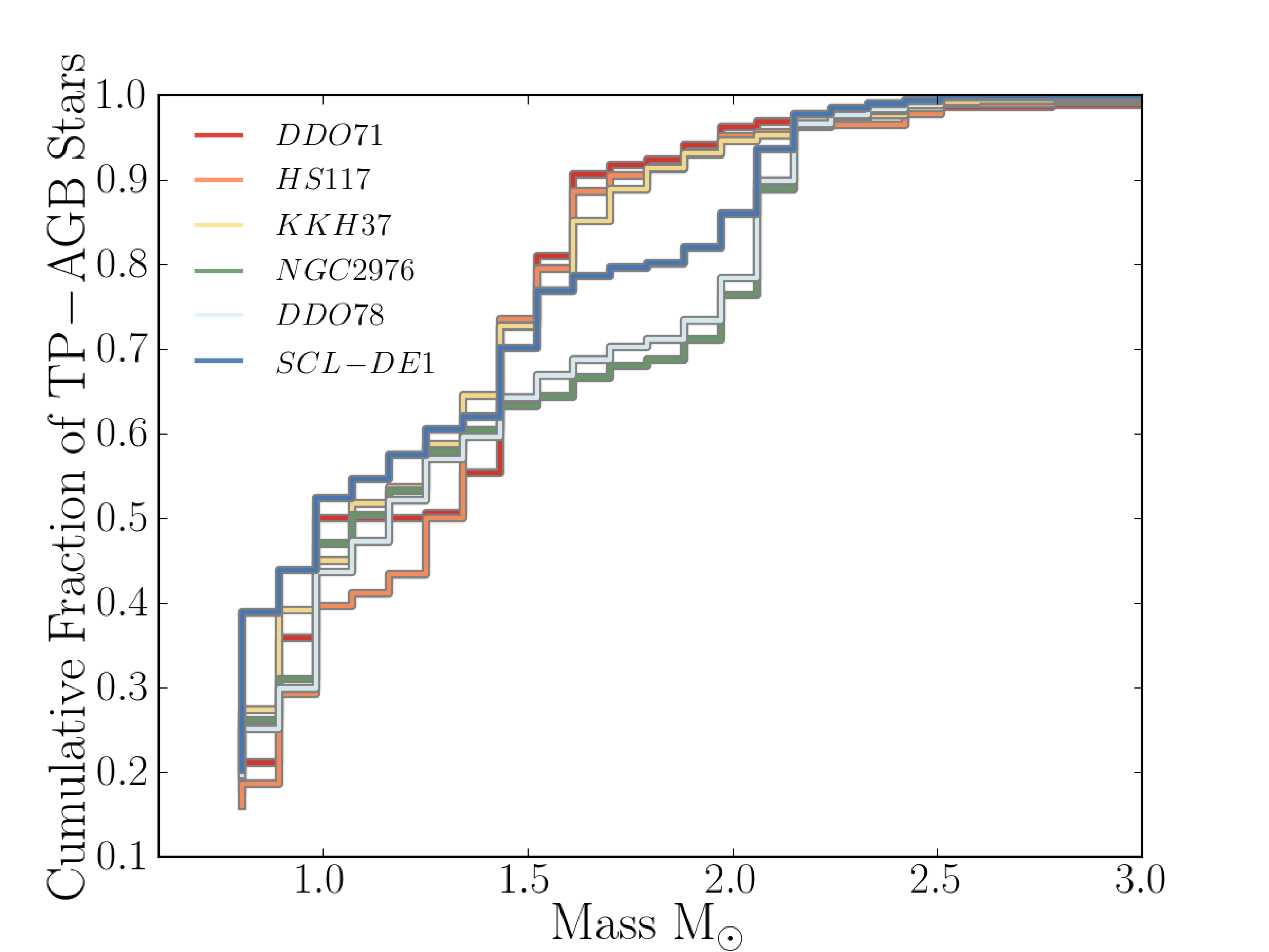}
\includegraphics[width=0.45\textwidth]{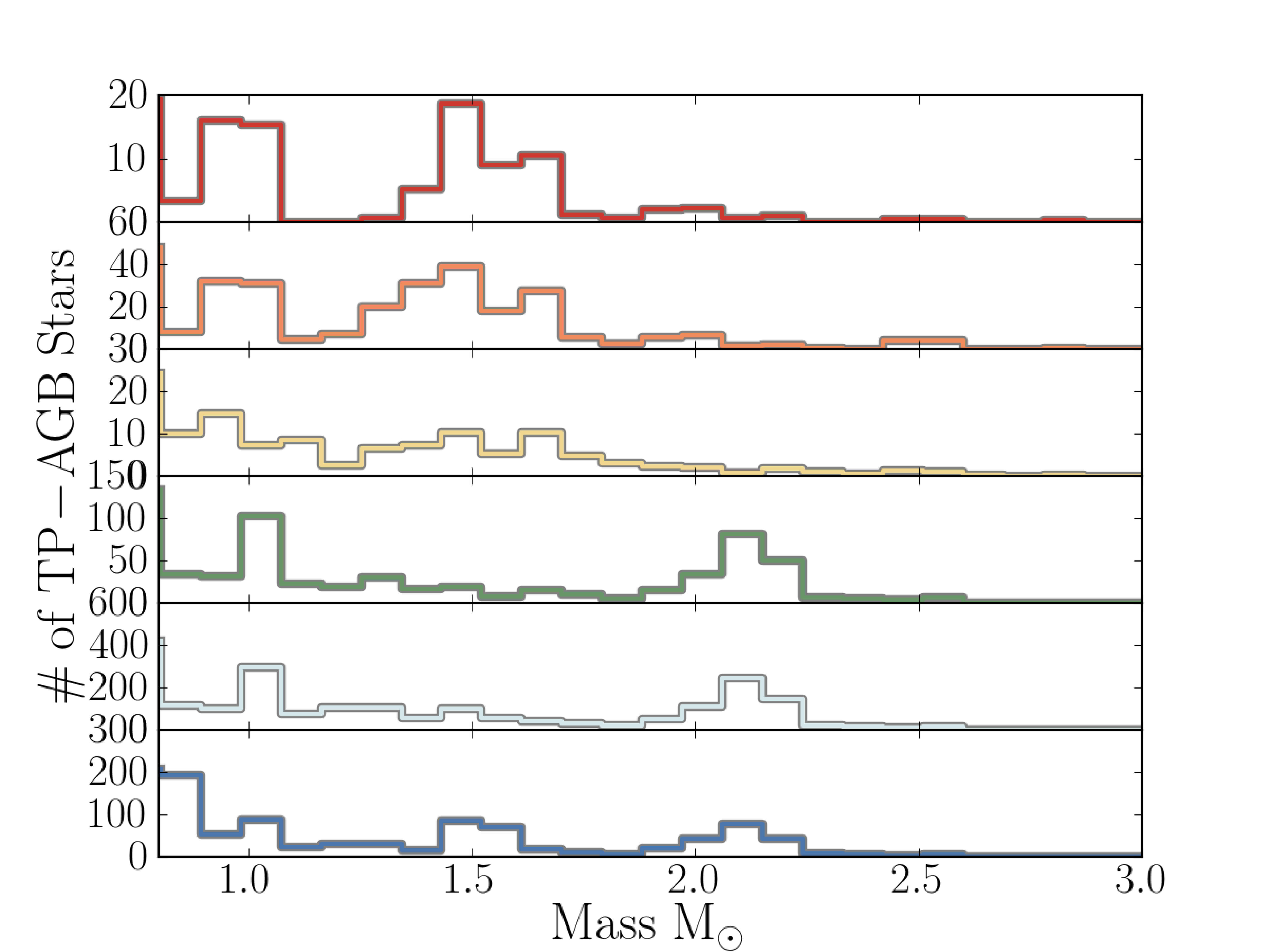}
\caption{Cumulative fraction (left) and histograms (right) of the initial masses of TP-AGB stars produced in the best fitting simulations for each galaxy with the \NOV\ model.}
\label{fig_mass_hist}
\end{figure*}

With the success in the \NOV\ mass-loss prescription we can now constrain the expected lifetimes of low metallicity, low mass TP-AGB stars. Figure \ref{fig_tpagb_lifetimes} shows the corresponding TP-AGB lifetimes for low and intermediate metallicities, for the \NOV\ model. The left panel shows the entire lifetime of the TP-AGB, including the amount of time spent below the TRGB. The right panel shows the expected lifetimes of TP-AGB stars above the TRGB. Figure \ref{fig_mass_hist} shows the distribution of TP-AGB masses from each of the best fitting LFs for each galaxy.  Therefore, based on the number of observed TP-AGB stars in six galaxies that are low metallicity and have little recent SFH, the lifetimes of the typical TP-AGB star in our sample ($\sim0.8<M<2.7\msun, Z<0.002$ or $[Fe/H] \lesssim -0.86$) will be less than 1.2 Myr. For stars of mass $\lesssim1\msun$, we expect the TP-AGB lifetime to be less than half a Myr, as the star will expel much of its atmosphere during the pre-dust driven phase of the TP-AGB.

\subsection{The Initial and Final Mass Relationship}

 As mentioned in G10, one of the clearest constraints on the evolution of low-mass low-metallicity TP-AGB stars is given by the (few) measured masses of white dwarfs (WDs) in globular clusters, and in particular those in M4, for which \citet{Kalirai2009} derived a mean mass of $M_{WD}=0.53\pm0.01 \Msun$. Assuming that M4 has a [Fe/H]=-1.07 and [$\alpha$/Fe]=0.39 dex \citep{Marino2008}, it should be well represented by PARSEC tracks of $Z\simeq0.002$. For this metallicity, the track which lifetime best fits the 12 Gyr age expected for globular clusters is the one with an initial mass of 0.85 \Msun, which takes 11.7 Gyr to evolve from the zero-age MS to the WD cooling ages of the observed M4 white dwarfs \citep[see Table 2 in][]{Kalirai2009}. This track finishes the TP-AGB with a remnant mass of 0.547 \Msun, which is just slightly larger than the mean value determined by \citeauthor{Kalirai2009}. The small difference in final mass can be easily explained by invoking a small additional amount of mass-loss on the RGB. We emphasize however that calibrating the RGB mass-loss to a precision of a few hundredths of solar masses, is certainly beyond the scope of this paper.

\begin{deluxetable*}{lrrrrrrr}
\tabletypesize{\scriptsize}
\tablecaption{Mean Optical \narratio\ Ratios}
\tablehead{
    \colhead{Target} &
    \colhead{$\frac{N_{\rm TP-AGB}}{N_{\rm RGB}}$ \NOV} &
    \colhead{Frac. Difference} &
    \colhead{$\frac{N_{\rm TP-AGB}}{N_{\rm RGB}}$ \NOVeta} & 
    \colhead{Frac. Difference} &
    \colhead{$\frac{N_{\rm TP-AGB}}{N_{\rm RGB}}$ \OCT} &
    \colhead{Frac. Difference} 
}
\startdata  
DDO71 & $0.020\pm0.002$ &  $1.117\pm0.215$ &  $0.068\pm0.004$ &  $3.765\pm0.552$ &  $0.057\pm0.003$ &  $3.147\pm0.459$ \\
DDO78 & $0.085\pm0.007$ &  $0.822\pm0.135$ &  $0.329\pm0.017$ &  $3.196\pm0.424$ &  $0.276\pm0.015$ &  $2.686\pm0.364$ \\
HS117 & $0.178\pm0.027$ &  $1.205\pm0.386$ &  $0.449\pm0.050$ &  $3.037\pm0.853$ &  $0.354\pm0.042$ &  $2.392\pm0.691$ \\
KKH37 & $0.140\pm0.020$ &  $0.666\pm0.180$ &  $0.453\pm0.044$ &  $2.161\pm0.485$ &  $0.363\pm0.037$ &  $1.730\pm0.400$ \\
NGC2976 & $0.021\pm0.002$ &  $0.538\pm0.089$ &  $0.081\pm0.004$ &  $2.061\pm0.249$ &  $0.050\pm0.003$ &  $1.274\pm0.167$ \\
SCL-DE1 & $0.030\pm0.006$ &  $0.392\pm0.133$ &  $0.167\pm0.017$ &  $2.201\pm0.534$ &  $0.135\pm0.015$ &  $1.779\pm0.446$ \\
Mean &  $0.079\pm0.011$ &  $0.790\pm0.190$ &  $0.258\pm0.023$ &  $2.737\pm0.516$ &  $0.206\pm0.019$ &  $2.168\pm0.421$
\enddata
\tablecomments{Mean optical TP-AGB to RGB model ratios and their fractional differences compared to the data. For each galaxy for each TP-AGB model, 50 TRILEGAL simulations produced a model stellar population from the best fit SFH and its uncertainties (see Section \ref{sec_sfh_err}). The only change from one set of simulations to the other is the TP-AGB model. The total fractional differences are calculated compared to the total \narratio\ ratio in the data listed in Table \ref{tab_data_numbers}.}
\label{tab_opt_narratio}
\end{deluxetable*}

\begin{deluxetable*}{lrrrrrrr}
\tabletypesize{\scriptsize}
\tablecaption{Mean NIR \narratio\ Ratios}
\tablehead{
    \colhead{Target} &
    \colhead{$\frac{N_{\rm TP-AGB}}{N_{\rm RGB}}$ \NOV} &
    \colhead{Frac. Difference} &
    \colhead{$\frac{N_{\rm TP-AGB}}{N_{\rm RGB}}$ \NOVeta} & 
    \colhead{Frac. Difference} &
    \colhead{$\frac{N_{\rm TP-AGB}}{N_{\rm RGB}}$ \OCT} &
    \colhead{Frac. Difference} 
}
\startdata
DDO71 & $0.105\pm0.010$ &  $1.292\pm0.271$ &  $0.295\pm0.020$ &  $3.634\pm0.653$ &  $0.257\pm0.018$ &  $3.168\pm0.579$ \\
DDO78 & $0.080\pm0.007$ &  $1.069\pm0.182$ &  $0.211\pm0.012$ &  $2.828\pm0.414$ &  $0.200\pm0.012$ &  $2.685\pm0.394$ \\
HS117 & $0.140\pm0.016$ &  $2.140\pm0.588$ &  $0.253\pm0.024$ &  $3.869\pm0.983$ &  $0.240\pm0.023$ &  $3.670\pm0.938$ \\
KKH37 & $0.108\pm0.014$ &  $0.795\pm0.205$ &  $0.334\pm0.030$ &  $2.459\pm0.530$ &  $0.279\pm0.027$ &  $2.057\pm0.453$ \\
NGC2976 & $0.154\pm0.012$ &  $0.573\pm0.086$ &  $0.561\pm0.030$ &  $2.089\pm0.255$ &  $0.405\pm0.024$ &  $1.509\pm0.193$ \\
SCL-DE1 & $0.038\pm0.008$ &  $0.488\pm0.178$ &  $0.186\pm0.021$ &  $2.360\pm0.643$ &  $0.152\pm0.018$ &  $1.932\pm0.540$ \\
Mean &  $0.104\pm0.011$ &  $1.059\pm0.252$ &  $0.307\pm0.023$ &  $2.873\pm0.580$ &  $0.256\pm0.020$ &  $2.504\pm0.516$ 
\enddata
\tablecomments{Same as Table \ref{tab_opt_narratio} for WFC3/IR data.}
\label{tab_ir_narratio}
\end{deluxetable*}

\section{Conclusions}
\label{sec_conc}

We have extended the analysis from G10, and confirmed that pre-dust mass-loss plays an important role in the TP-AGB evolution of low-mass metal-poor stars. We have shown that neglecting this phase of mass-loss altogether will overpredict the numbers of TP-AGB stars by a factor of $\sim 3$. We also showed that assuming Reimers' scaling relation to describe the pre-dust mass-loss phase will also overpredict the number to TP-AGB stars but by a factor of $\sim2$. Following recent results of detailed models that describe the chromospheric winds of red giants, we implemented a revised mass-loss prescription with a stronger dependence on the effective temperature. Using the \narratio\ ratio and comparing LFs, we showed this implementation reaches good to excellent agreement with the data. Moreover, the \NOV\ mass-loss prescription further lowers the expected TP-AGB lifetime for low-mass, low-metallicity TP-AGB stars.  

An interesting outcome of the our analysis is the dependence on the initial chemical composition: TP-AGB lifetimes of low-mass stars are expected to become shorter at decreasing metallicity. We also show that, given our calibrated mass-loss relation, the efficiency of the third dredge-up has little effect on TP-AGB lifetimes in this regime.

This paper represents the first step in a major calibration of the COLIBRI and PARSEC codes which will aid in the understanding of the physics involved in TP-AGB evolution. Upon the availability of PARSEC v2, which will include higher mass stellar evolution tracks, this study will be extended to include the 17 other galaxies in the AGB-SNAP sample. This addition will expand the metallicities available to test and increase the mass range of our sample to the complete sample of TP-AGB stars. 

These improvements to the TP-AGB models will be included in sets of isochrones and tools to synthesize stellar populations available on the CMD\footnote{http://stev.oapd.inaf.it/cmd} and TRILEGAL\footnote{http://stev.oapd.inaf.it/trilegal} websites.

\section{Acknowledgements}
P.R. acknowledges financial support from {\em Progetto di Ateneo 2012}, ID-CPDA125588/12, University of Padova and the HST-ANGST grant: GO-10915.
Support for DRW is provided by NASA through Hubble Fellowship grants HST-HF-51331.01 awarded by the Space Telescope Science Institute. 
P.M. acknowledges the research leading to these results has received funding from the European Research Council
under the European Union's Seventh Framework Programme (FP/2007-2013) / ERC Grant
Agreement n. 615604.

{\it Facilities:}  \facility{HST (ACS) (WFCP2) (WFC3)}.

\bibliography{tpagb2}

\end{document}